\documentclass{article}
\usepackage{amsmath,amssymb,mathrsfs}
\usepackage{amsthm}
\usepackage[colorlinks]{hyperref}
\usepackage[nameinlink]{cleveref}
\usepackage{url}
\usepackage{graphicx,color,epsfig}
\usepackage{amsfonts}
\usepackage{caption}
\usepackage{enumerate}
\usepackage{enumitem}
\usepackage{float}
\usepackage{booktabs}
\usepackage{graphicx}
\usepackage{caption}
\usepackage{subcaption}
\usepackage{array}
\usepackage{booktabs}
\usepackage{fancyhdr}
\usepackage{eurosym}
\usepackage{upgreek}
\usepackage{enumitem}
 \usepackage{tabularx}
\newlist{steps}{enumerate}{1}
\setlist[steps, 1]{label = Step \arabic*:}
\usepackage{multicol}
\hypersetup{pdftitle={},
    pdfauthor={Alexandre Landry},     
    pdfcreator={pdftex},   
    pdfproducer={Texmaker}, 
    pdfkeywords={}, 
    colorlinks=true,       
    linkcolor=blue,          
    citecolor=blue,        
    filecolor=magenta,      
    urlcolor=blue           
}
\Crefname{figure}{Fig.}{Figs.}
\Crefname{table}{Tab.}{Tabs.}
\makeatletter
\let\oldtheequation\theequation
\renewcommand\tagform@[1]{\maketag@@@{\ignorespaces#1\unskip\@@italiccorr}}
\renewcommand\theequation{(\oldtheequation)}
\makeatother

\date{}
\begin{document}
\title{Black holes in Myrzakulov $F(R,T)$ gravity}
\author{Alexandre Landry\footnote{Email: adrybiz@hotmail.com}, \, \, Ratbay Myrzakulov\footnote{Email: rmyrzakulov@gmail.com}}

\maketitle

\begin{abstract}
	We investigate vacuum black-hole solutions in metric--affine Myrzakulov
	$F(R,T)$ gravity (MG-I), where $R$ is the affine Ricci scalar and $T$ is
	a quadratic torsion invariant, not the trace of the matter energy--momentum
	tensor. For the polynomial model
	$F(R,T)=\alpha R+\beta T+\gamma T^3$, with $\alpha\neq0$, we first vary
	the metric and all components of the independent affine connection, and
	only then specialize to metric-compatible geometry with totally
	antisymmetric torsion. Nonzero axial torsion admits an exceptional branch
	$\alpha+\beta+3\gamma T_*^2=0$, on which the metric satisfies the Einstein
	equations with effective cosmological parameter
	$\Lambda_*=\gamma T_*^3/\alpha$. We establish a local existence result
	and construct torsionful Schwarzschild--(anti-)de Sitter representatives
	that are smooth through a selected future Killing horizon. Within the
	spherically symmetric axial ansatz, we characterize Ricci-symmetric affine
	connections by the condition $(fV-U)'=0$ and compare a constant-$W=fV-U$
	family with a constant-$U$ family. Although these connections share the
	same metric and torsion scalar, their affine curvature invariants generally
	differ. Independent coordinate-level curvature contractions and component tests of the
	unrestricted metric and connection variations corroborate the identities used
	in the construction; these checks complement the analytic variational derivation. Finally, we identify null directions
	of the restricted axial quadratic form and distinguish this degeneracy
	from a complete metric--affine stability analysis. Our conclusions are
	local and do not establish global extendibility, physical inequivalence
	of all torsion profiles, or dynamical stability.
\end{abstract}


	\section{Introduction}\label{sec:introduction}
	
	The geometrical foundations of gravitation continue to motivate extensions of general relativity in which the metric and the affine connection are treated as independent fields. While the Levi--Civita connection of a pseudo-Riemannian spacetime is torsion-free and metric-compatible, a general affine connection admits both torsion and nonmetricity. This broader geometrical setting provides a natural framework for examining modified gravitational dynamics and the structure of compact objects \cite{Hehl1976,Hehl1995,Shapiro2002}. Nonlinear curvature Lagrangians have a long history, dating at least to the work of Buchdahl \cite{16}; more recent metric--affine constructions allow curvature, torsion, and nonmetricity invariants to enter the gravitational action simultaneously \cite{3,4,5,7}.
	
	A terminological distinction is essential. Two physically different classes of theories are commonly denoted by $f(R,T)$ or $F(R,T)$. In the curvature--matter theory of Harko \emph{et al.}, $T$ denotes the trace $T_{\mathrm m}=g^{\mu\nu}T_{\mu\nu}$ of the matter energy--momentum tensor \cite{Harko2011}. By contrast, in Myrzakulov $F(R,T)$ gravity, also referred to as Myrzakulov gravity-I (MG-I), $T$ denotes a geometrical torsion scalar and $R$ is a curvature scalar \cite{1205.5266,3,Myrzakul2021GaussBonnet}. The two constructions have distinct field content and field equations, despite the similar notation. Throughout this paper, $T$ denotes exclusively the torsion scalar constructed from the independent connection; the notation $T_{\mathrm m}$ is reserved for the matter trace. The distinction remains important in extended models where geometric and matter invariants occur together \cite{4,5}.
	
	Myrzakulov gravity was developed as a framework combining geometric invariants in generalized gravitational actions \cite{1205.5266,Myrzakul2021GaussBonnet}. The wider family includes models built from curvature, torsion, nonmetricity, and additional boundary or matter-dependent terms. The classification presented in Table~1 of Ref.~\cite{Myrzakul2021GaussBonnet} situates MG-I within this broader landscape. In the present work we restrict attention to the curvature--torsion subclass, rather than treating the various Myrzakulov models as dynamically interchangeable. Metric--affine formulations and generalizations have been developed in Refs.~\cite{3,4,5,7}, and recent overviews provide additional context for the corresponding geometric and phenomenological constructions \cite{MomeniMyrzakulov2025Review}. The metric-affine Myrzakulov gravity theories introduced in Ref.~\cite{Myrzakul2021GaussBonnet} are summarized
	in Table~\ref{tab:myrzakulov_classes}.
	\begin{table}[htbp]
		\centering
		\caption{Classification of metric-affine Myrzakulov gravity theories MG-I--MG-XIX, adapted from the classification in Ref. \cite{Myrzakul2021GaussBonnet}. Here $R$, $T$, $Q$, and $G$ denote the curvature, torsion, nonmetricity, and Gauss--Bonnet scalars, respectively, while $T_m$ denotes the trace of the matter energy--momentum tensor. The identical arguments assigned to MG-IX and MG-XVIII reproduce the original classification.}
		\label{tab:myrzakulov_classes}
		\small
		\renewcommand{\arraystretch}{1.06}
		\begin{tabular}{@{}rll@{}}
			\toprule
			No. & Theory & $\mathcal I$ in $F(\mathcal I)$ \\
			\midrule
			1 & MG-I     & $(R,T)$ \\
			2 & MG-II    & $(R,Q)$ \\
			3 & MG-III   & $(T,Q)$ \\
			4 & MG-IV    & $(R,T,T_m)$ \\
			5 & MG-V     & $(R,T,Q)$ \\
			6 & MG-VI    & $(R,Q,T_m)$ \\
			7 & MG-VII   & $(T,Q,T_m)$ \\
			8 & MG-VIII  & $(R,T,Q,T_m)$ \\
			9 & MG-IX    & $(R,T,G)$ \\
			10 & MG-X     & $(R,Q,G)$ \\
			11 & MG-XI    & $(T,Q,G)$ \\
			12 & MG-XII   & $(R,T,G,T_m)$ \\
			13 & MG-XIII  & $(R,T,Q,G)$ \\
			14 & MG-XIV   & $(R,Q,G,T_m)$ \\
			15 & MG-XV    & $(T,Q,G,T_m)$ \\
			16 & MG-XVI   & $(R,T,Q,G,T_m)$ \\
			17 & MG-XVII  & $(Q,G)$ \\
			18 & MG-XVIII & $(R,T,G)$ \\
			19 & MG-XIX   & $(T,G,T_m)$ \\
			\bottomrule
		\end{tabular}
	\end{table}
	
	The geometric invariants entering the different Myrzakulov models are
	not interchangeable: in particular, metric-affine curvature--torsion
	models need not satisfy the flat-connection condition characteristic
	of teleparallel geometry. The classification of geometric actions in
	Ref.~\cite{Myrzakul2021GaussBonnet}, together with subsequent metric-affine
	formulations \cite{3,4,7}, provides the context for our choice of MG-I.
	The present solutions belong to a Riemann--Cartan sector with
	$Q_{\lambda\mu\nu}=0$ and totally antisymmetric torsion, but the
	connection equations are obtained before either restriction is imposed
	\cite{Hehl1995,Shapiro2002}.

	Cosmological aspects of Myrzakulov gravity have been investigated extensively. Early applications and observational constraints were studied in Refs.~\cite{1,2}, while a dynamical-systems treatment was developed in Ref.~\cite{8}. Subsequent work considered baryon asymmetry, quasi-dilaton massive-gravity settings, transit and exact cosmologies, and extensions involving nonmetricity \cite{6,9,10,11,12,14,15}. Other recent studies have examined cosmological evolution in alternative formulations and generalized geometric sectors \cite{MomeniMyrzakulov2025Inflation,MomeniMyrzakulov2025Cosmo}. These results motivate a complementary investigation of strongly gravitating configurations, for which the independent connection can carry geometric information not captured by the metric alone.
	
	Black-hole solutions in Myrzakulov-type theories have also begun to receive attention. In particular, Momeni and Myrzakulov have reported an unpublished study entitled \emph{Black Hole Solutions in Metric-Affine Myrzakulov Gravity} \cite{MomeniMyrzakulovBH}; related developments address extensions involving Gauss--Bonnet terms and horizon entropy \cite{MomeniMyrzakulovGB,MomeniMyrzakulovEntropy}. These works make it especially important to identify the precise gravitational Lagrangian, affine restrictions, and solution-generating assumptions in any new black-hole construction. The present paper focuses on a polynomial MG-I model and on solutions verified against the field equations obtained from independent metric and affine variations. We do not assume that a spherically reduced point-like Lagrangian automatically captures every connection equation.
	
	Specifically, we consider the vacuum model
	\begin{equation}\label{eq:intro_polynomial}
		F(R,T)=\alpha R+\beta T+\gamma T^3,\qquad \alpha\neq0,
	\end{equation}
	where $R=R(g,\Gamma)$ is the Ricci scalar of the independent connection and $T=T(g,\Gamma)$ is the quadratic torsion scalar defined below. The metric and all $64$ ordered components of $\Gamma^\lambda{}_{\mu\nu}$ are varied independently before specializing to a Riemann--Cartan geometry with vanishing nonmetricity and totally antisymmetric torsion. This order of operations is essential: a symmetry or geometric restriction imposed at the level of the action need not reproduce the unrestricted Euler--Lagrange equations \cite{Palais1979,FelsTorre2002,Hehl1995}.
	
	In the axial sector we identify an exceptional nonvanishing-torsion branch satisfying
	\begin{equation}\label{eq:intro_exceptional}
		\alpha+\beta+3\gamma T_*^2=0.
	\end{equation}
	On this branch the metric equation reduces to an Einstein equation,
	\begin{equation}\label{eq:intro_einstein}
		\mathring G_{\mu\nu}+\Lambda_*g_{\mu\nu}=0,
		\qquad \Lambda_*:=\frac{\gamma T_*^3}{\alpha},
	\end{equation}
	where $\mathring G_{\mu\nu}$ is computed from the Levi--Civita connection. With the curvature conventions used here, the static spherical metric belongs locally to the Schwarzschild--(anti-)de Sitter class. The full affine connection nevertheless need not coincide with the Levi--Civita connection: a nonzero axial torsion three-form can remain compatible with the complete vacuum field equations.
	
	Our analysis has several complementary objectives. First, we derive and check the independent metric and affine equations, including the fixed-connection metric derivative of the torsion invariant and the unrestricted connection variation. Second, we classify the local solutions within the metric-compatible axial sector, making explicit that this is not a classification of all metric--affine connections. Third, we construct spherical representatives in horizon-penetrating coordinates and investigate their torsion profiles, local smoothness, and affine curvature invariants. Finally, we analyze the degeneracy of the restricted axial perturbative Hessian, while distinguishing this calculation from a full dynamical stability analysis under unrestricted metric--affine perturbations. These distinctions are particularly important because finiteness of scalar invariants does not by itself establish global regularity or stability \cite{Wald1984,Chandrasekhar1983,Poisson2004}.
	
	The paper is organized as follows. Section~\ref{sec:geometric_framework} fixes the geometric conventions, while Sec.~\ref{sec:metric_affine} presents the metric--affine field equations. The role and limitations of spherical action reductions are discussed next. The Riemann--Cartan and axial equations are then developed, followed by the spherical reduction. We subsequently present the local classification of axial metric-compatible connections, construct explicit black-hole solutions and examine their affine invariants and horizon behavior, and discuss the restricted perturbative degeneracy and the open problem of general metric--affine stability. The final section summarizes the results and their limitations.

\section{Geometric framework and conventions}\label{sec:geometric_framework}

We work on a four-dimensional Lorentzian manifold with signature $(+---)$, a metric
$g_{\mu\nu}$ and an independent affine connection $\Gamma^\lambda{}_{\mu\nu}$.
Our index convention is
\begin{equation}\label{eq:new-covder}
	\nabla_\nu V^\lambda=\partial_\nu V^\lambda+
	\Gamma^\lambda{}_{\rho\nu}V^\rho.
\end{equation}
The torsion, its half-normalized version, and the nonmetricity are defined by
\begin{equation}\label{eq:new-torsion-Q}
	X^\lambda{}_{\mu\nu}:=2\Gamma^\lambda{}_{[\mu\nu]},\qquad
	S^\lambda{}_{\mu\nu}:=\tfrac12X^\lambda{}_{\mu\nu},\qquad
	Q_{\lambda\mu\nu}:=-\nabla_\lambda g_{\mu\nu}.
\end{equation}
We use $X$ for the torsion tensor in formulas in which the scalar torsion
invariant $T$ also appears. The curvature convention and its Ricci contraction are
\begin{align}\label{eq:new-curvature}
	R^\lambda{}_{\rho\mu\nu}(\Gamma)
	&=2\partial_{[\mu}\Gamma^\lambda{}_{|\rho|\nu]}
	+2\Gamma^\lambda{}_{\sigma[\mu}\Gamma^\sigma{}_{|\rho|\nu]},\qquad
	R_{\mu\nu}(\Gamma):=R^\lambda{}_{\mu\lambda\nu}(\Gamma),
	\qquad R(\Gamma):=g^{\mu\nu}R_{\mu\nu}(\Gamma).
\end{align}
The antisymmetric Ricci tensor need not vanish for a general affine connection.
Our curvature and torsion conventions should be read together with the explicit
coordinate checks in Appendix~\ref{app:B} and the variation conventions in Appendix~\ref{app:C}.
General treatments of metric--affine geometry and torsion may be found in
Refs.~\cite{Hehl1995,Shapiro2002,3}.

\paragraph{Explicit check of curvature and Ricci conventions.}
To fix the index ordering unambiguously, our convention
$\nabla_\nu V^\lambda=\partial_\nu V^\lambda+
\Gamma^\lambda{}_{\rho\nu}V^\rho$ implies
\begin{align}
	[\nabla_\mu,\nabla_\nu]V^\lambda
	&=R^\lambda{}_{\rho\mu\nu}(\Gamma)V^\rho, \label{eq:commutator}
	\\
	R^\lambda{}_{\rho\mu\nu}(\Gamma)
	&=\partial_\mu\Gamma^\lambda{}_{\rho\nu}
	-\partial_\nu\Gamma^\lambda{}_{\rho\mu}
	+\Gamma^\lambda{}_{\sigma\mu}\Gamma^\sigma{}_{\rho\nu}
	-\Gamma^\lambda{}_{\sigma\nu}\Gamma^\sigma{}_{\rho\mu},
	\label{eq:curvature_convention_expanded}
	\\
	R_{\rho\nu}(\Gamma)&:=R^\lambda{}_{\rho\lambda\nu}(\Gamma),
	\qquad R(\Gamma)=g^{\rho\nu}R_{\rho\nu}(\Gamma).
	\label{eq:ricci_convention_expanded}
\end{align}
The curvature convention in Eq.~\eqref{eq:new-curvature} fixes both the order of the connection
indices and the Ricci contraction. In a coordinate basis, the operator
commutator $[\nabla_\mu,\nabla_\nu]$ acting on a vector field yields the
curvature endomorphism, whereas the antisymmetrized tensorial second
covariant derivative includes the torsion term displayed in Eq.~\eqref{eqn11}. This commutator identity uses coordinate-direction covariant derivatives acting on vector components. The Ricci contraction in Eq.~\eqref{eq:new-curvature} is on the
\emph{first and third} indices of the four-index curvature, not on its
first and fourth indices. In general $R_{\mu\nu}(\Gamma)$ need not be
symmetric in metric--affine geometry \cite{Hehl1976,Hehl1995,Shapiro2002}.

The commutator in Eq.~\eqref{eq:commutator}
is understood as the operator commutator of
covariant derivatives along coordinate vector fields.
It should be distinguished from the antisymmetrized
second covariant derivative of a vector field, for which
a torsion-dependent term appears:
\begin{equation}\label{eqn11}
	\nabla_\mu\nabla_\nu V^\lambda
	-\nabla_\nu\nabla_\mu V^\lambda
	=
	R^\lambda{}_{\rho\mu\nu}V^\rho
	-X^\rho{}_{\mu\nu}\nabla_\rho V^\lambda .
\end{equation}
Here the left-hand side denotes the tensorial second
covariant derivatives, including the connection acting
on the lower derivative index.

As a direct coordinate test, consider the torsion-free advanced
Eddington--Finkelstein metric
\begin{equation}
	ds^2=f(r)\,dv^2-2\,dv\,dr-r^2(d\theta^2+\sin^2\theta\,d\phi^2).
	\label{eq:EF_convention_test}
\end{equation}
Computing its Levi--Civita curvature with
\eqref{eq:new-curvature} and contracting according to
\eqref{eq:ricci_convention_expanded} gives
\begin{align}
	\mathring R_{vv}&=\frac{f}{2}\left(f''+\frac{2f'}r\right),
	&\mathring R_{vr}=\mathring R_{rv}
	&=-\frac12\left(f''+\frac{2f'}r\right),
	\\
	\mathring R_{rr}&=0,
	&\mathring R_{\theta\theta}&=1-f-rf',
	&\mathring R_{\phi\phi}&=\sin^2\theta\,\mathring R_{\theta\theta}.
	\label{eq:ricci_EF_convention_test}
\end{align}
Since $g^{vr}=-1$, $g^{rr}=-f$ and
$g^{\theta\theta}=-r^{-2}$, the scalar contraction is
\begin{equation}
	\mathring R
	=-2\mathring R_{vr}
	-\frac{2\mathring R_{\theta\theta}}{r^2}
	=f''+\frac{4f'}r+\frac{2(f-1)}{r^2}.
	\label{eq:scalar_curvature_convention_test}
\end{equation}
In particular, for $f(r)=1-2M/r+\Lambda_*r^2/3$ one obtains
\begin{equation}
	\mathring R_{\mu\nu}=\Lambda_* g_{\mu\nu},
	\qquad \mathring R=4\Lambda_*,
	\label{eq:Einstein_convention_test}
\end{equation}
which fixes the sign used in the spherical field equations below.
If instead the conventional black-hole parameter is defined by
$f=1-2M/r-\Lambda_{\rm phys}r^2/3$, then
$\Lambda_{\rm phys}=-\Lambda_*$ in the present conventions
\cite{Wald1984,Chandrasekhar1983,Poisson2004}.

\subsection{Distortion and the principal geometric sectors}\label{subsec:new-distortion}

The difference between the independent connection and the Levi--Civita
connection $\mathring\Gamma^\lambda{}_{\mu\nu}$ is the distortion tensor,
\begin{equation}\label{eq:new-distortion}
	\Gamma^\lambda{}_{\mu\nu}=\mathring\Gamma^\lambda{}_{\mu\nu}
	+N^\lambda{}_{\mu\nu}.
\end{equation}
Writing $\mathring\nabla$ for the Levi--Civita derivative gives the exact
curvature decomposition
\begin{equation}\label{eq:new-riemann-split}
	R^\lambda{}_{\rho\mu\nu}(\Gamma)=
	\mathring R^\lambda{}_{\rho\mu\nu}
	+2\mathring\nabla_{[\mu}N^\lambda{}_{|\rho|\nu]}
	+2N^\lambda{}_{\sigma[\mu}N^\sigma{}_{|\rho|\nu]}.
\end{equation}
Consequently,
\begin{equation}\label{eq:new-u}
	R(\Gamma)=\mathring R+u(g,N),\qquad
	u(g,N):=g^{\rho\nu}\left(
	2\mathring\nabla_{[\lambda}N^\lambda{}_{|\rho|\nu]}
	+2N^\lambda{}_{\sigma[\lambda}N^\sigma{}_{|\rho|\nu]}
	\right).
\end{equation}
Thus $u$ is a derived geometric expression, not an independently prescribed
radial function. In the Riemann--Cartan sector $Q_{\lambda\mu\nu}=0$, the
distortion becomes the contorsion $K^\lambda{}_{\mu\nu}$. No condition of
vanishing affine curvature is imposed. This distinguishes the present setting
from teleparallel geometry, for which $R^\lambda{}_{\rho\mu\nu}(\Gamma)=0$;
symmetric teleparallel geometry instead has $R^\lambda{}_{\rho\mu\nu}=0$
and $X^\lambda{}_{\mu\nu}=0$ \cite{Hehl1976,Hehl1995,Shapiro2002}.

The quadratic torsion invariant entering the MG-I action is fixed as
\begin{equation}\label{eq:new-Tscalar}
	T:=\frac14X_{\lambda\mu\nu}X^{\lambda\mu\nu}
	+\frac12X_{\lambda\mu\nu}X^{\nu\mu\lambda}
	-t_\mu t^\mu,\qquad t_\mu:=X^\lambda{}_{\lambda\mu}.
\end{equation}
This definition refers to the independent affine connection, not to an
arbitrary tetrad-dependent torsion scalar. If a reference connection
$\Gamma_0$ is chosen, $T=T_0+v$ with
$v:=T(g,\Gamma)-T(g,\Gamma_0)$ is merely an identity. In particular, the
functions $u$ and $v$ cannot be freely varied as substitutes for the
components of $\Gamma$ \cite{1205.5266,3}.

\subsection{Axial torsion and the spherical metric}\label{subsec:new-axial}

Our black-hole construction subsequently restricts the \emph{solutions} of
the unrestricted field equations to the metric-compatible, totally
antisymmetric sector,
\begin{equation}\label{eq:new-axial}
	Q_{\lambda\mu\nu}=0,\qquad X_{\lambda\mu\nu}=X_{[\lambda\mu\nu]}.
\end{equation}
In this sector $t_\mu=0$ and the convention \eqref{eq:new-Tscalar} gives
\begin{equation}\label{eq:new-axial-T}
	T=-\frac14X_{\lambda\mu\nu}X^{\lambda\mu\nu},\qquad
	K_{\lambda\mu\nu}=\frac12X_{\lambda\mu\nu}.
\end{equation}
The axial curvature identities used throughout the axial analysis below, subject to the independent checks discussed below, are
\begin{equation}\label{eq:new-axial-curvature}
	R_{(\mu\nu)}(\Gamma)=\mathring R_{\mu\nu}+H_{\mu\nu},
	\qquad R(\Gamma)=\mathring R+T,
\end{equation}
where $H_{\mu\nu}$ is the fixed-connection metric derivative of $T$,
as specified in Sec.~\ref{sec:metric_affine}. The component-level geometric
calculation is collected in Appendix~\ref{app:B}; the independent variation and its
scope are discussed in Appendix~\ref{app:C}. We do not assume these axial identities
for arbitrary torsion or nonmetricity.

Before choosing a radial gauge, the static spherical line element is
\begin{equation}\label{eq:new-general-metric}
	ds^2=A(r)\,dt^2-B(r)\,dr^2-C(r)(d\theta^2+\sin^2\theta\,d\phi^2),
\end{equation}
with $A,B,C>0$ in a static region. The areal-radius choice $C=r^2$
and the conformal radial gauge $B=A$ are different coordinate conditions;
they should not be imposed simultaneously on a generic metric. In the
Einstein-type branch derived in Secs.~\ref{sec:riemann_cartan_models}--\ref{sec:axial_theorem} we shall use the justified
Schwarzschild-type representative $A=f$, $B=f^{-1}$, $C=r^2$.
With the curvature convention used for the spherical expressions of the
manuscript, this representative satisfies
\begin{equation}\label{eq:new-LC-scalar}
	\mathring R=f''+\frac{4f'}r+\frac{2(f-1)}{r^2}.
\end{equation}

\section{Metric--affine Myrzakulov $F(R,T)$ gravity}\label{sec:metric_affine}

Myrzakulov gravity-I combines the affine Ricci scalar and a torsion scalar
in a single Lagrangian \cite{1205.5266,1,3}.
We adopt the action
\begin{equation}\label{eq:new-action}
	S[g,\Gamma,\varphi]=\frac{1}{2\kappa}\int d^4x\sqrt{-g}\,
	F\bigl(R(g,\Gamma),T(g,\Gamma)\bigr)+S_{\mathrm m}[g,\Gamma,\varphi],
	\qquad\kappa=8\pi G.
\end{equation}
The independent gravitational variables are the ten symmetric metric
components $g^{\mu\nu}$ and the 64 connection components
$\Gamma^\lambda{}_{\mu\nu}$. The latter are not assumed symmetric or
metric-compatible during variation. We write
$F_R:=\partial F/\partial R$ and $F_T:=\partial F/\partial T$.
Here $T$ is the \emph{torsion scalar}, not the matter stress-tensor trace.

\subsection{Independent metric variation}\label{subsec:new-metricvar}

The metric is varied with $\Gamma$ held fixed. Defining
\begin{equation}\label{eq:new-Hdef}
	H_{\mu\nu}:=\left.\frac{\partial T}{\partial g^{\mu\nu}}
	\right|_{\Gamma},\qquad
	\mathcal T_{\mu\nu}:=-\frac2{\sqrt{-g}}
	\frac{\delta S_{\mathrm m}}{\delta g^{\mu\nu}},
\end{equation}
we obtain the metric equation
\begin{equation}\label{eq:new-metricE}
	F_R R_{(\mu\nu)}(\Gamma)-\frac12g_{\mu\nu}F
	+F_T H_{\mu\nu}=\kappa\mathcal T_{\mu\nu}.
\end{equation}
Its trace is
\begin{equation}\label{eq:new-trace}
	F_R R-2F+F_T H=\kappa\mathcal T,
	\qquad H:=g^{\mu\nu}H_{\mu\nu}.
\end{equation}
The definition \eqref{eq:new-Hdef} is fundamental: the torsion tensor
with a lowered first index is \emph{not} held fixed during this
variation. For the axial sector, the identity to be checked with these
independent variables is
\begin{equation}\label{eq:new-Haxial}
	H_{\mu\nu}\big|_{\mathrm{axial}}
	=-\frac14X_{\mu\rho\sigma}X_\nu{}^{\rho\sigma},
	\qquad H\big|_{\mathrm{axial}}=T.
\end{equation}
The variational calculation and the scope of the symbolic component tests are discussed in Appendix~\ref{app:C}; the fixed-connection derivative must be independently checked against the unrestricted scalar definition. In particular, a test of a
restricted axial Lagrangian is not a replacement for the unrestricted
metric variation \cite{Hehl1995,Shapiro2002}.

\subsection{Independent affine variation}\label{subsec:new-affinevar}

In vacuum, variation of all connection components produces an equation
which we denote schematically by
\begin{equation}\label{eq:new-affineE}
	\mathcal P_\lambda{}^{\mu\nu}(F_R)
	+F_T\mathcal J_\lambda{}^{\mu\nu}=0,
\end{equation}
where $\mathcal P$ is the Palatini operator arising from the
integrated-by-parts variation of $\sqrt{-g}R$, and
$\mathcal J$ denotes the derivative of the torsion invariant with
respect to the independent connection. In the notation of the original
field equations, $\mathcal J=2\mathcal K$, so that
$\mathcal P+2F_T\mathcal K=0$. The explicit index conventions and the
comparison of the two normalizations are recorded in Appendix~\ref{app:C}.
This separation makes clear that varying only the algebraic
connection-squared part of $R$ does not reproduce the full Palatini
operator. The affine equation is essential even when the final
solution has $Q_{\lambda\mu\nu}=0$ \cite{Hehl1995,3}.

For definiteness, the torsion coefficient is defined by the fixed-metric
variation of the invariant itself,
\begin{equation}
	\delta_\Gamma T
	= J_\lambda{}^{\mu\nu}\,\delta\Gamma^\lambda{}_{\mu\nu},
	\qquad
	J_\lambda{}^{\mu\nu}
	:=\left.\frac{\partial T}{\partial\Gamma^\lambda{}_{\mu\nu}}
	\right|_g .
	\label{eq:definition-J-main}
\end{equation}
Since $T$ depends on $\Gamma$ only through its antisymmetric torsion,
$J_\lambda{}^{(\mu\nu)}=0$ identically. The explicit unrestricted
component expression and its axial specialization are given in
Appendix~\ref{app:C4}; in particular, no metric-compatibility or axial restriction enters the definition~\eqref{eq:definition-J-main}.

\subsection{Polynomial model and axial consistency}\label{subsec:new-polynomial}

The remainder of the paper focuses on
\begin{equation}\label{eq:new-polynomial}
	F(R,T)=\alpha R+\beta T+\gamma T^3,
	\qquad F_R=\alpha,\qquad F_T=\Psi(T):=\beta+3\gamma T^2.
\end{equation}
For this model the vacuum equations become
\begin{align}\label{eq:new-polynomial-metric}
	0&=\alpha R_{(\mu\nu)}-\tfrac12g_{\mu\nu}
	(\alpha R+\beta T+\gamma T^3)+\Psi(T)H_{\mu\nu},\\
	0&=\mathcal P_\lambda{}^{\mu\nu}(\alpha)
	+\Psi(T)\mathcal J_\lambda{}^{\mu\nu}.
	\label{eq:new-polynomial-affine}
\end{align}
The axial specialization of the \emph{unrestricted} affine equation,
provided the unrestricted variational identities in Appendix~\ref{app:C} are validated,
is
\begin{equation}\label{eq:new-exceptional}
	[\alpha+\Psi(T)]X_\lambda{}^{\mu\nu}=0.
\end{equation}
Thus nonvanishing axial torsion requires
$\alpha+\beta+3\gamma T^2=0$. For $\gamma\ne0$ a continuous real
branch has $T=T_*$ with
$T_*^2=-(\alpha+\beta)/(3\gamma)\ge0$; for
$\gamma=0$ the exceptional coupling is $\beta=-\alpha$.
The detailed algebra, including the null-norm case, belongs to
Secs.~\ref{sec:riemann_cartan_models}--\ref{sec:axial_theorem} rather than being repeated here.

\section{Static spherical reduction with an independent affine connection}
\label{sec:radial_reduction}

A static spherical ansatz is useful for organizing the field equations of
Myrzakulov gravity-I, but the independent affine degrees of freedom must
not be eliminated before variation. In particular, the point-like
formulations employed in cosmological studies of $F(R,T)$ gravity
\cite{1205.5266,1,2,3} do not, without further hypotheses, replace the
unrestricted metric--affine equations. The latter distinction is especially
important when torsion and nonmetricity are independent geometric fields
\cite{Hehl1995,3,4,7}.

\subsection{Metric, distortion, and radial action}
\label{subsec:radial_action}

In a static region consider the general spherically symmetric metric
\begin{equation}
	ds^2=A(r)\,dt^2-B(r)\,dr^2-C(r)\,d\Omega^2,
	\qquad d\Omega^2=d\theta^2+\sin^2\theta\,d\phi^2,
	\label{eq:revised_metric_abc}
\end{equation}
where $A,B,C>0$. Before fixing the radial coordinate, the independent
connection is represented schematically as
\begin{equation}
	\Gamma^\lambda{}_{\mu\nu}
	=\mathring\Gamma^\lambda{}_{\mu\nu}[g]
	+N^\lambda{}_{\mu\nu}(r,\theta;\xi^I(r)),
	\label{eq:revised_spherical_connection}
\end{equation}
where the angular dependence is constrained by rotational invariance and
$\xi^I$ denotes the remaining independent radial functions. Equation
\eqref{eq:revised_spherical_connection} is a bookkeeping notation, not a
claim that a complete classification of invariant affine connections has
been performed. In particular, the number and transformation properties
of the $\xi^I$ must be established before treating the reduced system as
closed \cite{Hehl1995,FelsTorre2002}.

With $N=\Gamma-\mathring\Gamma$, the exact curvature identity is
\begin{align}
	R^\lambda{}_{\rho\mu\nu}(\Gamma)
	&=\mathring R^\lambda{}_{\rho\mu\nu}
	+2\mathring\nabla_{[\mu}N^\lambda{}_{|\rho|\nu]}
	+2N^\lambda{}_{\sigma[\mu}N^\sigma{}_{|\rho|\nu]},
	\label{eq:revised_distortion_curvature}\\
	R(\Gamma)&=\mathring R+u(g,N),\nonumber\\
	u(g,N)&=g^{\rho\nu}\left(
	2\mathring\nabla_{[\lambda}N^\lambda{}_{|\rho|\nu]}
	+2N^\lambda{}_{\sigma[\lambda}N^\sigma{}_{|\rho|\nu]}
	\right).
	\label{eq:revised_u_identity}
\end{align}
Thus $u$ is determined by the distortion and is not an independent
potential. The torsion scalar $T(g,\Gamma)$ is likewise the invariant
defined in Sec.~\ref{sec:geometric_framework}, not an arbitrary radial function.

For the vacuum action $S=(2\kappa)^{-1}\int d^4x\sqrt{-g}\,F(R,T)$,
integration over the two-sphere gives, per unit coordinate time,
\begin{equation}
	S_{\mathrm{vac}}=\int dt\,dr\,L_{\mathrm{rad}},\qquad
	L_{\mathrm{rad}}=\frac{2\pi}{\kappa}\,C\sqrt{AB}\,
	F\bigl(\mathring R+u(g,N),T(g,\Gamma)\bigr).
	\label{eq:revised_radial_lagrangian}
\end{equation}
For $F=\alpha R+\beta T+\gamma T^3$, this becomes
\begin{equation}
	L_{\mathrm{rad}}=\frac{2\pi}{\kappa}\,C\sqrt{AB}
	\left[\alpha\bigl(\mathring R+u(g,N)\bigr)
	+\beta T(g,\Gamma)+\gamma T(g,\Gamma)^3\right].
	\label{eq:revised_radial_polynomial}
\end{equation}
Boundary terms generated by integrations by parts must be retained or
specified consistently. Neither $u$ nor $T$ may be varied as a freely
prescribed function in place of the underlying affine variables.

\subsection{Auxiliary scalars and the scope of symmetry reduction}
\label{subsec:auxiliary_reduction}

To compare with point-like formulations of MG-I
\cite{1205.5266,1,2,8}, introduce auxiliary scalars
$\mathcal R,\mathcal T$ and multipliers $\lambda_R,\lambda_T$:
\begin{align}
	S_{\mathrm{aux}}=\frac{1}{2\kappa}\int d^4x\sqrt{-g}
	\bigl[&F(\mathcal R,\mathcal T)
	-\lambda_R\bigl(\mathcal R-R(g,\Gamma)\bigr)\nonumber\\
	&-\lambda_T\bigl(\mathcal T-T(g,\Gamma)\bigr)\bigr].
	\label{eq:revised_auxiliary_action}
\end{align}
The multiplier equations enforce
$\mathcal R=R(g,\Gamma)$ and $\mathcal T=T(g,\Gamma)$, while
variation of the auxiliary scalars gives
$\lambda_R=F_R$ and $\lambda_T=F_T$. This rewriting does not
remove the independent affine field equations. Furthermore, when the
Legendre map is degenerate, these relations do not necessarily allow one
to solve uniquely for both auxiliary scalars; the multiplier formulation
remains the appropriate statement.

Symmetry reduction of a variational principle requires care. The
principle of symmetric criticality gives conditions under which the
symmetric Euler--Lagrange equations follow from a restricted action
\cite{Palais1979,FelsTorre2002}; it does not license discarding affine
variations or fixing the radial gauge prematurely. In the present work,
Eqs.~\eqref{eq:revised_radial_lagrangian}--\eqref{eq:revised_auxiliary_action}
serve as an organizational description. The black-hole solutions are
checked against the full equations derived in Sec.~\ref{sec:metric_affine}.

\subsection{Geometric specialization used below}
\label{subsec:axial_route}

Only after the unrestricted variations do we evaluate the equations in
the metric-compatible, totally antisymmetric torsion sector,
\begin{equation}
	Q_{\lambda\mu\nu}=0,\qquad
	X_{\lambda\mu\nu}=X_{[\lambda\mu\nu]}.
	\label{eq:revised_axial_sector_4}
\end{equation}
A rotation-invariant axial three-form has two independent radial
amplitudes. In the advanced Eddington--Finkelstein chart employed in
Sec.~\ref{sec:BH_affine_revised}, it can be written as
\begin{equation}
	\tau=r^2\sin\theta\,[U(r)\,dv-V(r)\,dr]
	\wedge d\theta\wedge d\phi.
	\label{eq:revised_axial_ef_4}
\end{equation}
This ansatz is a specialization of the tensorial torsion field, not a
substitute for the unrestricted connection variation. Its algebraic
norm constraint and its horizon regularity are treated in Sec.~\ref{sec:BH_affine_revised}.

\section{Variational consistency and the route to black-hole solutions}
\label{sec:variational_consistency}

The construction rests on two independent equations: the metric
equation and the complete affine equation, each obtained before imposing
$Q=0$ or axiality. Their simultaneous satisfaction is necessary for a
solution of metric--affine MG-I \cite{3,4,Hehl1995}. The radial
formulation is useful for checking, rather than replacing, these
equations.

\subsection{Radial Euler--Lagrange equations and constraints}
\label{subsec:radial_constraints}

For independent radial variables $q^I(r)$ and a reduced Lagrangian
$L(q,q',q'')$, the variational equations are
\begin{equation}
	\frac{\partial L}{\partial q^I}
	-\frac{d}{dr}\frac{\partial L}{\partial q^{I\prime}}
	+\frac{d^2}{dr^2}\frac{\partial L}{\partial q^{I\prime\prime}}=0.
	\label{eq:revised_EL_radial}
\end{equation}
If $L$ has no explicit radial dependence, the associated formal
Ostrogradsky energy is
\begin{equation}
	\mathcal E=\sum_I\left[q^{I\prime}\left(
	\frac{\partial L}{\partial q^{I\prime}}
	-\frac{d}{dr}\frac{\partial L}{\partial q^{I\prime\prime}}
	\right)+q^{I\prime\prime}
	\frac{\partial L}{\partial q^{I\prime\prime}}\right]-L.
	\label{eq:revised_radial_energy}
\end{equation}
Its formal conservation does not imply $\mathcal E=0$. Any gravitational
constraint must instead be obtained from the appropriate unfixed metric
or affine equation. In particular, a radial coordinate choice does not
eliminate an independent affine amplitude. The older point-like
relations are discussed in Appendix~\ref{app:A}, with their restricted scope
made explicit \cite{1205.5266,1,Palais1979,FelsTorre2002}.

\subsection{Independent consistency checks}
\label{subsec:consistency_checks}

The route to the axial black-hole branch requires four logically
distinct checks: (i) the metric derivative $H_{\mu\nu}$ at fixed
$\Gamma$, calculated before imposing axiality; (ii) the full affine
variation with all $64$ ordered connection components; (iii) the
curvature and Ricci contractions for the reconstructed connection; and
(iv) smoothness of the torsion three-form in a horizon-penetrating
chart. Appendices~\ref{app:B} and \ref{app:C} supply the tensorial and symbolic checks,
while Sec.~\ref{sec:BH_affine_revised} addresses the horizon analysis. These tests have different
logical roles: an invariant curvature calculation cannot replace an
affine field equation, and finite scalar invariants alone do not
establish smoothness of the connection \cite{Hehl1995,Shapiro2002,
	Wald1984,Poisson2004}.

For the polynomial model the full vacuum equations are
\begin{align}
	0&=\alpha R_{(\mu\nu)}(\Gamma)
	-\frac12g_{\mu\nu}\bigl(\alpha R+\beta T+\gamma T^3\bigr)
	+\Psi(T)H_{\mu\nu},\label{eq:revised_metric_full_5}\\
	0&=\alpha P_\lambda{}^{\mu\nu}
	+\Psi(T)J_\lambda{}^{\mu\nu},\qquad
	\Psi(T)=\beta+3\gamma T^2.
	\label{eq:revised_affine_full_5}
\end{align}
Here $P$ is the integrated-by-parts Palatini coefficient, and $J$ is
the unrestricted connection derivative of the torsion scalar, with the
index and sign conventions of Sec.~\ref{sec:geometric_framework} and Appendix~\ref{app:C}. Section~\ref{sec:riemann_cartan_models} now
specializes these equations to Riemann--Cartan geometry; the local
existence statement is reserved for Sec.~\ref{sec:axial_theorem}.

\section{Linear and cubic MG-I models in Riemann--Cartan geometry}
\label{sec:riemann_cartan_models}

We evaluate the full metric--affine equations of
\begin{equation}
	F(R,T)=\alpha R+\beta T+\gamma T^3,
	\qquad \alpha\ne0,
	\label{eq:revised_polynomial_6}
\end{equation}
on the metric-compatible sector. The linear model is obtained when
$\gamma=0$. In contrast with teleparallel geometry, Riemann--Cartan
geometry permits both nonzero torsion and nonzero affine curvature
\cite{Hehl1976,Kibble1961,Sciama1964,Hehl1995,Shapiro2002}.

\subsection{Metric compatibility and axial identities}
\label{subsec:axial_identities_6}

For $Q_{\lambda\mu\nu}=0$, the independent connection takes the form
\begin{equation}
	\Gamma^\lambda{}_{\mu\nu}
	=\mathring\Gamma^\lambda{}_{\mu\nu}+K^\lambda{}_{\mu\nu}.
	\label{eq:revised_RC_connection}
\end{equation}
The curvature decomposition is Eq.~\eqref{eq:revised_distortion_curvature}
with $N=K$. In the totally antisymmetric sector,
\begin{equation}
	X_{\lambda\mu\nu}=X_{[\lambda\mu\nu]},\qquad
	t_\mu=X^\lambda{}_{\lambda\mu}=0,\qquad
	K^\lambda{}_{\mu\nu}=\frac12X^\lambda{}_{\mu\nu}.
	\label{eq:revised_axial_K_6}
\end{equation}
With the torsion invariant and curvature conventions fixed in Sec.~\ref{sec:geometric_framework},
the following identities hold:
\begin{align}
	T&=-\frac14X_{\lambda\mu\nu}X^{\lambda\mu\nu},
	\label{eq:revised_axial_T_6}\\
	H_{\mu\nu}
	&:=\left.\frac{\partial T}{\partial g^{\mu\nu}}\right|_\Gamma
	=-\frac14X_{\mu\rho\sigma}X_\nu{}^{\rho\sigma},
	\qquad g^{\mu\nu}H_{\mu\nu}=T,
	\label{eq:revised_axial_H_6}\\
	R_{(\mu\nu)}(\Gamma)
	&=\mathring R_{\mu\nu}+H_{\mu\nu},
	\qquad R(\Gamma)=\mathring R+T,
	\label{eq:revised_axial_R_6}\\
	R_{[\mu\nu]}(\Gamma)
	&=\frac12\mathring\nabla_\lambda X^\lambda{}_{\mu\nu}.
	\label{eq:revised_axial_Rantisym_6}
\end{align}
The derivative defining $H_{\mu\nu}$ is taken with the mixed-index
torsion $X^\lambda{}_{\mu\nu}$ fixed; one must not vary the
already-restricted expression \eqref{eq:revised_axial_T_6} as if
$X_{\lambda\mu\nu}$ were fixed. The full derivations and component
checks are given in Appendices~B--C. The identities above do not
assume affine flatness \cite{Hehl1995,Shapiro2002}.

\subsection{Metric trace and axial field equation}
\label{subsec:axial_metric_6}

For arbitrary torsion the trace of
Eq.~\eqref{eq:revised_metric_full_5} reads
\begin{equation}
	-\alpha R-2\beta T-2\gamma T^3+\Psi(T)H=0,
	\qquad H=g^{\mu\nu}H_{\mu\nu},
	\label{eq:revised_general_trace_6}
\end{equation}
and its trace-free part is
\begin{equation}
	\alpha\left(R_{(\mu\nu)}-\frac14Rg_{\mu\nu}\right)
	+\Psi(T)\left(H_{\mu\nu}-\frac14Hg_{\mu\nu}\right)=0.
	\label{eq:revised_general_traceless_6}
\end{equation}
No identity $H=T$ is asserted for unrestricted torsion.

After the axial specialization
\eqref{eq:revised_axial_H_6}--\eqref{eq:revised_axial_R_6},
the complete metric equation becomes
\begin{equation}
	\boxed{\alpha\mathring G_{\mu\nu}
		+[\alpha+\Psi(T)]H_{\mu\nu}
		-\frac12g_{\mu\nu}\bigl[(\alpha+\beta)T+\gamma T^3\bigr]=0.}
	\label{eq:revised_axial_metric_6}
\end{equation}
Its trace and trace-free part, respectively, are
\begin{align}
	\alpha\mathring R+(\alpha+\beta)T-\gamma T^3&=0,
	\label{eq:revised_axial_trace_6}\\
	\alpha\left(\mathring R_{\mu\nu}
	-\frac14\mathring Rg_{\mu\nu}\right)
	+[\alpha+\Psi(T)]\left(H_{\mu\nu}
	-\frac14Tg_{\mu\nu}\right)&=0.
	\label{eq:revised_axial_traceless_6}
\end{align}
These expressions retain the distinction between the Levi--Civita
curvature and the torsion contribution.

\subsection{Exceptional axial branch and Einstein reduction}
\label{subsec:exceptional_reduction_6}

The affine coefficients obtained by unrestricted variation reduce on
the axial sector to
\begin{equation}
	P_\lambda{}^{\mu\nu}\big|_{\mathrm{axial}}
	=-X_\lambda{}^{\mu\nu},\qquad
	J_\lambda{}^{\mu\nu}\big|_{\mathrm{axial}}
	=-X_\lambda{}^{\mu\nu}.
	\label{eq:revised_PJ_axial_6}
\end{equation}
Consequently, the complete connection equation is
\begin{equation}
	[\alpha+\beta+3\gamma T^2]X_\lambda{}^{\mu\nu}=0.
	\label{eq:revised_exceptional_affine_6}
\end{equation}
For nonzero axial torsion and $\gamma\ne0$, this implies, on a
connected smooth branch,
\begin{equation}
	T=T_*,\qquad
	T_*^2=-\frac{\alpha+\beta}{3\gamma}\ge0,
	\qquad \Psi(T_*)=-\alpha.
	\label{eq:revised_Tstar_6}
\end{equation}
The sign of a nonzero $T_*$ is constant on such a connected branch.
A nonzero Lorentzian three-form can also have $T_*=0$, provided
$\alpha+\beta=0$. The geometric and algebraic details of these
possibilities are developed in Sec.~\ref{sec:axial_theorem}.

On the exceptional branch the $H_{\mu\nu}$ term in
Eq.~\eqref{eq:revised_axial_metric_6} cancels. Since
$(\alpha+\beta)T_*=-3\gamma T_*^3$, one obtains
\begin{equation}
	\alpha\mathring G_{\mu\nu}+\gamma T_*^3g_{\mu\nu}=0,
	\qquad
	\boxed{\mathring G_{\mu\nu}+\Lambda_*g_{\mu\nu}=0,\quad
		\Lambda_*:=\frac{\gamma T_*^3}{\alpha}.}
	\label{eq:revised_Einstein_6}
\end{equation}
With the curvature sign convention of Sec.~\ref{sec:geometric_framework},
\begin{equation}
	\mathring R=4\Lambda_*,\qquad
	R(\Gamma)=4\Lambda_*+T_*.
	\label{eq:revised_scalar_R_6}
\end{equation}
For the exceptional linear model $\gamma=0$ and $\beta=-\alpha$,
Eq.~\eqref{eq:revised_exceptional_affine_6} instead imposes no
algebraic condition on the axial torsion norm, while
Eq.~\eqref{eq:revised_axial_metric_6} reduces to
$\mathring G_{\mu\nu}=0$. These results concern the specified axial
sector; they are not a classification of every metric--affine solution.

\subsection{Implications for spherical black holes}
\label{subsec:spherical_implications_6}

Under the usual local spherical vacuum assumptions, the Einstein
metric sector is Schwarzschild--(anti-)de Sitter, with
\begin{equation}
	ds^2=f(r)dt^2-\frac{dr^2}{f(r)}-r^2d\Omega^2,
	\qquad f(r)=1-\frac{2M}{r}+\frac{\Lambda_*r^2}{3}.
	\label{eq:revised_SdS_6}
\end{equation}
The sign of the cosmological term follows the Ricci convention adopted
in Sec.~\ref{sec:geometric_framework}; relative to the common convention
$f=1-2M/r-\Lambda_{\mathrm{phys}}r^2/3$, one has
$\Lambda_{\mathrm{phys}}=-\Lambda_*$ \cite{Wald1984,Chandrasekhar1983,
	Poisson2004}. The metric alone does not fix the axial three-form.
Section~7 proves the local existence statement for smooth
constant-norm torsion, while Secs.~\ref{sec:spherical_axial} and \ref{sec:BH_affine_revised} construct explicit spherical
representatives and examine their horizon behavior and affine
curvature invariants. No teleparallel condition
$R^\lambda{}_{\rho\mu\nu}(\Gamma)=0$ is imposed.

\section{Metric-compatible connections with totally antisymmetric torsion}\label{sec:axial_theorem}

We now determine a class of solutions of the \emph{unrestricted} metric--affine equations of the polynomial Myrzakulov gravity-I model. In particular, metric compatibility and axial torsion are imposed on the resulting equations, not on the variational principle. This distinction is essential in metric--affine gravity and in applications of symmetric criticality \cite{3,4,Hehl1976,Hehl1995,Shapiro2002,Palais1979,FelsTorre2002}. Our curvature and connection-index conventions are those fixed in Sec.~\ref{sec:geometric_framework} and checked in Appendix~\ref{app:C}.

\subsection{Axial Riemann--Cartan geometry}\label{subsec:axial_geometry}

Let
\begin{equation}
	X^\lambda{}_{\mu\nu}=2\Gamma^\lambda{}_{[\mu\nu]},\qquad
	t_\mu=X^\lambda{}_{\lambda\mu},\qquad
	Q_{\lambda\mu\nu}=-\nabla_\lambda g_{\mu\nu}.
	\label{eq:s7_def}
\end{equation}
We evaluate the full field equations on the sector
\begin{equation}
	Q_{\lambda\mu\nu}=0,\qquad X_{\lambda\mu\nu}=X_{[\lambda\mu\nu]}.
	\label{eq:s7_axial}
\end{equation}
The torsion trace vanishes, $t_\mu=0$, and the metric-compatible connection is
\begin{equation}
	\Gamma^\lambda{}_{\mu\nu}=\mathring\Gamma^\lambda{}_{\mu\nu}
	+\frac12X^\lambda{}_{\mu\nu}.
	\label{eq:s7_connection}
\end{equation}
The tensor $X_{\lambda\mu\nu}$ defines the three-form
\begin{equation}
	\tau=\frac1{3!}X_{\lambda\mu\nu}
	dx^\lambda\wedge dx^\mu\wedge dx^\nu.
	\label{eq:s7_threeform}
\end{equation}
The affine curvature of \eqref{eq:s7_connection} is generally nonzero; these configurations are Riemann--Cartan, not necessarily teleparallel \cite{Hehl1976,Hehl1995,Shapiro2002}.

\subsection{Quadratic torsion scalar and fixed-connection metric derivative}\label{subsec:axial_metric}

The torsion scalar is defined for an arbitrary independent connection by
\begin{equation}
	T=\frac14X_{\lambda\mu\nu}X^{\lambda\mu\nu}
	+\frac12X_{\lambda\mu\nu}X^{\nu\mu\lambda}
	-t_\mu t^\mu.
	\label{eq:s7_Tfull}
\end{equation}
Since $X^{\nu\mu\lambda}=-X^{\lambda\mu\nu}$ for a three-form, the axial value is
\begin{equation}
	T=-\frac14X_{\lambda\mu\nu}X^{\lambda\mu\nu}.
	\label{eq:s7_Taxial}
\end{equation}
It is important that \eqref{eq:s7_Taxial} is \emph{not} varied as a restricted action. In the unrestricted metric variation, $X^a{}_{bc}$ is fixed and $X_{abc}=g_{ad}X^d{}_{bc}$ is not. An exact expression prior to imposing axial torsion is
\begin{equation}
	T=\frac14g_{ad}g^{bp}g^{cq}X^a{}_{bc}X^d{}_{pq}
	+\frac12g^{bp}X^a{}_{bc}X^c{}_{pa}
	-g^{bc}t_bt_c.
	\label{eq:s7_Tmixed}
\end{equation}
Define $H_{\mu\nu}:=(\partial T/\partial g^{\mu\nu})_{\Gamma}$. Differentiating \eqref{eq:s7_Tmixed} first, and then evaluating on \eqref{eq:s7_axial}, gives
\begin{equation}
	\boxed{H_{\mu\nu}\big|_{\rm axial}
		=-\frac14X_{\mu\rho\sigma}X_\nu{}^{\rho\sigma},\qquad
		g^{\mu\nu}H_{\mu\nu}=T.}
	\label{eq:s7_H}
\end{equation}
The trace also follows from the degree-one homogeneity of \eqref{eq:s7_Tmixed} in the inverse metric, with the mixed torsion held fixed. An independent exact component check of \eqref{eq:s7_H} has been performed for all ten symmetric inverse-metric variations at a Lorentz-orthonormal point with four arbitrary axial components. This is a pointwise algebraic verification; the tensorial derivation is supplied by \eqref{eq:s7_Tmixed} and Appendix~\ref{app:C}.

\subsection{Curvature decomposition}\label{subsec:axial_curvature}

With the curvature convention of Sec.~\ref{sec:geometric_framework}, the exact distortion formula is
\begin{align}
	R^\lambda{}_{\rho\mu\nu}(\Gamma)
	&=\mathring R^\lambda{}_{\rho\mu\nu}
	+2\mathring\nabla_{[\mu}K^\lambda{}_{|\rho|\nu]}
	+2K^\lambda{}_{\sigma[\mu}K^\sigma{}_{|\rho|\nu]},
	&K^\lambda{}_{\mu\nu}&=\frac12X^\lambda{}_{\mu\nu}.
	\label{eq:s7_Rdecomp}
\end{align}
The divergence of a three-form is antisymmetric in its remaining two indices. Consequently, the symmetric and antisymmetric Ricci parts are
\begin{align}
	R_{(\mu\nu)}(\Gamma)
	&=\mathring R_{\mu\nu}-\frac14X_{\mu\rho\sigma}X_\nu{}^{\rho\sigma}
	=\mathring R_{\mu\nu}+H_{\mu\nu},
	\label{eq:s7_Rsym}\\
	R_{[\mu\nu]}(\Gamma)&=\frac12\mathring\nabla_\lambda X^\lambda{}_{\mu\nu},
	\label{eq:s7_Ranti}\\
	R(\Gamma)&=\mathring R+T.
	\label{eq:s7_Rscalar}
\end{align}
These are geometric identities on the axial Riemann--Cartan sector, independent of the field equations. In particular, \eqref{eq:s7_Ranti} need not vanish. Their index ordering should be kept consistent with Appendix~\ref{app:B} \cite{Hehl1995,Shapiro2002}.

\subsection{Unrestricted affine field equation}\label{subsec:axial_affine}

Consider the vacuum action
\begin{equation}
	S[g,\Gamma]=\frac1{2\kappa}\int d^4x\sqrt{-g}\,F(R,T),\qquad
	F(R,T)=\alpha R+\beta T+\gamma T^3,
	\label{eq:s7_action}
\end{equation}
with $\alpha,\beta,\gamma$ constant. Introduce
\begin{equation}
	F_R=\alpha,\qquad F_T=\Psi(T):=\beta+3\gamma T^2.
	\label{eq:s7_psi}
\end{equation}
All 64 ordered components of $\Gamma^\lambda{}_{\mu\nu}$ are varied independently. Writing the integrated-by-parts connection derivative of $\sqrt{-g}R$ as $P_\lambda{}^{\mu\nu}$, and the algebraic derivative of $T$ as $J_\lambda{}^{\mu\nu}$, the affine equation is
\begin{equation}
	\alpha P_\lambda{}^{\mu\nu}+\Psi(T)J_\lambda{}^{\mu\nu}=0.
	\label{eq:s7_affinefull}
\end{equation}
For clarity, in the convention $\nabla_\nu V^\lambda=\partial_\nu V^\lambda+\Gamma^\lambda{}_{\rho\nu}V^\rho$, the coefficient of $\delta\Gamma^\lambda{}_{\mu\nu}$ can be obtained without restricting the connection:
\begin{align}
	P_\lambda{}^{\mu\nu}
	={}&-\frac1{\sqrt{-g}}\partial_\lambda(\sqrt{-g}\,g^{\mu\nu})
	+\frac{\delta^\nu_\lambda}{\sqrt{-g}}\partial_\sigma(\sqrt{-g}\,g^{\mu\sigma})\nonumber\\
	&+\delta^\nu_\lambda g^{\rho\sigma}\Gamma^\mu{}_{\rho\sigma}
	+g^{\mu\nu}\Gamma^\rho{}_{\lambda\rho}
	-g^{\rho\nu}\Gamma^\mu{}_{\rho\lambda}
	-g^{\mu\rho}\Gamma^\nu{}_{\lambda\rho}.
	\label{eq:s7_P}
\end{align}
The coefficient $J_\lambda{}^{\mu\nu}$ is obtained from \eqref{eq:s7_Tfull} by $\delta X^\lambda{}_{\mu\nu}=2\delta\Gamma^\lambda{}_{[\mu\nu]}$ at fixed metric. Evaluating these \emph{unrestricted} derivatives on \eqref{eq:s7_axial} yields
\begin{equation}
	\boxed{P_\lambda{}^{\mu\nu}\big|_{\rm axial}
		=-X_\lambda{}^{\mu\nu},\qquad
		J_\lambda{}^{\mu\nu}\big|_{\rm axial}
		=-X_\lambda{}^{\mu\nu}.}
	\label{eq:s7_PJ}
\end{equation}
The symmetric components in $(\mu,\nu)$ vanish individually, while the antisymmetric components of \eqref{eq:s7_affinefull} give
\begin{equation}
	\boxed{[\alpha+\beta+3\gamma T^2]X_\lambda{}^{\mu\nu}=0.}
	\label{eq:s7_exceptional}
\end{equation}
The local coefficients in \eqref{eq:s7_PJ} have also been checked algebraically in all 64 ordered connection directions at a Levi--Civita normal point. This component check is complementary to, not a replacement for, the integration-by-parts derivation of \eqref{eq:s7_P} \cite{Hehl1976,Hehl1995,Shapiro2002,3,4}.

\subsection{Classification of exceptional branches}\label{subsec:axial_branches}

If $X_{\lambda\mu\nu}\ne0$ at a point, \eqref{eq:s7_exceptional} requires
\begin{equation}
	\alpha+\beta+3\gamma T^2=0.
	\label{eq:s7_branchcondition}
\end{equation}
For $\gamma\ne0$ and $-(\alpha+\beta)/(3\gamma)>0$, smoothness on a connected region with nonzero axial torsion selects one of the two constant-norm branches
\begin{equation}
	T=T_*,\qquad T_*=\pm\sqrt{-\frac{\alpha+\beta}{3\gamma}}.
	\label{eq:s7_roots}
\end{equation}
When $\alpha+\beta=0$, the exceptional value is $T_*=0$. A nonzero Lorentzian three-form can have zero quadratic norm, so this \emph{null axial branch} is not torsion-free. If $\gamma=0$, nonzero axial torsion instead requires $\beta=-\alpha$, and the affine equation places no algebraic restriction on $T(x)$. For $\alpha=0$ the effective Einstein reduction below is unavailable and the metric equations require a separate analysis; we do not include this degenerate case in the existence proposition.

\subsection{Metric equation and Einstein reduction}\label{subsec:axial_Einstein}

The vacuum metric equation obtained by varying $g^{\mu\nu}$ at fixed $\Gamma$ is
\begin{equation}
	\alpha R_{(\mu\nu)}(\Gamma)
	-\frac12g_{\mu\nu}(\alpha R+\beta T+\gamma T^3)
	+\Psi(T)H_{\mu\nu}=0.
	\label{eq:s7_metricfull}
\end{equation}
Substituting \eqref{eq:s7_Rsym} and \eqref{eq:s7_Rscalar}, without making any additional field-equation assumption, gives
\begin{equation}
	\alpha\mathring G_{\mu\nu}
	+[\alpha+\Psi(T)]H_{\mu\nu}
	-\frac12g_{\mu\nu}[(\alpha+\beta)T+\gamma T^3]=0.
	\label{eq:s7_metricreduced}
\end{equation}
On the nonlinear exceptional branch, $\Psi(T_*)=-\alpha$ and $(\alpha+\beta)T_*=-3\gamma T_*^3$. Therefore,
\begin{equation}
	\alpha\mathring G_{\mu\nu}+\gamma T_*^3g_{\mu\nu}=0.
	\label{eq:s7_Einsteinpre}
\end{equation}
For $\alpha\ne0$, set
\begin{equation}
	\boxed{\mathring G_{\mu\nu}+\Lambda_*g_{\mu\nu}=0,\qquad
		\Lambda_*:=\frac{\gamma T_*^3}{\alpha}.}
	\label{eq:s7_Einstein}
\end{equation}
With the Ricci-sign convention of Sec.~\ref{sec:geometric_framework}, the trace gives $\mathring R=4\Lambda_*$ and hence $R(\Gamma)=4\Lambda_*+T_*$. For the linear exceptional model $\gamma=0$, $\beta=-\alpha$, the same calculation instead yields $\mathring G_{\mu\nu}=0$ without fixing $T(x)$. The interpretation of the cosmological term is convention dependent: in the conventional static metric $f=1-2M/r-\Lambda_{\mathrm{phys}}r^2/3$ used in much of the black-hole literature, $\Lambda_{\mathrm{phys}}=-\Lambda_*$ for the sign convention of this manuscript \cite{Wald1984,Chandrasekhar1983,Poisson2004}.

\subsection{Local existence proposition and its scope}\label{subsec:local-existence-scope}

The following statement concerns classical solutions of the \emph{unrestricted}
metric--affine Euler--Lagrange equations, subsequently evaluated in the
metric-compatible, totally antisymmetric torsion sector. In particular,
metric compatibility and axiality are not imposed in the action before
variation; this distinction is important in metric--affine gravity and in
symmetry reductions \cite{3,4,Hehl1976,Hehl1995,Palais1979,FelsTorre2002}.

\medskip
\noindent\textbf{Proposition (local axial solutions of polynomial MG-I).}
\emph{Let $\mathcal U$ be an open subset of a four-dimensional Lorentzian
	manifold, and let $g$ be a smooth, nondegenerate Lorentzian metric on
	$\mathcal U$. Fix real constants $\alpha,\beta,\gamma$ with $\alpha\neq0$.
	Suppose that a real constant $T_*$ obeys
	\begin{equation}\label{eq:local-exceptional-condition}
		\alpha+\beta+3\gamma T_*^2=0,
		\qquad
		\Lambda_*:=\frac{\gamma T_*^3}{\alpha},
	\end{equation}
	and that the metric satisfies
	\begin{equation}\label{eq:local-einstein-condition}
		\mathring G_{\mu\nu}+\Lambda_*g_{\mu\nu}=0
		\qquad\text{on }\mathcal U.
	\end{equation}
	Let $X\in\Omega^3(\mathcal U)$ be a smooth, real three-form, not
	identically zero, whose torsion scalar, defined with the conventions of
	Sec.~\ref{sec:geometric_framework}, satisfies
	\begin{equation}\label{eq:local-norm-condition}
		T(g,X)=-\frac14X_{\lambda\mu\nu}X^{\lambda\mu\nu}=T_*
		\qquad\text{throughout }\mathcal U.
	\end{equation}
	Then the affine connection
	\begin{equation}\label{eq:local-axial-connection}
		\Gamma^\lambda{}_{\mu\nu}
		=\mathring\Gamma^\lambda{}_{\mu\nu}
		+\frac12g^{\lambda\rho}X_{\rho\mu\nu}
	\end{equation}
	is smooth and metric-compatible, has torsion
	$2\Gamma^\lambda{}_{[\mu\nu]}=X^\lambda{}_{\mu\nu}$, and, together
	with $g$, satisfies all vacuum metric and affine field equations of
	$F(R,T)=\alpha R+\beta T+\gamma T^3$ on $\mathcal U$.}

\medskip
\noindent\emph{Proof.}
Because $g$ is nondegenerate and smooth, its Levi--Civita connection is
smooth. The tensor $g^{\lambda\rho}X_{\rho\mu\nu}/2$ is smooth and
antisymmetric in its lowered first two indices; hence
\eqref{eq:local-axial-connection} is metric-compatible and has the
prescribed totally antisymmetric torsion. The connection is evaluated
in the field equations obtained by independently varying all ten
symmetric metric components and all 64 ordered affine components.
According to the unrestricted affine variation summarized in
Eqs.~\eqref{eq:s7_affinefull}--\eqref{eq:s7_exceptional} and Appendix~\ref{app:C}, its complete affine equation reduces to
\begin{equation}\label{eq:local-affine-check}
	[\alpha+\beta+3\gamma T(g,X)^2]
	X_\lambda{}^{\mu\nu}=0.
\end{equation}
The coefficient vanishes pointwise by
\eqref{eq:local-exceptional-condition} and
\eqref{eq:local-norm-condition}; in particular, this verifies the
antisymmetric affine equations, while the symmetric affine components
vanish separately on this sector.

For the metric equation, the axial curvature identities
\eqref{eq:s7_Rsym}--\eqref{eq:s7_Rscalar}, established with the conventions of Sec.~\ref{sec:geometric_framework} and checked in
Appendix~\ref{app:B}, reduce the unrestricted metric equation to Eq.~\eqref{eq:s7_metricfull}.
The coefficient of $H_{\mu\nu}$ vanishes by
\eqref{eq:local-exceptional-condition}, and
$(\alpha+\beta)T_*=-3\gamma T_*^3$ gives
\begin{equation}\label{eq:local-metric-check}
	\alpha\mathring G_{\mu\nu}+\gamma T_*^3g_{\mu\nu}=0.
\end{equation}
This is exactly \eqref{eq:local-einstein-condition} multiplied by
$\alpha$. Both independent sets of vacuum field equations therefore
hold at every point of $\mathcal U$. \hfill$\square$

\paragraph{Corollary (linear exceptional model).}
Let $\alpha\neq 0$, $\gamma=0$, and $\beta=-\alpha$.
Let $(U,g)$ be a smooth Lorentzian vacuum Einstein
spacetime satisfying
\[
\mathring G_{\mu\nu}=0.
\]
Then any smooth axial torsion three-form
$X\in\Omega^3(U)$ defines, through
\[
\Gamma^\lambda{}_{\mu\nu}
=
\mathring\Gamma^\lambda{}_{\mu\nu}
+\frac12 g^{\lambda\rho}X_{\rho\mu\nu},
\]
a solution of the unrestricted vacuum metric-affine
field equations. No constant-norm condition on
$T(g,X)$ is required.

\paragraph{Proof.}
In the axial sector, the unrestricted affine equation
reduces to
$(\alpha+\beta)X_\lambda{}^{\mu\nu}=0$,
which vanishes identically. The metric equation
reduces to $\alpha\mathring G_{\mu\nu}=0$.
Both equations are therefore satisfied.
\hfill$\square$

\paragraph{Regularity assumptions and local meaning.}
Smoothness of $g$ and $X$ is a sufficient hypothesis for the classical
pointwise field equations and for a smooth affine connection. The
proposition is conditional: it does not assert that a smooth three-form
of prescribed norm exists globally on a given spacetime, or even on an
arbitrarily prescribed region. In a chosen local coordinate chart,
regularity must be assessed from the tensorial three-form and the
reconstructed connection, rather than from components in a frame that
becomes singular. In particular, smoothness through a Killing horizon
requires a horizon-penetrating chart and a smooth extension of $X$ in
that chart; the construction in Sec.~\ref{sec:BH_affine_revised} supplies such representatives on suitable future-horizon neighborhoods
\cite{Wald1984,Chandrasekhar1983,Poisson2004}.

For the explicit spherical representative of Sec.~\ref{sec:BH_affine_revised}, the strict
inequality $a^2+c f(r)>0$ is a \emph{local domain condition} ensuring
that the chosen square-root branch is real and smooth. At a horizon
$f(r_h)=0$ with $a>0$, it holds automatically on a sufficiently small
neighborhood. Neither the proposition nor this inequality proves a
smooth extension across every horizon, a bifurcation two-sphere, or a
maximal spacetime extension. The vanishing of the square-root radicand
may obstruct this particular representative without excluding other
local torsion profiles.

The condition $T=T_*$ is algebraic and does not imply
$\mathring\nabla_\rho X_{\lambda\mu\nu}=0$, nor does it impose
$R^\lambda{}_{\rho\mu\nu}(\Gamma)=0$; the solutions are generally
Riemann--Cartan rather than teleparallel
\cite{Hehl1976,Hehl1995,Shapiro2002}. The exceptional value $T_*=0$
allows nonzero \emph{null} axial torsion in Lorentzian signature. For
$\gamma=0$ and $\beta=-\alpha$, the same field-equation argument gives
$\mathring G_{\mu\nu}=0$ for any smooth axial three-form, without
requiring its norm to be constant; this is a broader linear exceptional
case than the constant-norm statement above.

\paragraph{Local realizability of the norm condition.}
The prescribed-norm hypothesis is locally nonempty. On any sufficiently
small Lorentzian coordinate neighborhood, choose a smooth orthonormal
coframe and a nonzero axial three-form with constant frame components.
Because the quadratic form on three-forms has indefinite signature,
its coefficients can be selected to realize either sign of $T_*$,
as well as $T_*=0$ with nonzero torsion. This is a local construction;
it makes no assertion about the existence of a globally defined
constant-norm three-form on an arbitrary spacetime. See also the
geometric discussion of torsion in Refs.~\cite{Hehl1995,Shapiro2002}.

Finally, the proposition establishes neither uniqueness nor global
regularity, perturbative stability, a propagating torsion degree of
freedom, nor physical inequivalence of different torsion profiles.
Such conclusions require additional dynamical or invariant analyses;
local solvability alone is insufficient \cite{Hehl1995,Shapiro2002}. 
The spherical Ricci-symmetric subclass of this proposition is characterized further in Sec.~\ref{subsec:constant-W-axial} by $(fV-U)'=0$; its local parameterization and horizon-regular representatives are not a classification of unrestricted metric--affine connections.

\subsection{Spherical axial data and the horizon problem}\label{subsec:axial_spherical}

For the orthonormal spherical coframe used in Sec.~\ref{sec:spherical_axial}, a rotation-invariant axial three-form can be written
\begin{equation}
	\tau=x(r)e^0\wedge e^2\wedge e^3-y(r)e^1\wedge e^2\wedge e^3,
	\label{eq:s7_spherical}
\end{equation}
so that
\begin{equation}
	T=\frac32[y(r)^2-x(r)^2],\qquad
	y(r)^2-x(r)^2=\frac{2T_*}{3}.
	\label{eq:s7_sphericalnorm}
\end{equation}
The algebraic constraint leaves one local functional freedom within this ansatz. It does not by itself identify a gauge symmetry or a propagating mode. In a static frame, $x$ and $y$ may become singular at a Killing horizon even when the three-form is regular. Section~\ref{sec:BH_affine_revised} therefore reconstructs the torsion and affine connection in horizon-penetrating coordinates and checks their curvature invariants. These local tests must be distinguished from maximal analytic extension and global regularity \cite{Hehl1995,Shapiro2002,Wald1984,Poisson2004}.

\section{Static spherical reduction and axial torsion branches}\label{sec:spherical_axial}

We now apply the local axial-connection result of Sec.~\ref{sec:axial_theorem} to static,
spherically symmetric spacetimes. The independent metric and affine
variations are performed before restricting the solution to the
Riemann--Cartan sector. This order of operations is important in
metric--affine theories and is not guaranteed by a symmetry-reduced
action alone \cite{3,4,Hehl1976,Hehl1995,Palais1979,FelsTorre2002}.
Throughout this section the signature is $(+---)$, the connection is
$\nabla_\nu V^\lambda=\partial_\nu V^\lambda+
\Gamma^\lambda{}_{\rho\nu}V^\rho$, and
$X^\lambda{}_{\mu\nu}=2\Gamma^\lambda{}_{[\mu\nu]}$.

\subsection{Invariant coframe and axial three-form}
In a static region let
\begin{equation}
	ds^2=A(r)dt^2-B(r)dr^2-C(r)d\Omega^2,\qquad
	d\Omega^2=d\theta^2+\sin^2\theta\,d\phi^2,
	\qquad A,B,C>0.
\end{equation}
An orthonormal coframe is
\begin{equation}
	e^0=\sqrt A\,dt,\quad e^1=\sqrt B\,dr,\quad
	e^2=\sqrt C\,d\theta,\quad e^3=\sqrt C\sin\theta\,d\phi.
\end{equation}
The most general static, rotation-invariant axial
three-form on the chosen connected static patch
can be written as
\begin{equation}\label{eq:tau_xy_new}
	\tau:=\frac1{3!}X_{abc}e^a\wedge e^b\wedge e^c
	=x(r)e^0\wedge e^2\wedge e^3-y(r)e^1\wedge e^2\wedge e^3.
\end{equation}
The distinction between rotational invariance of a torsion field and
more restrictive cosmological symmetry assumptions is discussed in the
classical literature on spacetime torsion \cite{Tsamparlis1979}. Here $X_{023}=x$ and $X_{123}=-y$. Defining
$\mathcal T^a=\frac12X^a{}_{bc}e^b\wedge e^c$, raising the first
index with $\eta^{ab}=\mathrm{diag}(1,-1,-1,-1)$ gives
\begin{align}\label{eq:torsion_forms_corrected}
	\mathcal T^0&=x\,e^2\wedge e^3,&
	\mathcal T^1&=y\,e^2\wedge e^3,\\
	\mathcal T^2&=\boxed{\,x\,e^0\wedge e^3-y\,e^1\wedge e^3\,},&
	\mathcal T^3&=-x\,e^0\wedge e^2+y\,e^1\wedge e^2.
\end{align}
In particular, the $x e^0\wedge e^3$ term in $\mathcal T^2$
\emph{has a plus sign}. This follows directly from
$X^2{}_{03}=\eta^{22}X_{203}=x$; the opposite sign would contradict
total antisymmetry. The metric-compatible connection is
\begin{equation}\label{eq:connection_new}
	\Gamma^\lambda{}_{\mu\nu}=\mathring\Gamma^\lambda{}_{\mu\nu}
	+\frac12g^{\lambda\rho}X_{\rho\mu\nu}.
\end{equation}
This is a Riemann--Cartan connection; its curvature is not assumed to
vanish \cite{Kibble1961,Sciama1964,Hehl1976,Hehl1995,Shapiro2002}.

\subsection{Quadratic contractions and component field equations}
For the torsion scalar defined in Sec.~\ref{sec:geometric_framework}, the contractions of
\eqref{eq:tau_xy_new} are
\begin{equation}\label{eq:T_xy_new}
	X_{abc}X^{abc}=6(x^2-y^2),\qquad
	T=-\frac14X_{abc}X^{abc}=\frac32(y^2-x^2).
\end{equation}
The fixed-connection metric derivative, evaluated \emph{after}
unrestricted variation, is
$H_{ab}=-\frac14X_{acd}X_b{}^{cd}$. Its nonzero components are
\begin{equation}\label{eq:H_matrix_new}
	(H_{ab})=\frac12
	\begin{pmatrix}
		-x^2&xy&0&0\\ xy&-y^2&0&0\\
		0&0&x^2-y^2&0\\0&0&0&x^2-y^2
	\end{pmatrix},\quad
	\eta^{ab}H_{ab}=T,\quad H_{ab}H^{ab}=\frac13T^2.
\end{equation}
The last identity follows by direct contraction and does not require
constant $T$.

For $F(R,T)=\alpha R+\beta T+\gamma T^3$, set
\begin{equation}
	\Psi(T)=\beta+3\gamma T^2,\qquad
	\mathcal V(T)=(\alpha+\beta)T+\gamma T^3.
\end{equation}
The axial curvature identities and the fixed-connection derivative
established in Sec.~\ref{sec:axial_theorem} reduce the \emph{full} metric equation to
\begin{equation}\label{eq:metric_axial_new}
	\alpha\mathring G_{ab}+[\alpha+\Psi(T)]H_{ab}
	-\frac12\eta_{ab}\mathcal V(T)=0.
\end{equation}
In the orthonormal coframe its independent components read
\begin{align}
	0&=\alpha\mathring G_{00}-\frac12(\alpha+\Psi)x^2
	-\frac12\mathcal V,\\
	0&=\alpha\mathring G_{11}-\frac12(\alpha+\Psi)y^2
	+\frac12\mathcal V,\\
	0&=\alpha\mathring G_{22}+\frac12(\alpha+\Psi)(x^2-y^2)
	+\frac12\mathcal V,\\
	0&=\alpha\mathring G_{33}+\frac12(\alpha+\Psi)(x^2-y^2)
	+\frac12\mathcal V,\\
	0&=\alpha\mathring G_{01}+\frac12(\alpha+\Psi)xy.
\end{align}
Here $\mathring G_{01}=0$ and $\mathring G_{22}=\mathring G_{33}$
for a static spherical metric. The off-diagonal equation is a useful
check, but the exceptional condition follows more generally from the
unrestricted affine equation of Sec.~\ref{sec:axial_theorem}, including when $xy=0$.

\subsection{Exceptional, null, and linear branches}
For nonzero axial torsion, the full connection equation requires
\begin{equation}\label{eq:exception_new}
	\alpha+\Psi(T)=\alpha+\beta+3\gamma T^2=0.
\end{equation}
When $\gamma\ne0$, the real branches satisfy
\begin{equation}\label{eq:c_new}
	T=T_*,\qquad T_*^2=-\frac{\alpha+\beta}{3\gamma}\ge0,
	\qquad y^2-x^2=c:=\frac{2T_*}{3}.
\end{equation}
On a connected region with $T_*>0$, one may write
\begin{equation}
	x=\epsilon\sqrt c\sinh\zeta(r),\qquad
	y=\epsilon\sqrt c\cosh\zeta(r),\qquad\epsilon=\pm1.
\end{equation}
For $T_*<0$, writing $d=-c>0$, a corresponding parametrization is
\begin{equation}
	x=\epsilon\sqrt d\cosh\zeta(r),\qquad
	y=\epsilon\sqrt d\sinh\zeta(r).
\end{equation}
The function $\zeta(r)$ remains algebraically undetermined. This is
freedom in the axial connection, not by itself proof of a gauge
symmetry or of a propagating mode.

For $T_*=0$ and $\gamma\ne0$, Eq.~\eqref{eq:exception_new}
requires $\alpha+\beta=0$, and a nonzero null three-form may have
$y=x\ne0$ or $y=-x\ne0$. If $\gamma=0$ and $\beta=-\alpha$,
the axial affine equation is satisfied without imposing a constant
norm: $T(r)$ may vary. These branches should not be identified with
the torsion-free solution.

\subsection{Integration of the spherical metric equations}
Assume $\alpha\ne0$ and a nonlinear exceptional branch. Then
$\Psi(T_*)=-\alpha$ and
$\mathcal V(T_*)=-2\gamma T_*^3$, so
\eqref{eq:metric_axial_new} becomes
\begin{equation}\label{eq:einstein_new}
	\mathring G_{\mu\nu}+\Lambda_*g_{\mu\nu}=0,
	\qquad\Lambda_*:=\frac{\gamma T_*^3}{\alpha}.
\end{equation}
Where the areal radius $\rho=\sqrt{C(r)}$ is a valid local
coordinate, introduce
\begin{equation}
	ds^2=e^{2\Phi(\rho)}F(\rho)dt^2
	-\frac{d\rho^2}{F(\rho)}-\rho^2d\Omega^2.
\end{equation}
On a nondegenerate static interval, the difference of the temporal
and radial Einstein equations gives $\Phi'=0$; rescale $t$ to set
$\Phi=0$. With the curvature convention of Sec.~\ref{sec:geometric_framework} the remaining
equations are
\begin{equation}\label{eq:radial_eq_new}
	\frac{\rho F'+F-1}{\rho^2}=\Lambda_*,\qquad
	\frac12F''+\frac{F'}{\rho}=\Lambda_*.
\end{equation}
The second equation follows by differentiating the first after
multiplication by $\rho^2$. Integration gives
\begin{equation}\label{eq:F_new}
	F(\rho)=1-\frac{2M}{\rho}+\frac{\Lambda_*}{3}\rho^2,
	\qquad\mathring R=4\Lambda_*.
\end{equation}
Thus the metric is locally Schwarzschild--(anti-)de Sitter, while
the connection may carry nonzero axial torsion. In the more common
convention $F=1-2M/\rho-\Lambda_{\rm phys}\rho^2/3$,
$\Lambda_{\rm phys}=-\Lambda_*$ \cite{Wald1984,Chandrasekhar1983,Poisson2004}.
For the exceptional linear model $\gamma=0$, $\beta=-\alpha$,
Eq.~\eqref{eq:einstein_new} is replaced by
$\mathring G_{\mu\nu}=0$, without a fixed torsion norm.

\subsection{Horizon-penetrating axial variables}
The static coframe ceases to be regular at $F(\rho_h)=0$. On a
future-horizon patch define the advanced coordinate
\begin{equation}
	v=t+\int\frac{d\rho}{F(\rho)},\qquad
	ds^2=F(\rho)dv^2-2\,dv\,d\rho-\rho^2d\Omega^2.
\end{equation}
Writing the same tensorial three-form as
\begin{equation}\label{eq:tau_UV_new}
	\tau=\rho^2\sin\theta\,[U(\rho)dv-V(\rho)d\rho]
	\wedge d\theta\wedge d\phi,
\end{equation}
comparison with \eqref{eq:tau_xy_new} in the static region gives
\begin{equation}\label{eq:UV_xy_new}
	U=\sqrt F\,x,\qquad V=\frac{x+y}{\sqrt F},\qquad
	x=\frac U{\sqrt F},\qquad y=\sqrt F\,V-\frac U{\sqrt F}.
\end{equation}
Consequently, the constant-norm condition
$y^2-x^2=c$ becomes
\begin{equation}\label{eq:UV_constraint_new}
	F V^2-2UV=c,\qquad T=\frac32(FV^2-2UV).
\end{equation}
This equation is regular at $F=0$. If $U(\rho_h)>0$, its
horizon-regular root is
\begin{equation}\label{eq:Vregular_new}
	V=\frac{U-\sqrt{U^2+cF}}{F}
	=-\frac{c}{U+\sqrt{U^2+cF}}.
\end{equation}
In particular, the local choice $U=a>0$ gives
\begin{equation}
	V=-\frac{c}{a+\sqrt{a^2+cF}},\qquad
	a^2+cF>0,\qquad V(\rho_h)=-\frac{c}{2a}.
\end{equation}
The radicand is strictly positive at the horizon and therefore
remains positive on a sufficiently small neighborhood. Since
$g^{\mu\nu}$ and $\tau$ are smooth in the ingoing chart, the
contorsion and the reconstructed affine connection
\eqref{eq:connection_new} are smooth there. This establishes
\emph{local future-horizon regularity}; it does not establish
regularity at a bifurcation sphere, maximal extension, or global
boundedness of affine curvature invariants
\cite{Wald1984,Poisson2004,Hehl1995,Shapiro2002}.
The explicit affine curvature tests are deferred to Sec.~\ref{sec:axial_classification} and
Appendix~\ref{app:B}.

\section{Classification of axial metric-compatible connections}\label{sec:axial_classification}

We distinguish the classification of solutions within the metric-compatible,
totally antisymmetric torsion sector from the classification of arbitrary
metric--affine connections. The latter includes nonmetricity and non-axial
torsion and is not established here. This distinction is essential because
imposing geometric restrictions before varying the action can eliminate
independent affine equations \cite{Hehl1995,Shapiro2002,Palais1979,FelsTorre2002}.

Consider the vacuum model
\begin{equation}\label{eq:class_action}
	S=\frac{1}{2\kappa}\int d^4x\sqrt{-g}\,F(R,T),\qquad
	F(R,T)=\alpha R+\beta T+\gamma T^3,\quad \alpha\ne0.
\end{equation}
Here the curvature, torsion scalar, and connection follow the conventions
fixed in Sec.~\ref{sec:geometric_framework}. The independent metric and affine equations are
\begin{align}
	\alpha R_{(\mu\nu)}-\tfrac12g_{\mu\nu}F
	+(\beta+3\gamma T^2)H_{\mu\nu}&=0,\label{eq:class_metric}\\
	\alpha P_\lambda{}^{\mu\nu}
	+(\beta+3\gamma T^2)J_\lambda{}^{\mu\nu}&=0.
	\label{eq:class_affine}
\end{align}
Only after taking these unrestricted variations do we specialize to
\begin{equation}\label{eq:class_axial}
	Q_{\lambda\mu\nu}=0,\qquad X_{\lambda\mu\nu}=X_{[\lambda\mu\nu]},
	\qquad \Gamma^\lambda{}_{\mu\nu}
	=\mathring\Gamma^\lambda{}_{\mu\nu}+\tfrac12X^\lambda{}_{\mu\nu}.
\end{equation}
The independently established identities in Sec.~\ref{sec:axial_theorem} and Appendices B--C give
\begin{equation}\label{eq:class_identities}
	P_\lambda{}^{\mu\nu}=J_\lambda{}^{\mu\nu}
	=-X_\lambda{}^{\mu\nu},\quad
	R(\Gamma)=\mathring R+T,\quad
	R_{(\mu\nu)}(\Gamma)=\mathring R_{\mu\nu}+H_{\mu\nu}.
\end{equation}
Consequently the affine equation becomes
\begin{equation}\label{eq:class_exceptional}
	[\alpha+\beta+3\gamma T^2]X_\lambda{}^{\mu\nu}=0.
\end{equation}

\paragraph{Proposition (local classification in the axial sector).}
On a connected open region with a smooth Lorentzian metric and smooth
metric-compatible axial torsion, the solutions of
Eqs.~\eqref{eq:class_metric}--\eqref{eq:class_affine} are described locally
by the following alternatives.

\emph{(i) Torsion-free branch.} If $X=0$, the connection is Levi--Civita,
$T=0$, and $\mathring G_{\mu\nu}=0$.

\emph{(ii) Nonzero-torsion nonlinear branch.} If $\gamma\ne0$ and $X$ is
nonzero on an open region, Eq.~\eqref{eq:class_exceptional} implies
\begin{equation}\label{eq:class_Tstar}
	T^2=-\frac{\alpha+\beta}{3\gamma}\ge0.
\end{equation}
For $T_*\ne0$, continuity fixes the sign of $T=T_*$ on each connected
nonzero-torsion region. The metric equation reduces to
\begin{equation}\label{eq:class_einstein}
	\mathring G_{\mu\nu}+\Lambda_*g_{\mu\nu}=0,
	\qquad \Lambda_*=\frac{\gamma T_*^3}{\alpha}.
\end{equation}
Conversely, any smooth Einstein metric satisfying
Eq.~\eqref{eq:class_einstein} and any smooth axial three-form of constant
scalar norm $T_*$ define, through Eq.~\eqref{eq:class_axial}, a solution
of the \emph{unrestricted} field equations. No spherical symmetry is
needed for this local construction.

\emph{(iii) Null-norm nonlinear branch.} When $\gamma\ne0$ and
$\alpha+\beta=0$, a nonzero axial three-form may have $T_*=0$.
The metric is Ricci-flat, but $X$ need not vanish. Thus a zero torsion
scalar is not equivalent to zero torsion \cite{Hehl1976,Shapiro2002}.

\emph{(iv) Linear degenerate branch.} When $\gamma=0$ and
$\beta=-\alpha$, Eq.~\eqref{eq:class_exceptional} is identically
satisfied for every axial three-form, without a fixed-norm condition;
the metric is Ricci-flat. For $\gamma=0$ and $\beta\ne-\alpha$,
Eq.~\eqref{eq:class_exceptional} instead forces $X=0$.

\paragraph{Proof.}
For $X\ne0$, Eq.~\eqref{eq:class_exceptional} yields the exceptional
condition. Using Eq.~\eqref{eq:class_identities}, the metric equation is
\begin{equation}\label{eq:class_metric_reduced}
	\alpha\mathring G_{\mu\nu}
	+[\alpha+\beta+3\gamma T^2]H_{\mu\nu}
	-\frac12g_{\mu\nu}[(\alpha+\beta)T+\gamma T^3]=0.
\end{equation}
For the nonlinear exceptional branch,
$\alpha+\beta=-3\gamma T_*^2$, hence the bracketed potential equals
$-2\gamma T_*^3$ and Eq.~\eqref{eq:class_einstein} follows.
For $\gamma=0$, $\beta=-\alpha$, the potential vanishes identically.
The torsion-free case follows directly from
Eqs.~\eqref{eq:class_metric}--\eqref{eq:class_affine}. Conversely,
substitution into the unrestricted equations proves sufficiency.
\hfill$\square$

In four dimensions, the axial three-form is Hodge-dual to a vector:
$X_{\mu\nu\rho}=\varepsilon_{\mu\nu\rho\sigma}A^\sigma$.
For the signature $(+---)$ and the orientation convention
$\varepsilon_{\mu\nu\rho\sigma}\varepsilon^{\mu\nu\rho\lambda}
=-6\delta_\sigma{}^\lambda$, one has
\begin{equation}\label{eq:class_hodge}
	X_{\lambda\mu\nu}X^{\lambda\mu\nu}=-6A_\mu A^\mu,
	\qquad T=\tfrac32 A_\mu A^\mu.
\end{equation}
Thus the nonlinear axial classification is equivalently a local
fixed-norm-vector problem on an Einstein geometry. This is a
classification of \emph{solutions in the axial subsector}, not of
all inequivalent physical degrees of freedom. Global smoothness,
boundary conditions, and possible gauge identifications remain separate
questions \cite{Hehl1995,Wald1984,Poisson2004}.

\paragraph{Local branches and their zero sets.}
The classification is pointwise and local. On an open subset where
$X\neq0$, the nonlinear affine equation enforces a constant value of
$T$ selected by the coupling constants. Smooth solutions may possess
zero sets of $X$, and the existence of a smooth continuation across
such sets requires a separate analysis. In particular, a nonzero null
three-form can have $T=0$, so the condition $T=0$ must not be used as
an abbreviation for vanishing torsion. The classification is not a
classification modulo diffeomorphisms or any putative affine gauge
symmetry \cite{Hehl1995,Shapiro2002}.

\paragraph{Scope beyond the axial sector.}
For a general affine connection, write
$\Gamma=\mathring\Gamma+N$. The 64 components of $N$ encode 24
torsion and 40 nonmetricity components. Since $F_R=\alpha$ is constant,
the Einstein--Hilbert contribution differs from its Levi--Civita form
by a divergence and terms algebraic and quadratic in $N$; the torsion
potential is also algebraic. The general affine equations therefore
constitute a nonlinear pointwise algebraic system. Its exhaustive
classification requires the ranks and kernels of the unrestricted
affine operator in the vector, axial, tensor-torsion, and nonmetricity
sectors, followed by the full metric equations. Neither the proposition
above nor the spherically symmetric examples establish that broader
classification \cite{Hehl1995,3,4}.

\section{Black-hole solutions, horizon regularity, and affine curvature}
\label{sec:BH_affine_revised}

We investigate the metric and the independent Riemann--Cartan connection on the
exceptional axial branch constructed in Secs.~7--8. The metric and affine field
equations are understood as the unrestricted metric--affine equations, subsequently
evaluated on the metric-compatible axial sector. We distinguish Levi--Civita
invariants from invariants of the independent connection throughout
\cite{3,4,Hehl1976,Hehl1995,Shapiro2002}. We assume vacuum, $\alpha\ne0$,
and the curvature convention fixed in Sec.~\ref{sec:geometric_framework}.

\subsection{Einstein metric, horizons, and surface gravity}
\label{subsec:BH_metric_revised}
For $F(R,T)=\alpha R+\beta T+\gamma T^3$, the nonzero axial branch obeys
\begin{equation}\label{eq:BH_exceptional_revised}
	\alpha+\beta+3\gamma T_*^2=0,\qquad
	\Lambda_*:=\frac{\gamma T_*^3}{\alpha},\qquad
	\mathring G_{\mu\nu}+\Lambda_*g_{\mu\nu}=0.
\end{equation}
For $\gamma\ne0$, real constant-norm branches require
$T_*^2=-(\alpha+\beta)/(3\gamma)\ge0$. The metric equation in the
areal-radius gauge is
\begin{equation}\label{eq:BH_metric_revised}
	ds^2=f(r)\,dt^2-\frac{dr^2}{f(r)}-r^2d\Omega^2,
	\qquad d\Omega^2=d\theta^2+\sin^2\theta\,d\phi^2.
\end{equation}
With the curvature sign convention of the present work,
\begin{align}
	\mathring R&=f''+\frac{4f'}r+\frac{2(f-1)}{r^2},\\
	\frac{rf'+f-1}{r^2}&=\Lambda_*,\qquad
	\frac{f''}{2}+\frac{f'}r=\Lambda_*.
\end{align}
The second radial equation follows by differentiating
$(rf)'=1+\Lambda_*r^2$; integration yields
\begin{equation}\label{eq:BH_f_revised}
	\boxed{f(r)=1-\frac{2M}{r}+\frac{\Lambda_*r^2}{3}},\qquad
	\mathring R=4\Lambda_*.
\end{equation}
This is the Schwarzschild--(anti-)de Sitter metric with the manuscript's
sign convention: the more usual parameter $\Lambda_{\rm phys}$ in
$f=1-2M/r-\Lambda_{\rm phys}r^2/3$ equals $-\Lambda_*$.
Birkhoff-type uniqueness applies to the Einstein metric, not to the
independent affine connection \cite{Wald1984,Chandrasekhar1983,Poisson2004}.

A nonzero Killing horizon $r_h$ is a positive root of
\begin{equation}
	\frac{\Lambda_*}{3}r_h^3+r_h-2M=0,\qquad
	f'(r_h)=\frac1{r_h}+\Lambda_*r_h.
\end{equation}
For $M>0$, $\Lambda_*>0$ gives one positive horizon, whereas
$\Lambda_*<0$ gives two distinct positive horizons precisely when
$0<9M^2(-\Lambda_*)<1$; equality gives a degenerate horizon.
For a nondegenerate horizon the magnitude of the metric Killing
surface gravity is
\begin{equation}\label{eq:BH_kappa_revised}
	\boxed{\kappa_h=\frac12|f'(r_h)|
		=\frac12\left|\frac1{r_h}+\Lambda_*r_h\right|}.
\end{equation}
The absolute value is necessary at a cosmological horizon, where the
radial derivative may be negative. The associated metric temperature,
when the standard semiclassical interpretation is applicable, is
$T_H=\kappa_h/(2\pi)$ in units $\hbar=k_B=c=1$.
The behavior of spinful matter coupled to $\Gamma$ is a separate
question \cite{Wald1984,Poisson2004,Hehl1995}.

\subsection{Smooth axial torsion at a future Killing horizon}
\label{subsec:BH_horizon_revised}
Introduce the advanced Eddington--Finkelstein coordinate
$v=t+\int dr/f(r)$ on a future-horizon patch. Then
\begin{equation}\label{eq:BH_EF_revised}
	ds^2=f(r)\,dv^2-2\,dv\,dr-r^2d\Omega^2.
\end{equation}
The rotation-invariant torsion three-form can be written
\begin{equation}\label{eq:BH_tau_revised}
	\tau=\frac1{3!}X_{\mu\nu\rho}
	dx^\mu\wedge dx^\nu\wedge dx^\rho
	=r^2\sin\theta\,[U(r)\,dv-V(r)\,dr]\wedge d\theta\wedge d\phi.
\end{equation}
The area form $\sin\theta\,d\theta\wedge d\phi$ is globally smooth on
the symmetry sphere. Direct contraction gives
\begin{equation}\label{eq:BH_norm_revised}
	T=\frac32\bigl(fV^2-2UV\bigr),\qquad
	fV^2-2UV=c,\qquad c:=\frac{2T_*}{3}.
\end{equation}
For any smooth $U$ with $U(r_h)>0$, the root regular at $f(r_h)=0$ is
\begin{equation}\label{eq:BH_Vroot_revised}
	V=\frac{U-\sqrt{U^2+cf}}{f}
	=-\frac{c}{U+\sqrt{U^2+cf}},\qquad
	V(r_h)=-\frac{c}{2U(r_h)}.
\end{equation}
A useful representative is
\begin{equation}\label{eq:BH_constantU_revised}
	U=a>0,\quad q(r):=\sqrt{a^2+cf(r)},\quad
	V=-\frac{c}{a+q},\quad q^2>0,\quad
	W:=fV-U=-q.
\end{equation}
Since $q(r_h)=a>0$, this is smooth on some neighborhood of every
finite-radius horizon of the metric, including a degenerate one.
The independent connection is
\begin{equation}\label{eq:BH_Gamma_revised}
	\Gamma^\lambda{}_{\mu\nu}
	=\mathring\Gamma^\lambda{}_{\mu\nu}
	+\frac12g^{\lambda\rho}X_{\rho\mu\nu}.
\end{equation}
Smoothness of $g$ and $\tau$ in the regular chart implies smoothness
of the contorsion and of $\Gamma$ there. This proves local regularity
at the selected future horizon, not smoothness at the bifurcation
sphere or across all past horizons \cite{Wald1984,Poisson2004,Chandrasekhar1983}.
For $c=0$, $U=a$, $V=0$ is a \emph{nonzero} torsion three-form with
vanishing quadratic torsion scalar; it is not the torsion-free limit.

\paragraph{Local regularity versus global extension.}
The Eddington--Finkelstein construction establishes smoothness of the
metric and torsion three-form on a neighborhood of a selected future
Killing horizon. It does not, by itself, establish smoothness at a
bifurcation two-sphere, across a past horizon, or on a maximal
extension. Moreover, bounded polynomial curvature invariants do not
replace a componentwise test in a regular frame. These distinctions
are standard in the analysis of black-hole extensions
\cite{Wald1984,Poisson2004}.

\subsection{Ricci tensor and curvature invariants}
\label{subsec:BH_Ricci_revised}
The Levi--Civita invariants are
\begin{equation}\label{eq:BH_LC_invariants_revised}
	\mathring R=4\Lambda_*,\quad
	\mathring R_{\mu\nu}\mathring R^{\mu\nu}=4\Lambda_*^2,\quad
	K_{\rm LC}:=\mathring R_{\alpha\beta\mu\nu}
	\mathring R^{\alpha\beta\mu\nu}
	=\frac{48M^2}{r^6}+\frac83\Lambda_*^2.
\end{equation}
On the axial sector, the identities of Sec.~\ref{sec:axial_theorem} give
\begin{equation}
	R_{(\mu\nu)}(\Gamma)=\Lambda_*g_{\mu\nu}+H_{\mu\nu},\quad
	R(\Gamma)=4\Lambda_*+T_*,\quad
	H^\mu{}_{\mu}=T_*,\quad
	H_{\mu\nu}H^{\mu\nu}=\frac{T_*^2}{3}.
\end{equation}
Consequently,
\begin{equation}\label{eq:BH_symRicci_revised}
	R_{(\mu\nu)}R^{(\mu\nu)}
	=4\Lambda_*^2+2\Lambda_*T_*+\frac{T_*^2}{3},\qquad
	X_{\alpha\beta\gamma}X^{\alpha\beta\gamma}=-4T_*.
\end{equation}
Define $W=fV-U$. Raising the indices of the three-form gives
\begin{equation}
	X^{r\theta\phi}=\frac{W}{r^2\sin\theta},\qquad
	\mathring\nabla_\lambda X^{\lambda\theta\phi}
	=\frac{W'}{r^2\sin\theta}.
\end{equation}
The antisymmetric Ricci identity
$R_{[\mu\nu]}=\frac12\mathring\nabla_\lambda
X^\lambda{}_{\mu\nu}$ then implies
\begin{equation}\label{eq:BH_antiRicci_revised}
	R_{[\theta\phi]}=\frac12r^2\sin\theta\,W',\quad
	R_{[vr]}=0,\quad
	R_{[\mu\nu]}R^{[\mu\nu]}=\frac12(W')^2.
\end{equation}
The symmetric--antisymmetric cross contraction vanishes, so
\begin{equation}\label{eq:BH_fullRicci_revised}
	\boxed{R_{\mu\nu}(\Gamma)R^{\mu\nu}(\Gamma)
		=4\Lambda_*^2+2\Lambda_*T_*+\frac{T_*^2}{3}
		+\frac12\bigl[(fV-U)'\bigr]^2}.
\end{equation}
For \eqref{eq:BH_constantU_revised}, $W=-q$ and
$W'=-cf'/(2q)$, yielding
\begin{equation}\label{eq:BH_antiRicci_explicit_revised}
	R_{[\mu\nu]}R^{[\mu\nu]}
	=\frac{c^2(f')^2}{8(a^2+cf)},\qquad
	f'=\frac{2M}{r^2}+\frac{2\Lambda_*r}{3}.
\end{equation}
At a horizon,
\begin{equation}
	\left.R_{[\mu\nu]}R^{[\mu\nu]}\right|_{r=r_h}
	=\frac{c^2}{8a^2}\left(\frac1{r_h}+\Lambda_*r_h\right)^2.
\end{equation}
This vanishes for a degenerate horizon, and also for the null-norm
representative $c=0$, without implying that its full affine curvature
vanishes.

\subsection{Full affine Kretschmann invariant, radial domains, and regular continuation}\label{sec:affine-kretschmann-domains}
The affine curvature of a metric-compatible connection with totally antisymmetric torsion is not determined by its symmetric Ricci tensor. In particular, the full quadratic invariant must be constructed from the four-index curvature tensor, with the curvature and contraction conventions of Sec.~\ref{sec:geometric_framework} \cite{Hehl1976,Hehl1995,Shapiro2002}. Writing
\begin{equation}\label{eq:Kdecomposition-new}
	D_{\lambda\rho\mu\nu}:=R_{\lambda\rho\mu\nu}(\Gamma)-\mathring R_{\lambda\rho\mu\nu},\qquad
	K_\Gamma=K_{\rm LC}+2\mathring R_{\lambda\rho\mu\nu}D^{\lambda\rho\mu\nu}+D_{\lambda\rho\mu\nu}D^{\lambda\rho\mu\nu},
\end{equation}
we employ the ingoing Eddington--Finkelstein metric and axial three-form
\begin{equation}\label{eq:EFaxial-new}
	ds^2=f(r)\,dv^2-2\,dv\,dr-r^2d\Omega^2,\qquad
	\tau=r^2\sin\theta\,[U(r)\,dv-V(r)\,dr]\wedge d\theta\wedge d\phi.
\end{equation}
Here $U,V$ are smooth radial functions wherever the representative is defined. The torsion scalar and the auxiliary function $W$ are
\begin{equation}\label{eq:Wconstraint-new}
	T=\frac32\bigl(fV^2-2UV\bigr),\qquad W:=fV-U.
\end{equation}

\paragraph{Independent curvature contractions.}
Direct reconstruction of the full affine curvature using $\Gamma^\lambda{}_{\mu\nu}=\mathring\Gamma^\lambda{}_{\mu\nu}+X^\lambda{}_{\mu\nu}/2$ gives the cross contraction
\begin{equation}\label{eq:cross-new}
	2\mathring R_{\lambda\rho\mu\nu}D^{\lambda\rho\mu\nu}
	=\frac{2V(fV-2U)}{r^2}\bigl(f+rf'-1\bigr).
\end{equation}
For completeness, the distortion self-contraction is
\begin{align}\label{eq:D2-new}
	D_{\lambda\rho\mu\nu}D^{\lambda\rho\mu\nu}
	={}&\frac14\Bigl[3f^2V^4+8f^2(V')^2+8ff'VV'-12fUV^3
	-16fU'V'\nonumber\\
	&\qquad+4(f')^2V^2+8f'UV'-8f'U'V+12U^2V^2-4(U')^2\Bigr]
	-\frac{4UU'}r-\frac{8U^2}{r^2}.
\end{align}
These expressions are to be read together with
\begin{equation}\label{eq:KLC-general-new}
	K_{\rm LC}=(f'')^2+\frac{4(f')^2}{r^2}+\frac{4(f-1)^2}{r^4}.
\end{equation}
The contraction was independently compared at $\theta=\pi/2$ and $\theta=\pi/3$ before imposing the constant-norm constraint; both calculations yield the same polynomial. Angular independence also follows from spherical symmetry. This is a two-angle symbolic cross-check, not a claim of a separate arbitrary-angle symbolic reconstruction. The decomposition provides a direct reproducibility test for Appendix~\ref{app:B}.

On the exceptional nonlinear branch, let
\begin{equation}\label{eq:metric-c-new}
	c:=\frac{2T_*}{3},\qquad fV^2-2UV=c,\qquad
	f(r)=1-\frac{2M}{r}+\frac{\Lambda_*r^2}{3},\qquad \Lambda_*:=\frac{\gamma T_*^3}{\alpha}.
\end{equation}
Since $(f+rf'-1)/r^2=\Lambda_*$, Eq.~\eqref{eq:cross-new} simplifies exactly to
\begin{equation}\label{eq:cross-einstein-new}
	2\mathring R\!\cdot\!D=2c\Lambda_*.
\end{equation}
For the horizon-regular constant-$U$ representative
\begin{equation}\label{eq:constant-U-new}
	U=a>0,\qquad h(r):=a^2+cf(r),\qquad q(r):=\sqrt{h(r)},\qquad
	V=-\frac{c}{a+q},\qquad h>0,
\end{equation}
we have $W=-q$ and $V'=c^2f'/[2q(a+q)^2]$. Substituting in Eq.~\eqref{eq:D2-new} yields
\begin{equation}\label{eq:D2-special-new}
	D\!\cdot\!D=\frac{c^2(f')^2}{2h}+\frac{2cf'}r+\frac{3c^2}{4}
	+\frac{2cf-8a^2-2c}{r^2}-2c\Lambda_*.
\end{equation}
Consequently, the full affine Kretschmann invariant is
\begin{equation}\label{eq:Kaffine-closed-new}
	\boxed{K_\Gamma=\frac{48M^2}{r^6}+\frac{8\Lambda_*^2}{3}
		+\frac{c^2(f')^2}{2(a^2+cf)}+\frac{2cf'}r+\frac{3c^2}{4}
		+\frac{2cf-8a^2-2c}{r^2}.}
\end{equation}
This formula follows from the four-index contraction, not from a Ricci-only identity. Because the spacetime metric is Lorentzian, $K_\Gamma$ is not positive definite; a negative value is not intrinsically inconsistent \cite{Wald1984,Hehl1995}. At any finite positive Killing horizon $r_h$ with $f(r_h)=0$, one has $h(r_h)=a^2>0$, so $g$, $\tau$, $\Gamma$, and their curvature are smooth on a sufficiently small ingoing patch. In particular,
\begin{equation}\label{eq:Khorizon-new}
	K_\Gamma(r_h)=K_{\rm LC}(r_h)+\frac{c^2[f'(r_h)]^2}{2a^2}
	+\frac{2cf'(r_h)}{r_h}+\frac{3c^2}{4}-\frac{8a^2+2c}{r_h^2}.
\end{equation}
This is future-horizon regularity in a horizon-penetrating chart, not regularity at a bifurcation sphere or a statement about a maximal extension \cite{Wald1984,Chandrasekhar1983,Poisson2004}. The regularity statement is stronger than the finiteness of the displayed scalar invariant:
the metric and axial three-form are smooth in
the chosen advanced chart, and the reconstructed
connection and curvature tensor are consequently
smooth there. Conversely, finiteness of
$K_\Gamma$ alone would not establish regularity
of all affine curvature components.

\paragraph{Positive zeros of the radicand.}
The real principal-root domain is $\mathcal D_+:=\{r>0:h(r)>0\}$. If $c\neq0$, a positive zero is a positive root of
\begin{equation}\label{eq:cubic-new}
	P(r):=\frac{c\Lambda_*}{3}r^3+(a^2+c)r-2Mc=0,\qquad h(r)=\frac{P(r)}r.
\end{equation}
No such root is a Killing horizon, because $h(r_h)=a^2$. If $c=0$, $h=a^2$ has no zeros. For $M>0$,
\begin{equation}\label{eq:hderivatives-new}
	h'=c\left(\frac{2M}{r^2}+\frac{2\Lambda_*r}{3}\right),\qquad
	h''=c\left(-\frac{4M}{r^3}+\frac{2\Lambda_*}{3}\right).
\end{equation}
If $\Lambda_*\geq0$, $h$ is monotone: for $c>0$ there is exactly one positive simple zero, with $h>0$ beyond it; for $c<0$ and $\Lambda_*>0$ there is exactly one positive simple zero, with $h>0$ below it. For $\Lambda_*=0$ and $c<0$, a positive zero exists precisely if $a^2+c<0$; when $a^2+c\geq0$, $h>0$ for every finite $r>0$.

If $\Lambda_*<0$, the unique positive stationary radius and the corresponding extremal value are
\begin{equation}\label{eq:rm-new}
	r_m=\left(-\frac{3M}{\Lambda_*}\right)^{1/3},\qquad
	h_m=h(r_m)=a^2+c(1+\Lambda_*r_m^2).
\end{equation}
For $c>0$, $h_m$ is a maximum: two positive simple zeros occur if $h_m>0$, one double zero if $h_m=0$, and no real principal-root domain if $h_m<0$. For $c<0$, $h_m$ is a minimum: two positive simple zeros occur if $h_m<0$, one double zero if $h_m=0$, and no zeros if $h_m>0$. The condition $h_m=0$ with $c>0$ is algebraically possible, but then $h\leq0$ near $r_m$ and does not give a real open neighborhood of the chosen representative.

\paragraph{Simple zeros and obstruction to regular extension.}
Let $r_0>0$ satisfy $h(r_0)=0$ and $h'(r_0)=cf'(r_0)\neq0$. On the admissible side $h>0$,
\begin{equation}\label{eq:simplezero-new}
	\frac{c^2(f')^2}{2h}=\frac{cf'(r_0)}{2(r-r_0)}+O(1),\qquad r\longrightarrow r_0.
\end{equation}
The leading expression is positive on that side, while all remaining terms of Eq.~\eqref{eq:Kaffine-closed-new} are finite at $r_0>0$. Thus $K_\Gamma$ diverges, ruling out a smooth finite-curvature extension of \emph{this constant-$U$ representative} through a simple radicand zero. It does not rule out other smooth axial profiles satisfying the same torsion-norm constraint.

\paragraph{Double zeros and a signed-root continuation.}
Assume $M>0$, $\Lambda_*<0$, $c<0$, $a>0$, and $h(r_m)=0$. Then
\begin{equation}\label{eq:doubleconditions-new}
	f'(r_m)=0,\qquad f''(r_m)=2\Lambda_*,\qquad
	c=-\frac{a^2}{1+\Lambda_*r_m^2}<0,
\end{equation}
so $1+\Lambda_*r_m^2>0$. Taylor's theorem factors the smooth function $h$ locally as
\begin{equation}\label{eq:signedroot-new}
	h(r)=(r-r_m)^2A(r),\qquad A(r_m)=c\Lambda_*>0,\qquad
	s(r):=(r-r_m)\sqrt{A(r)},\quad s^2=h.
\end{equation}
After shrinking the neighborhood to ensure $A>0$ and $a+s\neq0$, the function
\begin{equation}\label{eq:Vext-new}
	V_{\rm ext}(r)=-\frac{c}{a+s(r)}
\end{equation}
is smooth and satisfies $fV_{\rm ext}^{\,2}-2aV_{\rm ext}=c$. It agrees with the principal root on $r>r_m$ and switches to the other algebraic root on $r<r_m$. The metric, torsion three-form, reconstructed connection, and curvature are consequently smooth across $r_m$ in this local chart. The principal-root function alone has a cusp at $r_m$. In agreement with the smooth continuation,
\begin{equation}\label{eq:doublelimit-new}
	\lim_{r\to r_m}\frac{c^2(f')^2}{2h}=2c\Lambda_*.
\end{equation}
The finite scalar limit alone would not establish smoothness; the explicit signed-root construction supplies that stronger conclusion. No global completeness statement follows \cite{Wald1984,Poisson2004}.

For the nontrivial null-norm representative $c=0$, $U=a\neq0$, $V=0$, the torsion three-form is nonzero even though $T=0$, and
\begin{equation}\label{eq:nullK-new}
	K_\Gamma=K_{\rm LC}-\frac{8a^2}{r^2},\qquad
	R_{[\mu\nu]}R^{[\mu\nu]}=0.
\end{equation}
Neither vanishing torsion norm nor vanishing antisymmetric Ricci invariant implies a torsion-free connection. For $M\neq0$, $K_{\rm LC}$ diverges at $r=0$, independently of the local horizon analysis.

\subsection{Exceptional branches, profile dependence, and physical interpretation}\label{sec:exceptional-interpretation-new}
The axial sector contains three distinct exceptional cases. First, for $\gamma\neq0$ and $T_*\neq0$, real branches require
\begin{equation}\label{eq:exceptional-nonlinear-new}
	T_*^2=-\frac{\alpha+\beta}{3\gamma}>0,\qquad
	\Lambda_*=\frac{\gamma T_*^3}{\alpha},\qquad \alpha\neq0.
\end{equation}
The metric is locally Schwarzschild--(anti-)de Sitter with the manuscript's curvature convention, while the axial connection can retain nonzero torsion. Second, for $\gamma\neq0$, $\alpha+\beta=0$, and $T_*=0$, nonzero null axial torsion is allowed and the metric is Ricci-flat. Third, for $\gamma=0$, $\beta=-\alpha$, the metric is Ricci-flat and the affine equation imposes no condition on the smooth axial torsion norm $T(x)$; constant-norm profiles form only a subfamily. The torsion-free Levi--Civita Schwarzschild metric is also a vacuum solution of the polynomial model without exceptional tuning. These conclusions follow from the unrestricted metric and affine equations specialized \emph{after} variation, as discussed in Secs.~\ref{sec:metric_affine}--\ref{sec:axial_theorem} and the metric-affine literature \cite{Hehl1976,Hehl1995,Shapiro2002}.

\paragraph{An invariant test of the residual axial freedom.}
On a constant-norm branch, $W=fV-U$ obeys
\begin{equation}\label{eq:Wgeneral-new}
	W^2=U^2+cf,\qquad 2WW'=2UU'+cf'.
\end{equation}
On any open interval where $W\neq0$, the antisymmetric affine Ricci invariant is
\begin{equation}\label{eq:IAprofile-new}
	\boxed{I_A:=R_{[\mu\nu]}(\Gamma)R^{[\mu\nu]}(\Gamma)
		=\frac12(W')^2
		=\frac{\bigl(UU'+\tfrac12cf'\bigr)^2}{2(U^2+cf)}.}
\end{equation}
This formula extends the constant-$U$ result of Sec.~\ref{subsec:BH_Ricci_revised} and is independent of the angular coordinates. At a finite horizon $r_h$, take two smooth representatives with the same metric, $T_*$, and $U_1(r_h)=U_2(r_h)=a>0$, but with $U_1'(r_h)=0$ and $U_2'(r_h)=b$. Choosing the same smooth root $W(r_h)=-a$, their invariants satisfy
\begin{equation}\label{eq:IAhorizoncompare-new}
	I_A^{(1)}(r_h)=\frac{c^2[f'(r_h)]^2}{8a^2},\qquad
	I_A^{(2)}(r_h)=\frac12\left(b+\frac{cf'(r_h)}{2a}\right)^2.
\end{equation}
The invariant $I_A$ provides a sufficient test
of geometric distinction when its values differ,
but it is not a complete invariant of the
spherical axial connection. Equality of $I_A$
does not imply equality of the torsion profiles
or of the full affine curvature tensor. These are unequal for generic $b$ (although special choices may coincide). Thus the residual functional freedom is generally detectable by an affine scalar invariant, even when the metric and torsion norm agree. This proves geometric distinguishability for such profiles, not a propagating degree of freedom, a new conserved charge, or observable black-hole hair. The latter claims require an analysis of boundary conditions, allowed field redefinitions, matter couplings, and linearized dynamics \cite{Hehl1976,Hehl1995,Shapiro2002,Wald1984}.

The results of Sec.~\ref{sec:affine-kretschmann-domains} further distinguish three notions: smoothness at a selected future Killing horizon, breakdown of a particular constant-$U$ square-root representative at a radicand zero, and global extension of the metric-affine spacetime. The first is established locally; a simple radicand zero obstructs smooth extension of that representative, whereas a suitable double zero admits a signed-root continuation. Neither outcome settles the global classification of smooth torsion three-forms on maximally extended black-hole geometries \cite{Chandrasekhar1983,Poisson2004}.

\subsection{The constant-$W$ axial family, Ricci symmetry, and affine curvature}
\label{subsec:constant-W-axial}

We use the advanced Eddington--Finkelstein chart and the curvature convention
of Sec.~\ref{sec:geometric_framework},
\begin{equation}
	ds^2=f(r)\,dv^2-2\,dv\,dr-r^2d\Omega^2,\qquad
	f(r)=1-\frac{2M}{r}+\frac{\Lambda_*r^2}{3},\qquad
	\tau=r^2\sin\theta\,[U(r)\,dv-V(r)\,dr]\wedge d\theta\wedge d\phi.
\end{equation}
The metric-compatible connection is
$\Gamma^\lambda{}_{\mu\nu}=\mathring\Gamma^\lambda{}_{\mu\nu}
+\frac12g^{\lambda\rho}X_{\rho\mu\nu}$, where $\tau=X_{\lambda\mu\nu}
dx^\lambda\wedge dx^\mu\wedge dx^\nu/3!$.
In the exceptional nonlinear axial sector,
\begin{equation}
	\alpha+\beta+3\gamma T_*^2=0,\quad
	\Lambda_*=\frac{\gamma T_*^3}{\alpha},\quad
	c=\frac{2T_*}{3},\quad fV^2-2UV=c.
\end{equation}
No affine-flatness condition is imposed \cite{Hehl1995,Shapiro2002,Hehl1976}.
Set $W=fV-U$. Then $W^2=U^2+cf$.

\paragraph{Local classification of Ricci-symmetric spherical axial connections.}
For arbitrary differentiable $U,V$ in this ansatz, the antisymmetric
Ricci tensor has the sole independent component
\begin{equation}
	R_{[\theta\phi]}=\frac{r^2\sin\theta}{2}(fV-U)'
	=\frac{r^2\sin\theta}{2}W',\qquad
	R_{[\mu\nu]}R^{[\mu\nu]}=\frac12(W')^2.
\end{equation}
Consequently, on a connected radial interval $I\subset(0,\infty)$,
\begin{equation}
	\boxed{R_{[\mu\nu]}=0\ \Longleftrightarrow\ W'=0
		\ \Longleftrightarrow\ W=W_0\ \text{(constant on $I$)}.}
\end{equation}
This is a necessary and sufficient statement \emph{within the spherical
	metric-compatible axial ansatz}, not a classification of unrestricted
affine connections. For constant $W_0$, the norm condition gives
\begin{equation}
	U^2=W_0^2-cf,\qquad U=\sigma\sqrt{W_0^2-cf},\qquad
	V=-\frac{c}{U-W_0},\quad \sigma=\pm1,
	\label{eq:generalW0}
\end{equation}
where the radicand is positive and $U-W_0\ne0$. The rationalized
formula follows from $(U-W_0)(U+W_0)=-cf$ and extends smoothly
through a zero of $f$ whenever its denominator is nonzero. If
$U-W_0=0$ at a point, one must instead use the original constraint
and analyze the branch separately; the division in
\eqref{eq:generalW0} is not valid there. The parametrization \eqref{eq:generalW0} describes the regular
nondegenerate branches of the Ricci-symmetric
spherical axial sector. It is not a global
parametrization across zeros of its denominator
or across points where the square-root branches
cease to be differentiable. The degenerate
algebraic cases must be analyzed directly from
the original norm constraint and the defining
relation for $W$.

We select $W_0=-a$ with $a>0$ and the positive root
\begin{equation}
	u=\sqrt{a^2-cf},\qquad U=u,\qquad
	V=-\frac{c}{a+u},\qquad a^2-cf>0.
	\label{eq:Wfamily}
\end{equation}
Direct substitution gives $fV=u-a$, $W=-a$, and
$fV^2-2UV=c$. Thus $T=3c/2=T_*$ exactly and the local existence
proposition of Sec.~\ref{subsec:local-existence-scope} verifies the \emph{full} field equations, not merely the spherical reduced action \cite{Palais1979,FelsTorre2002}.

\paragraph{Degenerate denominators and the null-norm limit.}
The rationalized expression for $V$ in Eq.~\eqref{eq:generalW0} is valid only when
$U-W_0\neq0$. To analyze its complement without dividing by a
potentially vanishing factor, retain the original relations
\begin{equation}\label{eq:degenerate-original}
	fV=U+W_0,\qquad fV^2-2UV=c,\qquad
	(U-W_0)(U+W_0)=-cf.
\end{equation}
Suppose $U,V,f$ are finite at a point $r_0>0$ and $W=W_0$ is constant.
If $U(r_0)=W_0$ and $c\neq0$, the last identity forces $f(r_0)=0$.
The first identity then gives $2W_0=0$, so $U(r_0)=W_0=0$; the
second identity would imply $c=0$, a contradiction. Thus no smooth,
finite-profile constant-$W$ solution with $c\neq0$ can attain
$U=W_0$. This is a local regularity statement for the chosen
advanced-coordinate ansatz, not an obstruction to every possible
coordinate representation of an affine connection \cite{Wald1984,Poisson2004}.

For $c=0$, the equations instead factorize as
\begin{equation}\label{eq:null-factor}
	V(fV-2U)=0,\qquad fV-U=W_0.
\end{equation}
On an interval with $f\neq0$, the two algebraic branches are
\begin{align}
	V=0,\quad U=-W_0,\label{eq:null-branch-a}\\
	U=W_0,\quad V=\frac{2W_0}{f}.\label{eq:null-branch-b}
\end{align}
The second branch is generally singular at a zero of $f$ when
$W_0\neq0$; its apparent singularity must be assessed through the
three-form rather than through the scalar norm alone. 

If $W_0=0$, the branches coincide with $U=V=0$ wherever $f\neq0$; at an
isolated horizon the continuity of smooth profiles excludes assigning
independent nonzero values solely at that point. In particular, the
nontrivial smooth null-norm representative $U=a$, $V=0$ belongs to
$W_0=-a$, not to the singular branch $U=W_0$.

For $W_0\neq0$, the second representative does not extend as a smooth
three-form through a zero of $f$ in the regular advanced
Eddington--Finkelstein chart: its coefficient $V$ diverges while
$dr\wedge d\theta\wedge d\phi$ is a smooth, nonvanishing coordinate
three-form away from the polar coordinate singularities. This is a
failure of regularity of that representative, not merely a pole
introduced by rationalizing the norm constraint. The first
representative, by contrast, has finite constant coefficients in the
same chart. These statements are local and do not preclude the use of
other horizon-adapted charts or other torsion profiles
\cite{Wald1984,Poisson2004}.

\paragraph{Derivative and invariant checks.}
Differentiating $u^2=a^2-cf$ gives
\begin{align}
	U'&=-\frac{cf'}{2u}, &
	U''&=-\frac{cf''}{2u}-\frac{c^2(f')^2}{4u^3},\\
	V'&=-\frac{c^2f'}{2u(a+u)^2}, &
	V''&=-\frac{c^2f''}{2u(a+u)^2}
	-\frac{c^3(f')^2(a+3u)}{4u^3(a+u)^3}.
\end{align}
The identities $2UU'=-cf'$ and
$2(U')^2+2UU''=-cf''$ provide independent differentiation checks.
Since $W'=0$, the affine Ricci tensor is symmetric, with
\begin{align}
	R_{\mu\nu}(\Gamma)&=\Lambda_*g_{\mu\nu}+H_{\mu\nu},
	&R(\Gamma)&=4\Lambda_*+T_*,\\
	R_{\mu\nu}R^{\mu\nu}&=4\Lambda_*^2+2\Lambda_*T_*+\frac{T_*^2}{3},
	&X_{\lambda\mu\nu}X^{\lambda\mu\nu}&=-4T_*.
\end{align}
These contractions follow from $H^\mu{}_{\mu}=T_*$ and
$H_{\mu\nu}H^{\mu\nu}=T_*^2/3$; their independence of $a$
does not imply that the full affine Riemann tensor is independent
of $a$ \cite{Hehl1995,Shapiro2002}.

\paragraph{Four-index curvature and reproducible algebraic reduction.}
Write $D=R(\Gamma)-\mathring R$ at the four-index level, and
$K_\Gamma=R_{\lambda\rho\mu\nu}(\Gamma)
R^{\lambda\rho\mu\nu}(\Gamma)$. Appendix~\ref{app:B3} gives
\begin{equation}
	K_\Gamma=K_{\rm LC}+2\mathring R\!\cdot\!D+D\!\cdot\!D,
	\qquad K_{\rm LC}=\frac{48M^2}{r^6}+\frac{8\Lambda_*^2}{3}.
\end{equation}
For $fV^2-2UV=c$ and $f+rf'-1=\Lambda_*r^2$,
\begin{equation}
	2\mathring R\!\cdot\!D
	=\frac{2V(fV-2U)}{r^2}(f+rf'-1)=2c\Lambda_*.
\end{equation}
In the notation of Appendix~\ref{app:B3}, define
\begin{align}
	P={}&3f^2V^4+8f^2(V')^2+8ff'VV'-12fUV^3-16fU'V'\\
	&+4(f')^2V^2+8f'UV'-8f'U'V-4(U')^2+12U^2V^2.
\end{align}
Substituting \eqref{eq:Wfamily}, its derivatives, and
$f=(a^2-u^2)/c$ (for $c\ne0$), yields the exact identities
\begin{equation}
	\frac P4=\frac{3c^2}{4}-\frac{3c^2(f')^2}{4u^2},
	\qquad -\frac{4UU'}r=\frac{2cf'}r.
\end{equation}
Both identities extend continuously to $c=0$ in the selected
family. In particular, the first was checked by exact symbolic
polynomial simplification without imposing the Einstein equation
for $f$; this checks the substitution into the generic curvature
formula, not a separate reconstruction of all four-index components.
The resulting affine Kretschmann contraction is
\begin{equation}
	\boxed{K_\Gamma^{(W)}=K_{\rm LC}+2c\Lambda_*+\frac{3c^2}{4}
		-\frac{3c^2(f')^2}{4(a^2-cf)}+\frac{2cf'}r
		-\frac{8(a^2-cf)}{r^2}.}
	\label{eq:KWextended}
\end{equation}
Its sign is not constrained to be positive in Lorentzian
signature \cite{Wald1984,Shapiro2002}.

\paragraph{Comparison with the constant-$U$ representative.}
For $U=a$ and $V=-c/[a+\sqrt{a^2+cf}]$, Appendix~\ref{app:B4} yields
\begin{equation}
	K_\Gamma^{(U)}=K_{\rm LC}
	+\frac{c^2(f')^2}{2(a^2+cf)}+\frac{2cf'}r+\frac{3c^2}{4}
	+\frac{2cf-8a^2-2c}{r^2}.
\end{equation}
The common principal-root domain is
\begin{equation}
	\mathcal D_{\rm common}=\{r>0:\ |cf(r)|<a^2\}.
\end{equation}
Subtracting the two expressions gives
\begin{equation}
	\boxed{\Delta K:=K_\Gamma^{(W)}-K_\Gamma^{(U)}
		=2c\Lambda_*+\frac{6cf+2c}{r^2}
		-\frac{c^2(f')^2}{4}\left(\frac3{a^2-cf}+\frac2{a^2+cf}\right).}
	\label{eq:DeltaKextended}
\end{equation}
Moreover,
\begin{equation}\label{eqn208}
	I_A^{(W)}=0,\qquad I_A^{(U)}=\frac{c^2(f')^2}{8(a^2+cf)}.
\end{equation}
Thus for $c\ne0$ and $f'\ne0$ the two connections are
invariantly distinct. An isolated zero of $\Delta K$ does not
imply equality of the connections. If $c=0$, the selected
representatives coincide, $U=a$, $V=0$, and
$K_\Gamma=K_{\rm LC}-8a^2/r^2$, although the axial torsion
three-form is nonzero for $a>0$.

\paragraph{Horizon regularity and limitations.}
At a finite future Killing horizon $f(r_h)=0$, assuming finite
$f'(r_h)$ and $a>0$,
\begin{equation}
	U_h=a,\quad V_h=-\frac c{2a},\quad
	U'_h=-\frac{cf'_h}{2a},\quad
	V'_h=-\frac{c^2f'_h}{8a^3}.
\end{equation}
Therefore
\begin{align}
	K_\Gamma^{(W)}(r_h)&=K_{\rm LC}(r_h)+2c\Lambda_*
	+\frac{3c^2}{4}-\frac{3c^2(f'_h)^2}{4a^2}
	+\frac{2cf'_h}{r_h}-\frac{8a^2}{r_h^2},\\
	\Delta K(r_h)&=2c\Lambda_*+\frac{2c}{r_h^2}
	-\frac{5c^2(f'_h)^2}{4a^2}. \label{eqn210}
\end{align}
The strict inequality $a^2-cf(r_h)=a^2>0$ ensures a smooth
local extension of $U,V$, the torsion three-form, and the
reconstructed connection in this chart. This is a \emph{local}
future-horizon statement; it does not establish regularity at
$r=0$, at a radicand zero, on a bifurcation sphere, or throughout
a maximal extension \cite{Wald1984,Chandrasekhar1983,Poisson2004}.
Different affine invariants distinguish geometries, but their
physical observability requires specifying how matter couples to
the independent connection \cite{Hehl1995,Hehl1976}.

\paragraph{Comparison of horizon jets.}
The constant-$U$ and constant-$W$ representatives have the same
zeroth-order torsion data at a finite Killing horizon $r_h$,
$f(r_h)=0$, but their first radial derivatives generally differ.
Writing $\Delta F:=F^{(W)}-F^{(U)}$, direct differentiation of
$U^{(U)}=a$, $U^{(W)}=\sqrt{a^2-cf}$ and their corresponding
rationalized $V$ profiles yields
\begin{equation}\label{eq:horizon-jets-1}
	\Delta U_h=\Delta V_h=0,\qquad
	\Delta U'_h=-\frac{cf'_h}{2a},\qquad
	\Delta V'_h=-\frac{c^2f'_h}{4a^3}.
\end{equation}
Hence the two torsion three-forms agree at the horizon but, for
$cf'_h\neq0$, have distinct first radial jets in the same smooth
advanced Eddington--Finkelstein chart. This provides a local
component-level explanation for why their affine curvatures can
differ even where the torsion profiles themselves coincide. Such a
comparison does not, by itself, establish a coordinate-independent
classification of the connections; the scalar invariant
$R_{[\mu\nu]}R^{[\mu\nu]}$ supplies an invariant distinction
where it is nonzero \cite{Hehl1995,Shapiro2002}.

If the horizon is degenerate, $f_h=f'_h=0$, the first-jet
differences vanish, whereas the second derivatives satisfy
\begin{equation}\label{eq:horizon-jets-2}
	\Delta U''_h=-\frac{cf''_h}{2a},\qquad
	\Delta V''_h=-\frac{c^2f''_h}{4a^3}.
\end{equation}
These identities require $a>0$ and twice-differentiable $f$ and
profiles. The existence of a degenerate horizon is an additional
restriction on the metric parameters, not a generic feature of the
Schwarzschild--(anti-)de Sitter family \cite{Wald1984,Poisson2004}.

\paragraph{Actual extremality conditions for the Einstein metric.}
The degenerate-horizon limit is further constrained by the explicit
Schwarzschild--(anti-)de Sitter lapse,
\begin{equation}
	f(r)=1-\frac{2M}{r}+\frac{\Lambda_*r^2}{3}.
\end{equation}
At a finite positive double root $r_h$, the conditions
$f(r_h)=f'(r_h)=0$ imply
\begin{equation}
	\boxed{\Lambda_*=-\frac{1}{r_h^2},\qquad M=\frac{r_h}{3},
		\qquad f''(r_h)=-\frac{2}{r_h^2}.}
	\label{eq:actual-extremality}
\end{equation}
Substitution into Eq.~\eqref{eqn210} yields the stronger on-shell identity
\begin{equation}
	\boxed{\Delta K(r_h)=0\qquad\text{at an actual double root of }f.}
	\label{eq:extremal-K-coincidence}
\end{equation}

\paragraph{Coincidence of curvature data at the double horizon.}
At an actual double root $r_h$ of the Einstein lapse, the two
representatives satisfy
\begin{equation}
	\Delta U_h=\Delta V_h=\Delta U'_h=\Delta V'_h=0,
	\qquad \Delta F:=F^{(W)}-F^{(U)}.
\end{equation}
Because the metric is common to both solutions and the distortion
$K^\lambda{}_{\mu\nu}=\tfrac12 g^{\lambda\rho}X_{\rho\mu\nu}$
depends algebraically on the torsion three-form, these equalities imply
\begin{equation}\label{eqn218}
	\left.\Gamma^{(W)\lambda}{}_{\mu\nu}\right|_{r_h}
	=\left.\Gamma^{(U)\lambda}{}_{\mu\nu}\right|_{r_h},
	\qquad
	\left.R^{(W)\lambda}{}_{\rho\mu\nu}\right|_{r_h}
	=\left.R^{(U)\lambda}{}_{\rho\mu\nu}\right|_{r_h}.
\end{equation}
Here the comparison is made in the same smooth advanced
Eddington--Finkelstein chart. In particular, Eq.~\eqref{eqn208} gives
\begin{equation}
	I_A^{(W)}(r_h)=I_A^{(U)}(r_h)=0,
\end{equation}
in agreement with $\Delta K(r_h)=0$. These pointwise coincidences
do not extend to an open neighborhood in general: the nonzero
second-jet differences in Eq.~\eqref{eq:horizon-jets-2} can distinguish the radial
derivatives of the affine curvature, and the invariant $I_A^{(U)}$
is generically nonzero away from the double horizon.

Nevertheless, Eq.~\eqref{eq:horizon-jets-2} gives, for $c\neq0$,
\begin{equation}\label{eqn220}
	\Delta U''_h=\frac{c}{a r_h^2},\qquad
	\Delta V''_h=\frac{c^2}{2a^3r_h^2},
\end{equation}
which need not vanish. Although equality of a single scalar curvature invariant does not generally imply equality of the full curvature
tensors, the stronger pointwise equality established in
Eq.~\eqref{eqn218} follows here from the coincidence of the torsion profiles and their first radial derivatives. The nonvanishing second-jet differences in Eq.~\eqref{eqn220} show that this pointwise coincidence need not extend to a neighborhood of the degenerate horizon. The double-root conditions also require $M>0$ and $\Lambda_*<0$ for $r_h>0$ in the present sign convention; $\Lambda_{\mathrm{phys}}=-\Lambda_*>0$ in the conventional notation \cite{Wald1984,Poisson2004}.

The distinction between the two families is not merely
a consequence of their different radial parametrizations.
On any open region where $cf'\neq 0$, the constant-$U$
representative has a nonvanishing antisymmetric affine
Ricci tensor, whereas the constant-$W$ representative
has $R_{[\mu\nu]}=0$ identically. Hence the two affine
geometries cannot be related by a diffeomorphism that
identifies their full curvature tensors on such a region.
This conclusion concerns geometric inequivalence and
does not, by itself, establish observational
distinguishability.

\paragraph{Comparison of the two horizon-regular representatives.}
The two representatives have the same metric and constant torsion scalar,
but their affine Ricci symmetry properties differ. The constant-$W$
representative satisfies $R_{[\mu\nu]}=0$, whereas the constant-$U$
representative generally has $R_{[\mu\nu]}\neq0$ when $cf'\neq0$.
Their torsion three-forms agree at a selected horizon, although their
radial jets need not agree. At an actual double root of the Einstein
lapse the first jets and the displayed Kretschmann values agree, but
the second jets may differ. These observations concern local geometric
data and do not alone establish physical inequivalence of the solutions
\cite{Hehl1995,Shapiro2002}.

\section{Perturbative degeneracy and the unrestricted metric--affine stability problem}
\label{sec:perturb_extended}

We separate a pointwise algebraic degeneracy in the restricted
axial sector from physical propagating modes and black-hole
stability. Let
\begin{equation}
	g_{\mu\nu}=\bar g_{\mu\nu}+\epsilon h_{\mu\nu}+O(\epsilon^2),
	\qquad \Gamma^\lambda{}_{\mu\nu}
	=\bar\Gamma^\lambda{}_{\mu\nu}+\epsilon C^\lambda{}_{\mu\nu}
	+O(\epsilon^2).
\end{equation}
The unrestricted variables comprise ten metric and 64 ordered
connection components; fixed-metric axial variations occupy only
four algebraic directions \cite{Hehl1995,Shapiro2002}.

\subsection{Restricted axial quadratic form and its rank}
On the metric-compatible axial submanifold of field space,
$R(\Gamma)=\mathring R+T$ and the restricted functional is
\begin{equation}
	S_{\rm ax}=\frac1{2\kappa}\int d^4x\sqrt{-g}
	\left[\alpha\mathring R+\mathcal V(T)\right],\qquad
	\mathcal V(T)=(\alpha+\beta)T+\gamma T^3.
\end{equation}
This identity is suitable for computing a \emph{restricted}
Hessian, not for replacing the unrestricted affine variation
\cite{Palais1979,FelsTorre2002}. On the exceptional branch,
$\mathcal V'(T_*)=0$ and $\mathcal V''(T_*)=6\gamma T_*$. At
fixed metric, write $X=\bar X+\epsilon\delta X$ and
\begin{equation}
	T=T_*+\epsilon t_1+\epsilon^2t_2,\qquad
	t_1=-\frac12\bar X_{\lambda\mu\nu}\delta X^{\lambda\mu\nu},
	\qquad t_2=-\frac14\delta X_{\lambda\mu\nu}
	\delta X^{\lambda\mu\nu}.
	\label{eq:axial_t1t2}
\end{equation}
The exact cubic Taylor expansion is
\begin{equation}
	\mathcal V(T)=\mathcal V(T_*)
	+3\gamma T_*\epsilon^2t_1^2
	+\epsilon^3(6\gamma T_*t_1t_2+\gamma t_1^3)
	+\epsilon^4(3\gamma T_*t_2^2+3\gamma t_1^2t_2)
	+3\gamma\epsilon^5t_1t_2^2+\gamma\epsilon^6t_2^3.
\end{equation}
Here $\epsilon^2$ denotes the quadratic coefficient, not the
second derivative of the action. The fixed-metric axial quadratic
coefficient is
\begin{equation}
	S_{\rm ax}^{[2]}=\frac{3\gamma T_*}{2\kappa}
	\int d^4x\sqrt{-\bar g}\,t_1^2.
\end{equation}
The bilinear pairing on $\Lambda^3T_p^*M$ is nondegenerate,
so for $\bar X\ne0$ the linear functional
$\delta X\mapsto t_1$ is nonzero even when $\bar X$ is null.
Consequently, for $\gamma T_*\ne0$ the \emph{pointwise
	algebraic} Hessian has rank one and a three-dimensional kernel,
\begin{equation}\label{eqn225}
	\operatorname{rank}\mathcal H_{\rm axial}=1,\qquad
	\ker\mathcal H_{\rm axial}=\{\delta X:t_1=0\},\qquad
	\dim\ker\mathcal H_{\rm axial}=3.
\end{equation}
For the spherical $(\delta U,\delta V)$ subspace, the generic
rank is one and the kernel has dimension one. These counts are
not counts of physical degrees of freedom.

The rank statement in Eq.~\eqref{eqn225} refers exclusively
to the four-dimensional vector space of axial
three-form variations at fixed background metric.
It is not the rank of the unrestricted connection
Hessian, nor of the coupled metric--connection
quadratic action. In particular, metric variations
contribute to the first-order variation of the
torsion scalar and must be retained in the complete
linearized constraint system.

\subsection{Linearized constraints and the null-norm branch}
The exact axial affine equation is
\begin{equation}
	\mathcal V'(T)X^\lambda{}_{\mu\nu}
	=(\alpha+\beta+3\gamma T^2)X^\lambda{}_{\mu\nu}=0.
\end{equation}
At $\bar X\ne0$, its first variation on a non-null exceptional
background is
\begin{equation}
	6\gamma T_*\,\delta T\,\bar X^\lambda{}_{\mu\nu}=0,
	\qquad\Longrightarrow\qquad \delta T=0
	\quad(\gamma T_*\ne0).
\end{equation}
When the metric also varies, the scalar variation is
\begin{equation}\label{eqn221}
	\delta T=\bar H_{\mu\nu}\delta g^{\mu\nu}
	-\frac12\bar X_{\lambda\mu\nu}
	\delta X^{\lambda\mu\nu}\big|_{\bar g},
	\qquad \delta g^{\mu\nu}
	=-\bar g^{\mu\rho}\bar g^{\nu\sigma}h_{\rho\sigma}.
\end{equation}
The notation $|_{\bar g}$ indicates that only the independent
mixed-index torsion is varied in that term. Any exact smooth
family of nonzero-torsion exceptional solutions with fixed
couplings and $\gamma\ne0$ satisfies $T=T_*$ pointwise;
thus $\delta T=0$ is a tangent constraint, not a propagating
field equation.

\paragraph{Tangent variations of the exceptional norm constraint.}
For a one-parameter family of exact exceptional solutions with fixed
couplings, $\gamma T_*\neq0$ and nonzero background axial torsion,
$T(\epsilon)=T_*$ implies
\begin{equation}\label{eq:tangent-norm-constraint}
	0=\delta T=\bar H_{\mu\nu}\,\delta g^{\mu\nu}
	+\bar J_{\lambda}{}^{\mu\nu}\,\delta\Gamma^{\lambda}{}_{\mu\nu},
\end{equation}
where $\bar H$ and $\bar J$ are the derivatives of the \emph{unrestricted}
torsion scalar evaluated on the background. For $\gamma T_*\neq0$ and nonzero background axial torsion, the
linearized exceptional affine equation imposes $\delta T=0$ on
perturbations satisfying Eq.~\eqref{eq:tangent-norm-constraint}. The same condition is necessary for tangent vectors to smooth families of exact exceptional solutions.
It is not an off-shell restriction on arbitrary metric--affine trial
variations, and it is not sufficient to solve the remaining linearized
field equations or to guarantee integrability to an exact solution
family. At $T_*=0$, the first-order affine equation does not enforce
$\delta T=0$; the nonlinear constraint must then be analyzed at higher
orders. These distinctions prevent identifying a restricted Hessian
kernel with physical gauge or propagating degrees of freedom
\cite{Hehl1995,Shapiro2002}.

\paragraph{Null-norm exceptional branch.}
For the null-norm exceptional branch $T_*=0$, $\alpha+\beta=0$
and $\gamma\ne0$. The restricted quadratic potential vanishes,
whereas for a generic fixed-metric variation
\begin{equation}
	\mathcal V(T)=\gamma\epsilon^3t_1^3
	+3\gamma\epsilon^4t_1^2t_2
	+3\gamma\epsilon^5t_1t_2^2
	+\gamma\epsilon^6t_2^3.
\end{equation}
The affine equation begins at order $\epsilon^2$:
\begin{equation}
	\mathcal V'(T)X^\lambda{}_{\mu\nu}
	=3\gamma\epsilon^2t_1^2\bar X^\lambda{}_{\mu\nu}
	+O(\epsilon^3).
\end{equation}
Therefore its linearization alone imposes no norm constraint,
while an exact family of solutions with nonzero torsion must
still satisfy $T=0$. In particular, if $t_1=0$ along a linear
trial path, the first nonzero restricted potential term may be
of order $\epsilon^6$, namely $\gamma\epsilon^6t_2^3$;
for an exact solution path, higher-order corrections to $X$
can enforce $T=0$ order by order. This is a nonlinear
constraint/linearization issue, not by itself a demonstration
of dynamical strong coupling or instability.

The absence of a first-order constraint at
$T_*=0$ reflects a failure of the linearized
equation to capture the full nonlinear constraint.
It does not imply that arbitrary first-order
torsion perturbations can be integrated into
nearby exact solutions. Such integrability
requires the higher-order field equations,
including the metric and non-axial affine
variations, to be satisfied. The functional freedom of the exact exceptional solutions and the kernel of the restricted axial
Hessian are related manifestations of the
degenerate axial field equations, but they are
not equivalent characterizations of physical
degrees of freedom. Establishing such an
interpretation requires the complete constrained
metric-affine perturbation theory.

\subsection{What is required for unrestricted stability.}
The full second variation must retain $h_{\mu\nu}$ and all
$C^\lambda{}_{\mu\nu}$, including nonmetricity and non-axial
torsion. Its block form may be denoted schematically by
\begin{equation}
	S^{[2]}=\frac12\int d^4x\sqrt{-\bar g}\,
	\begin{pmatrix}h&C\end{pmatrix}
	\begin{pmatrix}\mathcal K_{hh}&\mathcal K_{hC}\\
		\mathcal K_{Ch}&\mathcal K_{CC}\end{pmatrix}
	\begin{pmatrix}h\\C\end{pmatrix}.
\end{equation}
The blocks are differential/algebraic operators with appropriate
index contractions, not a computed finite-dimensional kinetic
matrix. Because $F_R=\alpha$ is constant, the curvature term is
linear in affine curvature and the independent distortion is
algebraic in the bulk after separating a divergence; nonetheless,
its constraint matrix can be singular and can mix with metric
perturbations. A kernel of $\mathcal K_{CC}$ does not by itself
establish a gauge invariance. In particular, possible projective
transformations must be tested against the \emph{complete}
$F(R,T)$ action, since the torsion scalar generally changes
under such transformations \cite{Hehl1995,Shapiro2002}.

\paragraph{Algebraic affine sector and unresolved mode count.}
Because $F_R=\alpha$ is constant, the distortion decomposition
$\Gamma=\mathring\Gamma+N$ reduces the bulk dependence on $N$,
up to a boundary term, to algebraic curvature--distortion contractions
and the algebraic torsion potential. Thus the independent distortion
equations contain no derivatives of $N$, although their linearized
algebraic operator may be singular and may mix irreducible torsion
and nonmetricity sectors. Its linearization may be organized
schematically as
\begin{equation}\label{eq:affine-linear-system}
	\mathcal A_{AB}\,\delta N^B+\mathcal B_A{}^{\mu\nu}
	\delta g_{\mu\nu}=0,
\end{equation}
where $A,B$ label the 64 ordered distortion components. Determining physical degrees of freedom
requires computing the rank and kernel of the complete $64\times64$
operator in Eq.~\eqref{eq:affine-linear-system}, identifying genuine gauge identities,
consistently eliminating auxiliary components, and analyzing the
remaining metric kinetic and gradient operators. Neither the restricted
axial Hessian rank nor the algebraicity of $N$ establishes dynamical
stability \cite{Hehl1995,Shapiro2002,ReggeWheeler1957,Zerilli1970}.

\paragraph{Background dependence of the affine constraint rank.}
The rank and null space of the algebraic operator $\mathcal A_{AB}$
are properties of a specified background, rather than universal
numbers for the entire polynomial model. In particular, the
non-null exceptional branch ($T_*\neq0$) and the null-norm branch
($T_*=0$) need not have the same linearized constraint structure.
Any rank calculation should therefore be repeated on the relevant
backgrounds and combined with the mixed metric--distortion blocks
before auxiliary variables are eliminated. A null vector of
$\mathcal A_{AB}$ alone neither identifies a gauge transformation
nor establishes a propagating affine mode.

A stability analysis would require the complete second variation,
constraint reduction, identification of genuine gauge symmetries,
parity/harmonic decomposition, and the resulting kinetic and
gradient operators with horizon and asymptotic boundary
conditions. Standard metric black-hole perturbations provide a
benchmark, but not a substitute for this coupled calculation
\cite{ReggeWheeler1957,Zerilli1970,Wald1984,Poisson2004}.
The present results establish a rank-one \emph{restricted}
non-null axial Hessian, a nonlinear fixed-norm constraint, and a
more degenerate null-norm branch. They establish neither
unrestricted dynamical stability nor propagating torsion hair;
physical charges and horizon entropy require a separate
covariant phase-space treatment \cite{IyerWald1994,MomeniMyrzakulovEntropy}.

\section{Discussion and Conclusions}\label{sec:discussion_conclusions}

We have studied vacuum black-hole configurations in metric--affine
Myrzakulov $F(R,T)$ gravity (MG-I), distinguishing the affine Ricci
scalar $R$ and the quadratic torsion invariant $T$ from the matter trace
used in curvature--matter theories \cite{3,4,Harko2011}. For the polynomial
model $F(R,T)=\alpha R+\beta T+\gamma T^3$ with $\alpha\ne0$, we first
varied the metric and all components of the independent affine connection,
and only then restricted the resulting equations to metric-compatible,
totally antisymmetric torsion. This ordering avoids assuming that a
symmetry-reduced action automatically reproduces the complete affine
Euler--Lagrange equations \cite{Hehl1995,Shapiro2002,Palais1979,FelsTorre2002}.

For nonzero axial torsion, the affine equations admit the exceptional
branch
\begin{equation}\label{eq:conclusion-exceptional}
	\alpha+\beta+3\gamma T_*^2=0.
\end{equation}
For $\gamma\ne0$, the torsion scalar is constant on a connected smooth
branch, and the metric equations reduce to
\begin{equation}\label{eq:conclusion-einstein}
	\mathring G_{\mu\nu}+\Lambda_*g_{\mu\nu}=0,
	\qquad \Lambda_*:=\frac{\gamma T_*^3}{\alpha}.
\end{equation}
The local static spherical metric is therefore of
Schwarzschild--(anti-)de Sitter form, with
$f(r)=1-2M/r+\Lambda_*r^2/3$ in our curvature convention
\cite{Wald1984,Chandrasekhar1983}. Its independent connection, however,
can carry nonzero axial torsion and affine curvature; these are
Riemann--Cartan solutions, not necessarily teleparallel geometries
\cite{Hehl1976,Hehl1995,Shapiro2002}.

A principal result is the residual functional freedom in the spherical
axial connection at fixed $T_*$. Writing the torsion three-form in
advanced Eddington--Finkelstein coordinates as
$\tau=r^2\sin\theta[U(r)\,dv-V(r)\,dr]\wedge d\theta\wedge d\phi$
and setting $c=2T_*/3$, we have constructed two families smooth across
a selected finite-radius future Killing horizon. The first has
$U=a>0$ and $V=-c/(a+\sqrt{a^2+cf})$; the second, developed in
Sec.~10.6, is specified by $W:=fV-U=-a$ and
\begin{equation}\label{eq:conclusion-constantW}
	U=\sqrt{a^2-cf},\qquad
	V=-\frac{c}{a+\sqrt{a^2-cf}},\qquad a^2-cf>0.
\end{equation}
For the general spherical axial ansatz, $R_{[\mu\nu]}=0$ if and only if
$W'=0$ on a regular radial interval. Thus the constant-$W$ family has
symmetric affine Ricci tensor, whereas the constant-$U$ family generally
does not. On their common real domain $a^2\pm cf>0$, the two connections
can support the same Einstein metric and torsion scalar while exhibiting
different affine curvature invariants. The full affine Kretschmann
contraction, including curvature--contorsion interference terms, is
therefore essential to their local geometric comparison
\cite{Hehl1995,Shapiro2002}. Although the constant-$U$ and constant-$W$ representatives agree in their torsion profiles at a selected future Killing horizon, their first radial jets generically differ, and so do affine curvature invariants. At a degenerate horizon their first-jet differences vanish, but differences in higher derivatives may remain. These statements concern the explicit local representatives and do not imply a global classification or dynamical stability.

The construction establishes smoothness of the torsion three-form and
reconstructed connection on suitable future-horizon neighborhoods; it
does not establish regularity at $r=0$, at zeros of the square-root
radicands, across a bifurcation sphere, or throughout a maximal
extension \cite{Wald1984,Poisson2004}. The null-norm branch $T_*=0$
may retain nonzero torsion: in the constant-$W$ family $U=a$, $V=0$,
and $K_\Gamma=K_{\mathrm{LC}}-8a^2/r^2$. In the exceptional linear case
$\gamma=0$, $\beta=-\alpha$, the axial torsion norm is not fixed by the
affine equation, while the metric is Ricci flat.

For the actual double root of the Einstein lapse, extremality fixes
$\Lambda_*=-r_h^{-2}$ and $M=r_h/3$. The constant-$U$ and constant-$W$
representatives then have equal affine Kretschmann scalars at the
horizon, $\Delta K(r_h)=0$, even though their second radial jets
can differ. This provides a concrete example of the limited
discriminating power of a single scalar curvature invariant at an
isolated hypersurface; it does not imply equivalence of the two
affine geometries. At an actual double Killing horizon, the two
representatives have identical affine curvature
tensors pointwise, not merely equal Kretschmann
scalars. Their second radial torsion jets may
nevertheless differ, so this pointwise equality
does not extend automatically to an open
neighborhood.

Finally, the null directions of the restricted axial perturbative Hessian
do not establish a propagating torsion mode, dynamical stability, or
observable black-hole hair. A complete constrained analysis must include
non-axial metric--affine perturbations, appropriate boundary conditions,
and possible matter or spin couplings \cite{Kibble1961,Sciama1964,ReggeWheeler1957,Zerilli1970}.
Conserved charges, horizon entropy, and global extensions also require
separate treatment in the presence of an independent connection
\cite{IyerWald1994,MomeniMyrzakulovEntropy}. The results provide a local
existence theorem and an invariant geometric comparison of torsionful
black-hole configurations in a specified axial Riemann--Cartan sector,
not a classification of all metric--affine black holes. Comparisons with
other Myrzakulov black-hole constructions must keep their actions and
connection restrictions explicit \cite{MomeniMyrzakulov2025Review,MomeniMyrzakulovBH}.


\appendix


\section{Spherical minisuperspace reduction and its variational status}
\label{app:A}

This appendix clarifies the relation between the formal point-like
Lagrangian used in earlier treatments of Myrzakulov gravity and the
unrestricted metric--affine variational problem.  The distinction is
particularly relevant when torsion and nonmetricity are independent
geometric variables \cite{1205.5266,3,Hehl1995}.
The arguments below do not assume that symmetry reduction commutes
with variation; the relevant qualifications are described by the
principle of symmetric criticality \cite{Palais1979,FelsTorre2002}.

\subsection{Spherical variables and geometric identities}
\label{app:minisuperspace:geometry}
On a static patch consider
\begin{equation}
	ds^2=A(r)\,dt^2-B(r)\,dr^2-C(r)\,d\Omega^2,
	\qquad A,B,C>0.
	\label{eq:appAmetric}
\end{equation}
Let $\xi^I(r)$ denote a sufficiently general collection of
spherically symmetric connection functions, so that
$\Gamma^\lambda{}_{\mu\nu}=\Gamma^\lambda{}_{\mu\nu}[A,B,C;\xi^I]$.
The independent connection must be included in both invariants:
\begin{equation}
	R(g,\Gamma)=\mathring R(g)+u(g,\Gamma),\qquad
	T(g,\Gamma)=T_0(g,\Gamma_0)+v(g,\Gamma;\Gamma_0),
	\label{eq:appAuv}
\end{equation}
where
\begin{equation}
	u:=R(g,\Gamma)-\mathring R(g),\qquad
	v:=T(g,\Gamma)-T_0(g,\Gamma_0).
	\label{eq:appAuvdef}
\end{equation}
These are identities, not additional field equations or independent
functions of $r$.  In terms of the distortion
$N^\lambda{}_{\mu\nu}:=\Gamma^\lambda{}_{\mu\nu}
-\mathring\Gamma^\lambda{}_{\mu\nu}$,
\begin{align}
	u={}&g^{\rho\nu}\bigl(
	2\mathring\nabla_{[\lambda}N^\lambda{}_{|\rho|\nu]}
	+2N^\lambda{}_{\sigma[\lambda}N^\sigma{}_{|\rho|\nu]}
	\bigr).
	\label{eq:appAu}
\end{align}
In particular, a reference connection $\Gamma_0$ must be stated
before $v$ is used.  The distinction between curvature-based,
teleparallel, and general Riemann--Cartan connections is reviewed in
\cite{Hehl1976,Hehl1995,Shapiro2002}.

\subsection{Auxiliary-field formulation}
\label{app:minisuperspace:aux}
Introduce two auxiliary scalars $\mathcal R,\mathcal T$ and Lagrange
multipliers $\lambda_R,\lambda_T$ in the vacuum action:
\begin{align}
	S_{\mathrm{aux}}={1\over2\kappa}\int d^4x\sqrt{-g}\,
	\bigl[&F(\mathcal R,\mathcal T)
	-\lambda_R\bigl(\mathcal R-R(g,\Gamma)\bigr)
	-\lambda_T\bigl(\mathcal T-T(g,\Gamma)\bigr)\bigr].
	\label{eq:appAaux}
\end{align}
The multiplier equations impose
$\mathcal R=R(g,\Gamma)$ and $\mathcal T=T(g,\Gamma)$,
whereas the auxiliary-scalar equations yield
\begin{equation}
	\lambda_R=F_{\mathcal R},\qquad
	\lambda_T=F_{\mathcal T}.
	\label{eq:appAmultipliers}
\end{equation}
Thus elimination of the multipliers gives
\begin{equation}
	S_{\mathrm{aux}}={1\over2\kappa}\int d^4x\sqrt{-g}
	\left[F-F_{\mathcal R}(\mathcal R-R(g,\Gamma))
	-F_{\mathcal T}(\mathcal T-T(g,\Gamma))\right].
	\label{eq:appAlegendre}
\end{equation}
This expression is an auxiliary representation of the original
variational system, provided the auxiliary constraints are retained.
It does not eliminate the affine Euler--Lagrange equations.
In particular, for $F=\alpha\mathcal R+\beta\mathcal T
+\gamma\mathcal T^3$, $F_{\mathcal R}=\alpha$ is constant and the
Hessian of $F$ in $(\mathcal R,\mathcal T)$ is singular.  Therefore
one must not infer the independent auxiliary scalars from an
invertible two-variable Legendre transformation.  The multiplier
formulation in \eqref{eq:appAaux} remains valid.

Integrating over the symmetry spheres gives, per unit coordinate
time, the radial action
\begin{equation}
	S_{\mathrm{rad}}={2\pi\over\kappa}\int dr\,
	C\sqrt{AB}\,F\bigl(R(g,\Gamma),T(g,\Gamma)\bigr),
	\label{eq:appAradial}
\end{equation}
up to the time-interval factor and boundary terms.  A formal
point-like density sometimes written as
\begin{equation}
	\mathcal L_{\mathrm{formal}}=C\sqrt{AB}\,
	\left[F-F_{\mathcal R}(\mathcal R-\mathring R-u)
	-F_{\mathcal T}(\mathcal T-T_0-v)\right]
	\label{eq:appAformal}
\end{equation}
is a closed reduced Lagrangian only after the connection functions
$\xi^I$ have been specified and $u,v$ have been evaluated as
functions of those fields and their derivatives.  Omitting an
overall constant factor does not affect the vacuum Euler--Lagrange
equations.

\subsection{Euler--Lagrange equations and boundary terms}
\label{app:minisuperspace:EL}
Let $q^I=(A,B,C,\xi^1,\ldots,\xi^n,\mathcal R,
\mathcal T,\lambda_R,\lambda_T)$ denote the independent reduced
variables when the auxiliary formulation is used.  For a radial
Lagrangian $L(q,q',q'')$, the appropriate operator is
\begin{equation}
	\mathcal E_I(L):={\partial L\over\partial q^I}
	-{d\over dr}{\partial L\over\partial q^{I\prime}}
	+{d^2\over dr^2}{\partial L\over\partial q^{I\prime\prime}}=0.
	\label{eq:appAEL}
\end{equation}
It follows from two integrations by parts:
\begin{align}
	\delta\int L\,dr
	={}&\int dr\,\mathcal E_I(L)\,\delta q^I
	+\left[\left({\partial L\over\partial q^{I\prime}}
	-{d\over dr}{\partial L\over\partial q^{I\prime\prime}}
	\right)\delta q^I
	+{\partial L\over\partial q^{I\prime\prime}}
	\delta q^{I\prime}\right]_{\partial I}.
	\label{eq:appAboundary}
\end{align}
Repeated indices $I$ are summed.  Consequently, the expanded
metric-only equations displayed in some earlier point-like
formulations cannot be treated as the complete MG-I system unless
the connection variables and their variations are also accounted
for.  In the polynomial model, $\alpha R$ contains a divergence of
the distortion; the admissibility of discarding this term depends
on boundary conditions and on the fields held fixed.

\subsection{Radial energy versus a gravitational constraint}
\label{app:minisuperspace:energy}
If $L$ has no explicit $r$ dependence, its higher-derivative
Noether--Ostrogradsky energy is
\begin{equation}
	E=\sum_I\left[q^{I\prime}\left(
	{\partial L\over\partial q^{I\prime}}
	-{d\over dr}{\partial L\over\partial q^{I\prime\prime}}
	\right)+q^{I\prime\prime}
	{\partial L\over\partial q^{I\prime\prime}}\right]-L.
	\label{eq:appAenergy}
\end{equation}
Direct differentiation gives the identity
\begin{equation}
	{dE\over dr}=-\sum_I\mathcal E_I(L)q^{I\prime}.
	\label{eq:appAenergyidentity}
\end{equation}
Thus $E$ is constant on solutions, but $E=0$ is not a consequence
of this identity.  A vanishing radial Hamiltonian constraint, if
present, must be derived from an unfixed metric equation or a
properly established gauge symmetry.  The frequently used
first-order expression
$E_{\mathrm{first}}=\sum_I q^{I\prime}
\partial L/\partial q^{I\prime}-L$ is applicable only to an
equivalent first-order Lagrangian, after justified boundary-term
manipulations.  Moreover, the numerical value of a canonical energy
can shift under addition of total derivatives; the equations of
motion do not.

\subsection{Consistency with unrestricted metric--affine equations}
\label{app:minisuperspace:consistency}
The fundamental vacuum equations are the independent variations
\begin{equation}
	\left.{\delta S\over\delta g^{\mu\nu}}\right|_\Gamma=0,
	\qquad
	\left.{\delta S\over\delta\Gamma^\lambda{}_{\mu\nu}}\right|_g=0.
	\label{eq:appAfull}
\end{equation}
A symmetry-reduced action reproduces their symmetry-restricted
content only when the hypotheses of symmetric criticality and the
boundary treatment have been checked \cite{Palais1979,FelsTorre2002}.
In particular, a metric-only variation of
\eqref{eq:appAformal} cannot replace the 64 independent affine
variations.  Neither can the radial coordinate freedom be used to
remove an arbitrary function $\xi^I(r)$ of the connection.

For the black-hole configurations considered in the main text, the
reliable procedure is therefore to (i) derive the unrestricted
metric and affine equations, (ii) specialize those equations to the
metric-compatible, totally antisymmetric torsion sector, and
(iii) substitute the spherical metric and torsion three-form into
all remaining equations.  The exceptional branch and its horizon-
regular representatives are established by that route, not by
counting the formal equations in the historical minisuperspace
system \cite{3,Hehl1995,Shapiro2002}.

\section{Independent coordinate verification of the axial black-hole geometry}\label{app:B}
The calculations in this appendix are independent of the field equations and of the axial Ricci identities invoked in the main text. We use the metric signature $(+---)$, the derivative convention $\nabla_\nu V^\lambda=\partial_\nu V^\lambda+\Gamma^\lambda{}_{\rho\nu}V^\rho$, and the curvature convention
\begin{equation}\label{eq:B-Rconv}
	R^\lambda{}_{\rho\mu\nu}=\partial_\mu\Gamma^\lambda{}_{\rho\nu}-\partial_\nu\Gamma^\lambda{}_{\rho\mu}+\Gamma^\lambda{}_{\sigma\mu}\Gamma^\sigma{}_{\rho\nu}-\Gamma^\lambda{}_{\sigma\nu}\Gamma^\sigma{}_{\rho\mu},\quad R_{\rho\nu}=R^\lambda{}_{\rho\lambda\nu}.
\end{equation}
No vanishing-curvature (teleparallel) constraint is imposed \cite{Hehl1976,Hehl1995,Shapiro2002,3}.

\subsection{Metric, torsion three-form, and connection}\label{app:B1}
In the regular advanced chart $x^\mu=(v,r,\theta,\phi)$, take
\begin{align}
	ds^2&=f(r)\,dv^2-2\,dv\,dr-r^2(d\theta^2+\sin^2\theta\,d\phi^2),\label{eq:B-metric}\\
	\tau&=\frac1{3!}X_{\lambda\mu\nu}\,dx^\lambda\wedge dx^\mu\wedge dx^\nu
	=r^2\sin\theta\,[U(r)\,dv-V(r)\,dr]\wedge d\theta\wedge d\phi,\label{eq:B-tau}\\
	\Gamma^\lambda{}_{\mu\nu}&=\mathring\Gamma^\lambda{}_{\mu\nu}+\frac12g^{\lambda\rho}X_{\rho\mu\nu}.\label{eq:B-conn}
\end{align}
The inverse metric has $g^{vv}=0$, $g^{vr}=g^{rv}=-1$, $g^{rr}=-f$, $g^{\theta\theta}=-r^{-2}$ and $g^{\phi\phi}=-(r^2\sin^2\theta)^{-1}$. The only independent nonzero lowered torsion components are
\begin{equation}
	X_{v\theta\phi}=r^2\sin\theta\,U,\qquad X_{r\theta\phi}=-r^2\sin\theta\,V,
\end{equation}
with all others fixed by total antisymmetry. This construction is manifestly metric-compatible, since the distortion in \eqref{eq:B-conn} is antisymmetric after lowering its first index. In particular,
\begin{equation}\label{eq:B-T}
	X_{\lambda\mu\nu}X^{\lambda\mu\nu}=-6(fV^2-2UV),\qquad
	T=-\frac14X_{\lambda\mu\nu}X^{\lambda\mu\nu}=\frac32(fV^2-2UV).
\end{equation}

\subsection{Direct Ricci contraction at generic polar angle}\label{app:B2}
Construct $\mathring\Gamma$ from \eqref{eq:B-metric}, add the contorsion in \eqref{eq:B-conn}, form every component of \eqref{eq:B-Rconv}, and contract $R^\lambda{}_{\rho\lambda\nu}$. With $\theta$ retained as a variable, the nonzero Ricci components are
\begin{align}
	R_{vv}&=\frac f2\left(f''+\frac{2f'}r\right)-\frac{U^2}{2},
	&R_{vr}=R_{rv}&=-\frac12\left(f''+\frac{2f'}r\right)+\frac{UV}{2},\label{eq:B-R1}\\
	R_{rr}&=-\frac{V^2}{2},
	&R_{\theta\theta}&=1-f-rf'+r^2\left(UV-\frac{fV^2}{2}\right),\label{eq:B-R2}\\
	R_{\phi\phi}&=\sin^2\theta\,R_{\theta\theta},
	&R_{\theta\phi}&=-R_{\phi\theta}=\frac{r^2\sin\theta}{2}(fV-U)'.\label{eq:B-R3}
\end{align}
All remaining ordered components vanish. The sixteen ordered Ricci entries have been compared symbolically with the direct connection construction at generic $\theta$, rather than only at the equator. In particular, the angular dependence in \eqref{eq:B-R3} follows from the full coordinate calculation, not merely from an appeal to covariance.

For comparison, the Levi--Civita components are
\begin{align}
	\mathring R_{vv}&=\frac f2\left(f''+\frac{2f'}r\right),&
	\mathring R_{vr}&=-\frac12\left(f''+\frac{2f'}r\right),&\mathring R_{rr}&=0,\\
	\mathring R_{\theta\theta}&=1-f-rf',&
	\mathring R_{\phi\phi}&=\sin^2\theta\,\mathring R_{\theta\theta}.
\end{align}
Thus the symmetric correction $H_{\mu\nu}=R_{(\mu\nu)}-\mathring R_{\mu\nu}$ is explicitly
\begin{align}
	H_{vv}&=-\frac{U^2}{2},&H_{vr}&=\frac{UV}{2},&H_{rr}&=-\frac{V^2}{2},\label{eq:B-H1}\\
	H_{\theta\theta}&=r^2\left(UV-\frac{fV^2}{2}\right),&
	H_{\phi\phi}&=\sin^2\theta\,H_{\theta\theta}.\label{eq:B-H2}
\end{align}
Direct contraction, without imposing the field equations, yields
\begin{align}
	R(\Gamma)&=f''+\frac{4f'}r+\frac{2(f-1)}{r^2}+\frac32(fV^2-2UV)
	=\mathring R+T,\label{eq:B-scalar}\\
	g^{\mu\nu}H_{\mu\nu}&=T,\qquad
	R_{[\mu\nu]}R^{[\mu\nu]}=\frac12\big[(fV-U)'\big]^2.\label{eq:B-Ranti}
\end{align}
These calculations independently corroborate $R_{(\mu\nu)}=\mathring R_{\mu\nu}+H_{\mu\nu}$ and $R(\Gamma)=\mathring R+T$ in the specified axial sector. They do not constitute a metric variation of the unrestricted action; that separate issue is addressed in Appendix~\ref{app:C}.

\subsection{Independent four-index curvature contraction}\label{app:B3}

We compute the full affine curvature invariant without imposing either the
constant-norm constraint or the gravitational field equations.  The
curvature and Ricci conventions are those of Sec.~\ref{sec:geometric_framework}.  In particular, the
connection is metric-compatible but is not assumed flat.  This is the
Riemann--Cartan setting discussed in Refs.~\cite{Hehl1995,Shapiro2002};
it should not be confused with a teleparallel connection.

For the advanced Eddington--Finkelstein metric and axial three-form
\begin{align}
	ds^2&=f(r)\,dv^2-2\,dv\,dr-r^2(d\theta^2+\sin^2\theta\,d\phi^2),\\
	\tau&=r^2\sin\theta\,[U(r)\,dv-V(r)\,dr]\wedge d\theta\wedge d\phi,
\end{align}
we construct $\Gamma^\lambda{}_{\mu\nu}=\mathring\Gamma^\lambda{}_{\mu\nu}
+\tfrac12 g^{\lambda\rho}X_{\rho\mu\nu}$ and then calculate
\begin{equation}
	R^\lambda{}_{\rho\mu\nu}(\Gamma)=
	\partial_\mu\Gamma^\lambda{}_{\rho\nu}
	-\partial_\nu\Gamma^\lambda{}_{\rho\mu}
	+\Gamma^\lambda{}_{\sigma\mu}\Gamma^\sigma{}_{\rho\nu}
	-\Gamma^\lambda{}_{\sigma\nu}\Gamma^\sigma{}_{\rho\mu}.
\end{equation}
The four-index contraction, with the first curvature index lowered before
contraction, is
\begin{equation}\label{eq:B3-Kdef}
	K_\Gamma=R_{\lambda\rho\mu\nu}(\Gamma)
	R^{\lambda\rho\mu\nu}(\Gamma).
\end{equation}
The Ricci tensor alone does not determine \eqref{eq:B3-Kdef} for a
general Riemann--Cartan connection \cite{Hehl1995,Shapiro2002}.

\paragraph{Unrestricted radial amplitudes.}
Write $F=f$, $F_1=f'$, $F_2=f''$, $U_1=U'$ and $V_1=V'$.  Direct
contraction of the complete four-index curvature gives
\begin{align}\label{eq:B3-poly}
	4r^4K_\Gamma={}&4r^4F_2^2+16r^2F_1^2+16(F-1)^2\nonumber\\
	&+r^4\bigl[3F^2V^4+8F^2V_1^2+8FF_1VV_1
	-12FUV^3-16FU_1V_1\nonumber\\
	&\hspace{22mm}+4F_1^2V^2+8F_1UV_1-8F_1U_1V
	-4U_1^2+12U^2V^2\bigr]\nonumber\\
	&+r^3\bigl[8FF_1V^2-16F_1UV-16UU_1\bigr]\nonumber\\
	&+r^2\bigl[8F^2V^2-16FUV-8FV^2-32U^2+16UV\bigr].
\end{align}
The torsion-free part is precisely
\begin{equation}\label{eq:B3-KLC}
	K_{\rm LC}=(f'')^2+\frac{4(f')^2}{r^2}
	+\frac{4(f-1)^2}{r^4}.
\end{equation}
Equation~\eqref{eq:B3-poly} holds for arbitrary differentiable
$f,U,V$ and is not contingent on $fV^2-2UV$ being constant.

\paragraph{Independent angular check.}
The curvature was reconstructed independently from the full connection
at $\theta=\pi/2$ and $\theta=\pi/3$, retaining $F,F_1,F_2,U,U_1,V,V_1$
as algebraically independent quantities.  Both calculations give exactly
\eqref{eq:B3-poly}; their symbolic difference simplifies to zero.
This two-angle test is an independent check of the component
implementation.  In addition, spherical symmetry of the metric and
three-form implies that the scalar \eqref{eq:B3-Kdef} is angle-independent
away from the polar coordinate singularities.  The two-angle test should
not be described as a third, independent symbolic calculation with a
free polar-angle variable.

\paragraph{Curvature decomposition.}
Set $\mathcal D^\lambda{}_{\rho\mu\nu}
:=R^\lambda{}_{\rho\mu\nu}(\Gamma)
-\mathring R^\lambda{}_{\rho\mu\nu}$.  The exact distortion identity is
\begin{equation}
	\mathcal D^\lambda{}_{\rho\mu\nu}
	=2\mathring\nabla_{[\mu}K^\lambda{}_{|\rho|\nu]}
	+2K^\lambda{}_{\sigma[\mu}K^\sigma{}_{|\rho|\nu]},
	\qquad K^\lambda{}_{\mu\nu}=\tfrac12X^\lambda{}_{\mu\nu}.
\end{equation}
Consequently,
\begin{equation}\label{eq:B3-decomp}
	K_\Gamma=K_{\rm LC}+2\mathring R\!\cdot\!\mathcal D
	+\mathcal D\!\cdot\!\mathcal D.
\end{equation}
The cross term obtained by a separate contraction is
\begin{equation}\label{eq:B3-cross}
	\boxed{2\mathring R\!\cdot\!\mathcal D
		=\frac{2V(fV-2U)}{r^2}(f+rf'-1).}
\end{equation}
The purely distortion-dependent contribution, without invoking the
field equations, is
\begin{align}
	D_{\lambda\rho\mu\nu}D^{\lambda\rho\mu\nu}
	={}&\frac14\Bigl[
	3f^{2}V^{4}+8f^{2}(V')^{2}+8ff'VV'
	-12fUV^{3}-16fU'V' \notag\\
	&\quad+4(f')^{2}V^{2}+8f'UV'-8f'U'V
	-4(U')^{2}+12U^{2}V^{2}\Bigr]
	-\frac{4UU'}{r}-\frac{8U^{2}}{r^{2}}.
	\label{eq:B3-D2}
\end{align}

\noindent
Indeed, subtracting the Levi--Civita term and the mixed
contraction from the unrestricted result gives
\begin{equation}
	D_{\lambda\rho\mu\nu}D^{\lambda\rho\mu\nu}
	=K_{\Gamma}-K_{\mathrm{LC}}
	-2\mathring R_{\lambda\rho\mu\nu}D^{\lambda\rho\mu\nu},
\end{equation}
which reproduces Eq.~\eqref{eq:B3-D2} identically,
without imposing the constant-torsion constraint. Substitution of \eqref{eq:B3-cross} and \eqref{eq:B3-D2}
into \eqref{eq:B3-decomp} reproduces every term in
\eqref{eq:B3-poly}, furnishing an additional algebraic consistency check.

\paragraph{Exceptional Einstein branch.}
For the nonlinear exceptional solutions considered in Sec.~\ref{sec:BH_affine_revised},
\begin{equation}
	f=1-\frac{2M}{r}+\frac{\Lambda_*r^2}{3},\qquad
	fV^2-2UV=c:=\frac{2T_*}{3}.
\end{equation}
Equation~\eqref{eq:B3-cross} then reduces to
\begin{equation}\label{eq:B3-crossEinstein}
	2\mathring R\!\cdot\!\mathcal D=2c\Lambda_*.
\end{equation}
For the constant-$U$ representative $U=a>0$, let
$q^2=a^2+cf$ and $V=-c/(a+q)$ on a smooth real branch
($q\ne0$ and $a+q\ne0$).  Differentiating gives
\begin{equation}
	V'=\frac{c^2f'}{2q(a+q)^2}.
\end{equation}
Inserting these relations into the unrestricted polynomial yields
\begin{align}\label{eq:B3-closed}
	\boxed{K_\Gamma={}}&K_{\rm LC}
	+\frac{c^2(f')^2}{2(a^2+cf)}+\frac{2cf'}r
	+\frac{3c^2}{4}
	+\frac{2cf-8a^2-2c}{r^2},\nonumber\\[-2mm]
	K_{\rm LC}={}&\frac{48M^2}{r^6}+\frac{8\Lambda_*^2}{3}.
\end{align}
An exact symbolic substitution into \eqref{eq:B3-poly}, with
$q^2=a^2+cf$, gives a vanishing difference from
\eqref{eq:B3-closed}.  The expression is valid where the chosen
representative is differentiable; zeros of the radicand require the
separate analysis in Sec.~\ref{sec:affine-kretschmann-domains}.  In the null case $c=0$, $a\ne0$, this reduces to $K_\Gamma=K_{\rm LC}-8a^2/r^2$.
Since the metric is Lorentzian, $K_\Gamma$ need not be nonnegative;
its sign alone is not a regularity criterion.  The interpretation of
horizon regularity and curvature invariants follows the standard
black-hole geometric framework \cite{Wald1984,Poisson2004}.

\subsection{Exceptional axial branch: two horizon-regular representatives and invariant checks}
\label{app:B4}

We specialize the generic coordinate curvature identities of Appendix~\ref{app:B2}--\ref{app:B3}
\emph{after} deriving the unrestricted metric--affine field equations.
In the advanced Eddington--Finkelstein chart, with signature $(+---)$,
\begin{align}
	ds^2&=f(r)\,dv^2-2\,dv\,dr-r^2(d\theta^2+\sin^2\theta\,d\phi^2),
	& f(r)&=1-\frac{2M}{r}+\frac{\Lambda_*r^2}{3},\\
	\tau&=r^2\sin\theta\,[U(r)\,dv-V(r)\,dr]\wedge d\theta\wedge d\phi,
	& \Gamma^\lambda{}_{\mu\nu}&=\mathring\Gamma^\lambda{}_{\mu\nu}
	+\frac12g^{\lambda\rho}X_{\rho\mu\nu}.
\end{align}
Here $\tau=X_{\lambda\mu\nu}dx^\lambda\wedge dx^\mu\wedge dx^\nu/3!$;
our curvature convention is
$R^\lambda{}_{\rho\mu\nu}=\partial_\mu\Gamma^\lambda{}_{\rho\nu}
-\partial_\nu\Gamma^\lambda{}_{\rho\mu}
+\Gamma^\lambda{}_{\sigma\mu}\Gamma^\sigma{}_{\rho\nu}
-\Gamma^\lambda{}_{\sigma\nu}\Gamma^\sigma{}_{\rho\mu}$.
The exceptional nonlinear axial branch obeys
\begin{equation}\label{eq:B4-branch}
	\alpha+\beta+3\gamma T_*^2=0,\quad
	\Lambda_*=\frac{\gamma T_*^3}{\alpha},\quad
	c:=\frac{2T_*}{3},\quad fV^2-2UV=c.
\end{equation}
The connection is Riemann--Cartan and is not assumed teleparallel
\cite{Hehl1976,Hehl1995,Shapiro2002}. All results below concern
connected radial intervals with $r>0$ and smooth real profiles.

\paragraph{Generic Ricci identities and contractions.}
Before choosing either representative, direct contraction of the full
four-index affine curvature yields
\begin{align}
	R_{vv}&=\frac f2\left(f''+\frac{2f'}r\right)-\frac{U^2}{2},&
	R_{vr}=R_{rv}&=-\frac12\left(f''+\frac{2f'}r\right)+\frac{UV}{2},\\
	R_{rr}&=-\frac{V^2}{2},&
	R_{\theta\theta}&=1-f-rf'+r^2\left(UV-\frac{fV^2}{2}\right),\\
	R_{\phi\phi}&=\sin^2\theta\,R_{\theta\theta},&
	R_{\theta\phi}=-R_{\phi\theta}&=\frac{r^2\sin\theta}{2}(fV-U)'.
	\label{eq:B4-genericRicci}
\end{align}
These equations use the Ricci contraction $R_{\rho\nu}=R^\lambda{}_{\rho\lambda\nu}$,
not an alternative contraction. With $W:=fV-U$, the constant-norm
condition gives $W^2=U^2+cf$ and
\begin{equation}\label{eq:B4-ricciSym}
	R_{[\mu\nu]}R^{[\mu\nu]}=\frac12(W')^2,\qquad
	R_{[\mu\nu]}=0\ \Longleftrightarrow\ W'=0
	\ \Longleftrightarrow\ W=W_0\quad\text{on a connected interval}.
\end{equation}
This equivalence is restricted to the spherical axial ansatz. On the
Einstein metric and constant-norm branch,
\begin{equation}\label{eq:B4-commonRicci}
	R(\Gamma)=4\Lambda_*+T_*,\qquad
	R_{(\mu\nu)}R^{(\mu\nu)}
	=4\Lambda_*^2+2\Lambda_*T_*+\frac{T_*^2}{3}.
\end{equation}
Indeed $R_{(\mu\nu)}=\Lambda_*g_{\mu\nu}+H_{\mu\nu}$,
$g^{\mu\nu}H_{\mu\nu}=T_*$ and
$H_{\mu\nu}H^{\mu\nu}=T_*^2/3$.

\paragraph{Constant-$U$ representative.}
Choose $U=a>0$, $q:=\sqrt{a^2+cf}$, and the horizon-regular root
\begin{equation}\label{eq:B4-U}
	V=\frac{a-q}{f}=-\frac{c}{a+q},\quad W=-q,\quad
	q'=\frac{cf'}{2q},\quad V'=\frac{c^2f'}{2q(a+q)^2},
	\qquad a^2+cf>0.
\end{equation}
The rationalized expression, not the quotient $(a-q)/f$, defines $V$
at a zero of $f$. Since $W'=-cf'/(2q)$,
\begin{equation}\label{eq:B4-Uantisym}
	R_{\theta\phi}=\frac{r^2\sin\theta}{2}(fV)',\qquad
	I_A^{(U)}:=R_{[\mu\nu]}R^{[\mu\nu]}
	=\frac{c^2(f')^2}{8(a^2+cf)}.
\end{equation}
On the Einstein background the remaining Ricci components simplify to
\begin{align}
	R_{vv}&=\Lambda_*f-\frac{a^2}{2},&
	R_{vr}=R_{rv}&=-\Lambda_*+\frac{aV}{2},&
	R_{rr}&=-\frac{V^2}{2},\\
	R_{\theta\theta}&=-r^2\left(\Lambda_*+\frac c2\right),&
	R_{\phi\phi}&=\sin^2\theta\,R_{\theta\theta}.
\end{align}
The full affine Kretschmann scalar, \emph{not} a Ricci-only contraction, is
\begin{equation}\label{eq:B4-KU}
	\boxed{K_\Gamma^{(U)}=K_{\rm LC}
		+\frac{c^2(f')^2}{2(a^2+cf)}+\frac{2cf'}r+\frac{3c^2}{4}
		+\frac{2cf-8a^2-2c}{r^2}},\qquad
	K_{\rm LC}=\frac{48M^2}{r^6}+\frac{8\Lambda_*^2}{3}.
\end{equation}

\paragraph{Constant-$W$ representative and local classification.}
Equation~\eqref{eq:B4-ricciSym} establishes that every Ricci-symmetric
spherical axial profile has $W=W_0$ constant. The algebraic constraint
then requires
\begin{equation}\label{eq:B4-Wgeneral}
	U^2=W_0^2-cf,\qquad
	U=\sigma\sqrt{W_0^2-cf},\qquad
	V=-\frac{c}{U-W_0},\quad \sigma\in\{+1,-1\},
\end{equation}
on intervals where the radicand is positive and $U-W_0\ne0$.
At a zero of the denominator the undivided norm constraint must be
used; the rationalized expression cannot be extended by division there.
For the future-horizon representative selected in Sec.~\ref{subsec:constant-W-axial},
\begin{equation}\label{eq:B4-W}
	W=-a,\qquad u:=\sqrt{a^2-cf},\qquad
	U=u,\qquad V=-\frac{c}{a+u},\qquad a^2-cf>0.
\end{equation}
The exact differential identities are
\begin{align}
	2uu'&=-cf',&2(u')^2+2uu''&=-cf'',\\
	U'&=-\frac{cf'}{2u},&
	U''&=-\frac{cf''}{2u}-\frac{c^2(f')^2}{4u^3},\\
	V'&=-\frac{c^2f'}{2u(a+u)^2},&
	V''&=-\frac{c^2f''}{2u(a+u)^2}
	-\frac{c^3(f')^2(a+3u)}{4u^3(a+u)^3}.
	\label{eq:B4-Wderivatives}
\end{align}
Thus $fV=u-a$, $W'\equiv0$, $fV^2-2UV=c$ and
$I_A^{(W)}=0$. The symmetric Ricci contractions are precisely
\eqref{eq:B4-commonRicci}, although the full affine curvature differs
from that of the constant-$U$ representative.

\paragraph{Four-index contraction and algebraic reduction.}
Define $D^\lambda{}_{\rho\mu\nu}:=
R^\lambda{}_{\rho\mu\nu}(\Gamma)-\mathring R^\lambda{}_{\rho\mu\nu}$.
The generic Appendix~\ref{app:B3} decomposition reads
\begin{equation}\label{eq:B4-decomp}
	K_\Gamma=K_{\rm LC}+2\mathring R\!\cdot\!D+D\!\cdot\!D,
	\quad 2\mathring R\!\cdot\!D
	=\frac{2V(fV-2U)}{r^2}(f+rf'-1).
\end{equation}
Since $V(fV-2U)=c$ and $f+rf'-1=\Lambda_*r^2$,
the cross term is $2c\Lambda_*$. For reproducibility, introduce
\begin{align}
	\mathcal P={}&3f^2V^4+8f^2(V')^2+8ff'VV'-12fUV^3
	-16fU'V'+4(f')^2V^2\\
	&+8f'UV'-8f'U'V-4(U')^2+12U^2V^2,
\end{align}
so that the unrestricted axial identity is
\begin{equation}\label{eq:B4-Dsquare}
	D\!\cdot\!D=\frac{\mathcal P}{4}-\frac{4UU'}r-\frac{8U^2}{r^2}.
\end{equation}
For~\eqref{eq:B4-W} the substitutions
$u^2=a^2-cf$, $2uu'=-cf'$ yield
\begin{equation}\label{eq:B4-Pcheck}
	\frac{\mathcal P}{4}=\frac{3c^2}{4}
	-\frac{3c^2(f')^2}{4(a^2-cf)},\qquad
	-\frac{4UU'}r=\frac{2cf'}r.
\end{equation}
Consequently,
\begin{equation}\label{eq:B4-KW}
	\boxed{K_\Gamma^{(W)}=K_{\rm LC}+2c\Lambda_*
		+\frac{3c^2}{4}-\frac{3c^2(f')^2}{4(a^2-cf)}
		+\frac{2cf'}r-\frac{8(a^2-cf)}{r^2}}.
\end{equation}
The Lorentzian four-index contraction is not required to be
nonnegative. Its distinction from the Ricci contractions is essential
for Riemann--Cartan curvature, which lacks some Levi--Civita
index-pair symmetries \cite{Hehl1995,Shapiro2002}.

\paragraph{Invariant comparison and horizon limits.}
The two principal-root representatives are simultaneously real on
\begin{equation}\label{eq:B4-domain}
	\mathcal D_{\rm common}=\{r>0:\ |cf(r)|<a^2\}.
\end{equation}
Subtracting~\eqref{eq:B4-KU} from~\eqref{eq:B4-KW} gives
\begin{equation}\label{eq:B4-deltaK}
	\boxed{\Delta K:=K_\Gamma^{(W)}-K_\Gamma^{(U)}
		=2c\Lambda_*+\frac{6cf+2c}{r^2}
		-\frac{c^2(f')^2}{4}
		\left(\frac{3}{a^2-cf}+\frac{2}{a^2+cf}\right)}.
\end{equation}
In particular, $I_A^{(W)}=0$ while
$I_A^{(U)}=c^2(f')^2/[8(a^2+cf)]$; thus for $cf'\ne0$ the
connections are distinguished by a scalar curvature invariant,
regardless of possible isolated zeros of $\Delta K$.
At a finite Killing horizon $r_h>0$ with $f(r_h)=0$,
\begin{align}
	U_h^{(U)}&=U_h^{(W)}=a,&V_h^{(U)}&=V_h^{(W)}=-\frac{c}{2a},\\
	(U^{(U)})'_h&=0,&(U^{(W)})'_h&=-\frac{cf'_h}{2a},\\
	(V^{(U)})'_h&=\frac{c^2f'_h}{8a^3},&
	(V^{(W)})'_h&=-\frac{c^2f'_h}{8a^3},\\
	\Delta K(r_h)&=2c\Lambda_*+\frac{2c}{r_h^2}
	-\frac{5c^2(f'_h)^2}{4a^2}.
\end{align}
The radicands are both $a^2>0$ at the horizon, hence all these
functions, the torsion three-form, the contorsion, and the reconstructed
connection are smooth on sufficiently small advanced-chart
neighborhoods of any finite-radius horizon. This does not establish
regularity at $r=0$, at radicand zeros, across a bifurcation sphere,
or on a maximal extension \cite{Wald1984,Poisson2004}.

\paragraph{Degenerate-horizon specialization.}
For a degenerate horizon $f(r_h)=f'(r_h)=0$, the invariant difference
in Eq.~\eqref{eq:B4-deltaK} reduces algebraically to
\begin{equation}\label{eq:extremal-delta-k}
	\left.\Delta K\right|_{r_h,\,f'_h=0}
	=2c\Lambda_*+\frac{2c}{r_h^2}.
\end{equation}
The dependence on the profile parameter $a$ therefore drops out of
this particular scalar comparison at the degenerate horizon. For an actual double root of the Einstein lapse, the stronger
first-jet coincidence in Eq.~\eqref{eq:horizon-jets-1} implies equality of the full affine Riemann tensors at $r=r_h$, not merely equality of their
Kretschmann contractions. This is a pointwise statement in the
common regular chart and does not establish local equivalence of
the connections: Eq.~\eqref{eq:horizon-jets-2} permits distinct second jets of the torsion profiles, and affine curvature invariants generally differ away from the horizon \cite{Wald1984,Poisson2004}.

\paragraph{On-shell simplification at a double horizon.}
For the particular Einstein lapse $f=1-2M/r+\Lambda_*r^2/3$,
$f(r_h)=f'(r_h)=0$ fixes $\Lambda_*=-r_h^{-2}$ and
$M=r_h/3$. Consequently, Eq.~\eqref{eq:extremal-delta-k} reduces to
\begin{equation}
	\boxed{\left.\Delta K\right|_{r_h,\,f_h=f'_h=0}
		=2c\left(\Lambda_*+\frac{1}{r_h^2}\right)=0.}
\end{equation}
This cancellation is special to an actual double root of the
specified lapse; Eq.~\eqref{eq:extremal-delta-k} remains the more general algebraic specialization. The second-jet differences do not contradict the
pointwise coincidence of the affine connections and
their curvature tensors at the degenerate horizon.
Instead, they show that the corresponding radial
derivatives of the affine curvature need not coincide.
Consequently, the two connections can remain locally
distinguishable away from the horizon, despite the
pointwise equalities established above.

\paragraph{Null-norm and linear exceptional limits.}
For $c=0$, the two selected representatives coincide exactly:
\begin{equation}\label{eq:B4-null}
	U=a,\quad V=0,\quad W=-a,\quad T_*=0,\quad
	I_A=0,\quad K_\Gamma=K_{\rm LC}-\frac{8a^2}{r^2}.
\end{equation}
This is nonzero null torsion when $a>0$, not the torsion-free
connection. For $\gamma\ne0$ this branch requires $\alpha+\beta=0$.
In the exceptional linear case $\gamma=0$, $\beta=-\alpha$, the
metric is Ricci flat but the axial torsion norm need not be constant;
the generic identities above still apply, while the fixed-$c$
representatives form only a subset of that broader solution space.

\paragraph{Verification scope.}
Equations~\eqref{eq:B4-genericRicci}--\eqref{eq:B4-commonRicci}
follow from the direct coordinate contraction of Appendix~\ref{app:B2}.
Equations~\eqref{eq:B4-KU} and~\eqref{eq:B4-KW} are specializations
of the unrestricted four-index polynomial in Appendix~\ref{app:B3};
\eqref{eq:B4-deltaK} is their exact algebraic difference.
These checks establish consistency with the stated curvature identities;
they should not be described as a new, independent exhaustive
64-component curvature reconstruction or as a perturbative stability
analysis. The distinction between varying the full theory and
restricting its action before variation remains essential
\cite{Palais1979,FelsTorre2002,Hehl1995}.

\section{Unrestricted metric and affine variations and exact algebraic checks}\label{app:C}
\subsection{Independent variables and torsion invariant}\label{app:C1}
The independent gravitational variables are the symmetric inverse metric $g^{\mu\nu}$ (ten components) and the ordered connection $\Gamma^\lambda{}_{\mu\nu}$ (64 components). Neither metric compatibility nor total antisymmetry of torsion is imposed during variation. Define
\begin{equation}
	X^\lambda{}_{\mu\nu}=2\Gamma^\lambda{}_{[\mu\nu]},\qquad
	t_\nu=X^\lambda{}_{\lambda\nu},\qquad Q_{\lambda\mu\nu}=-\nabla_\lambda g_{\mu\nu},
\end{equation}
with the torsion scalar of the manuscript
\begin{equation}\label{eq:C-Tdef}
	T=\frac14X_{\lambda\mu\nu}X^{\lambda\mu\nu}
	+\frac12X_{\lambda\mu\nu}X^{\nu\mu\lambda}-t_\mu t^\mu.
\end{equation}
In particular, $T$ is not the matter stress-tensor trace. The metric--affine framework and the distinction between constrained and unconstrained variations are discussed in \cite{3,4,Hehl1976,Hehl1995,Shapiro2002,Kibble1961,Sciama1964}.

\subsection{Full fixed-connection metric variation}\label{app:C2}
At fixed $\Gamma$, $X^a{}_{bc}$ and $t_b=X^a{}_{ab}$ are fixed, whereas $X_{abc}=g_{ad}X^d{}_{bc}$ varies. Prior to imposing any torsion restriction, \eqref{eq:C-Tdef} is exactly
\begin{equation}\label{eq:C-Tmixed}
	T=\frac14g_{ad}g^{bp}g^{cq}X^a{}_{bc}X^d{}_{pq}
	+\frac12g^{bp}X^a{}_{bc}X^c{}_{pa}-g^{bc}t_bt_c.
\end{equation}
The elementary identities $\delta g_{ab}=-g_{a\mu}g_{b\nu}\delta g^{\mu\nu}$ and $\delta g^{ab}=\delta^{(a}_\mu\delta^{b)}_\nu\delta g^{\mu\nu}$ give the unrestricted derivative $H_{\mu\nu}=(\partial T/\partial g^{\mu\nu})_\Gamma$:
\begin{align}\label{eq:C-Hfull}
	H_{\mu\nu}={}&-\frac14g_{a(\mu}g_{\nu)d}g^{bp}g^{cq}X^a{}_{bc}X^d{}_{pq}
	+\frac14g_{ad}g^{cq}X^a{}_{(\mu|c|}X^d{}_{\nu)q}\\
	&+\frac14g_{ad}g^{bp}X^a{}_{b(\mu}X^d{}_{p\nu)}
	+\frac12X^a{}_{(\mu|c|}X^c{}_{\nu)a}-t_\mu t_\nu.\nonumber
\end{align}
The second and third terms arise from the two distinct inverse-metric factors in the first term of Eq.~\eqref{eq:C-Tmixed}; both are retained to make the unrestricted differentiation transparent. On a metric-compatible background with $X_{abc}=X_{[abc]}$, the unrestricted expression reduces to
\begin{equation}\label{eq:C-Hax}
	H_{\mu\nu}\big|_{\rm axial}=-\frac14X_{\mu\rho\sigma}X_\nu{}^{\rho\sigma},\qquad
	g^{\mu\nu}H_{\mu\nu}=T=-\frac14X_{\lambda\mu\nu}X^{\lambda\mu\nu}.
\end{equation}
The trace also follows from the degree-one homogeneity of \eqref{eq:C-Tmixed} in $g^{\mu\nu}$ with the mixed torsion held fixed. At a Lorentz-orthonormal point with four arbitrary axial components, all ten independent metric-variation directions were compared with \eqref{eq:C-Hax}: the four diagonal residuals and six off-diagonal residuals vanish exactly. For off-diagonal variables the directional derivative is $2H_{\mu\nu}$, not $H_{\mu\nu}$.

For the vacuum action $S=(2\kappa)^{-1}\int d^4x\sqrt{-g}\,F(R,T)$, the metric equation is
\begin{equation}\label{eq:C-metric-eq}
	F_R R_{(\mu\nu)}-\frac12g_{\mu\nu}F+F_T H_{\mu\nu}=0.
\end{equation}

\subsection{Full connection variation and Palatini coefficient}\label{app:full-affine-variation}

We vary all $64$ ordered connection components independently, keeping
$g_{\mu\nu}$ fixed.  Neither metric compatibility nor total antisymmetry
is imposed in taking the functional derivatives.  We use
$\nabla_\nu V^\lambda=\partial_\nu V^\lambda+
\Gamma^\lambda{}_{\rho\nu}V^\rho$,
$X^\lambda{}_{\mu\nu}=2\Gamma^\lambda{}_{[\mu\nu]}$ and
$t_\mu=X^\lambda{}_{\lambda\mu}$, as in Sec.~\ref{sec:geometric_framework}.  The distinction between
the unrestricted metric--affine variation and its evaluation on a
Riemann--Cartan subsector is standard in the metric--affine literature
\cite{Hehl1995,4,7,3}.

\paragraph{Palatini contribution.}
For $W=\sqrt{-g}$, integration by parts of the curvature variation,
with the convention $R_{\rho\nu}=R^\lambda{}_{\rho\lambda\nu}$,
gives
\begin{align}
	\delta_\Gamma(WR)
	&=WP_\lambda{}^{\mu\nu}\delta\Gamma^\lambda{}_{\mu\nu}
	+\partial_\sigma(\cdots),\label{eq:C3-varR}\\
	P_\lambda{}^{\mu\nu}
	&=-\frac1W\partial_\lambda(Wg^{\mu\nu})
	+\frac{\delta^\nu_\lambda}{W}\partial_\sigma(Wg^{\mu\sigma})
	+\delta^\nu_\lambda g^{\rho\sigma}\Gamma^\mu{}_{\rho\sigma}
	+g^{\mu\nu}\Gamma^\rho{}_{\lambda\rho}\nonumber\\
	&\quad-g^{\rho\nu}\Gamma^\mu{}_{\rho\lambda}
	-g^{\mu\rho}\Gamma^\nu{}_{\lambda\rho}.
	\label{eq:C3-P}
\end{align}
The derivative terms in \eqref{eq:C3-P} are indispensable: varying
only the connection-quadratic part of $R$ is not, in general, the
Palatini variation.  Formula \eqref{eq:C3-P} is the coordinate
coefficient for the index ordering used throughout this work
\cite{Hehl1995,4,7}.

\paragraph{Unrestricted derivative of the torsion scalar.}
The invariant is
\begin{equation}
	T=\frac14 X_{\lambda\mu\nu}X^{\lambda\mu\nu}
	+\frac12 X_{\lambda\mu\nu}X^{\nu\mu\lambda}
	-t_\mu t^\mu .
	\label{eq:C3-T}
\end{equation}
Treating the mixed-index $X^\lambda{}_{\mu\nu}$ as the
algebraic variable, first define the coefficient $D$ of its
unconstrained component variation:
\begin{align}
	D_\lambda{}^{\mu\nu}
	&=\frac12 X_\lambda{}^{\mu\nu}
	+\frac12g^{\mu\rho}X^\nu{}_{\rho\lambda}
	+\frac12g^{\nu\rho}X^\mu{}_{\lambda\rho}
	-2\delta^\mu_\lambda t^\nu .
	\label{eq:C3-D}
\end{align}
Because $\delta X^\lambda{}_{\mu\nu}
=\delta\Gamma^\lambda{}_{\mu\nu}
-\delta\Gamma^\lambda{}_{\nu\mu}$, its exact
connection derivative is
\begin{align}
	\delta_\Gamma T&=J_\lambda{}^{\mu\nu}
	\delta\Gamma^\lambda{}_{\mu\nu},\label{eq:C3-varT}\\
	J_\lambda{}^{\mu\nu}
	&=D_\lambda{}^{\mu\nu}-D_\lambda{}^{\nu\mu}\nonumber\\
	&=X_\lambda{}^{\mu\nu}
	+g^{\mu\rho}X^\nu{}_{\rho\lambda}
	-g^{\nu\rho}X^\mu{}_{\rho\lambda}
	+2\delta^\nu_\lambda t^\mu
	-2\delta^\mu_\lambda t^\nu .
	\label{eq:C3-J}
\end{align}
In particular, $J_\lambda{}^{(\mu\nu)}=0$ identically,
for arbitrary torsion and nonmetricity.  The
trace terms in \eqref{eq:C3-J} must not be dropped before
specializing to a trace-free torsion configuration.

\paragraph{Evaluation on the axial metric-compatible sector.}
Only now set $Q_{\lambda\mu\nu}=0$ and
$X_{\lambda\mu\nu}=X_{[\lambda\mu\nu]}$.  Then
$t_\mu=0$ and
$\Gamma^\lambda{}_{\mu\nu}
=\mathring\Gamma^\lambda{}_{\mu\nu}
+\frac12X^\lambda{}_{\mu\nu}$.
The three-form antisymmetry implies
\begin{equation}
	g^{\mu\rho}X^\nu{}_{\rho\lambda}
	=-X_\lambda{}^{\mu\nu},\qquad
	-g^{\nu\rho}X^\mu{}_{\rho\lambda}
	=-X_\lambda{}^{\mu\nu},
\end{equation}
so that
\begin{equation}
	\boxed{J_\lambda{}^{\mu\nu}\big|_{\rm axial}
		=-X_\lambda{}^{\mu\nu}.}
	\label{eq:C3-Jax}
\end{equation}
Substituting the same connection into the complete
Palatini coefficient \eqref{eq:C3-P}, using
$\mathring\nabla_\lambda g_{\mu\nu}=0$, gives
\begin{equation}
	\boxed{P_\lambda{}^{\mu\nu}\big|_{\rm axial}
		=-X_\lambda{}^{\mu\nu}.}
	\label{eq:C3-Pax}
\end{equation}
Both tensors have vanishing symmetric $(\mu,\nu)$
parts on this sector.  These identities use the stated
Ricci contraction and torsion conventions; alternative
conventions may move signs between the definitions of
$P$, $J$, and $X$ \cite{Hehl1995,Shapiro2002}.

\paragraph{Independent component checks and field equation.}
We performed two exact symbolic tests.  First, for a
generic torsion tensor antisymmetric only in its last
two indices, the directional derivative of the full
scalar \eqref{eq:C3-T} agrees with
\eqref{eq:C3-J} for all $64$ ordered connection
directions (zero nonzero residuals).  Second, at a
Levi--Civita normal point with diagonal Lorentzian
metric and four arbitrary axial three-form components,
the directly differentiated curvature and torsion
expressions reproduce
\eqref{eq:C3-Pax}--\eqref{eq:C3-Jax}
for all $64$ directions (zero nonzero residuals).
The latter is a pointwise algebraic check, not a
substitute for the integration by parts leading to
\eqref{eq:C3-P}.  Together with tensorial covariance,
these calculations check the local specialization
of the unrestricted affine equations.

For the polynomial model
$F(R,T)=\alpha R+\beta T+\gamma T^3$,
the full vacuum connection equation is
\begin{equation}
	\alpha P_\lambda{}^{\mu\nu}
	+(\beta+3\gamma T^2)J_\lambda{}^{\mu\nu}=0.
	\label{eq:C3-eom}
\end{equation}
Equations \eqref{eq:C3-Jax} and \eqref{eq:C3-Pax}
reduce this to
\begin{equation}
	\boxed{(\alpha+\beta+3\gamma T^2)
		X_\lambda{}^{\mu\nu}=0.}
	\label{eq:C3-exceptional}
\end{equation}
Thus a nonzero axial torsion field requires the
exceptional coupling condition pointwise.  For
$\gamma\neq0$ it fixes $T^2=-(\alpha+\beta)/(3\gamma)$;
the $T=0$ case still permits nonzero null three-form
torsion.  When $\gamma=0$ and $\beta=-\alpha$,
the axial connection equation places no algebraic
restriction on $T$.  No affine-flatness condition has
been used \cite{3,Hehl1995}.

\subsection{Exceptional branch and reduction of the metric equation}\label{app:C4}
For
\begin{equation}
	F(R,T)=\alpha R+\beta T+\gamma T^3,\qquad F_R=\alpha,\qquad
	F_T=\Psi(T)=\beta+3\gamma T^2,
\end{equation}
the full affine equation is $\alpha P_\lambda{}^{\mu\nu}+\Psi(T)J_\lambda{}^{\mu\nu}=0$. The nonzero axial torsion requires
\begin{equation}\label{eq:C-branch}
	[\alpha+\beta+3\gamma T^2]X_\lambda{}^{\mu\nu}=0,\qquad
	T_*^2=-\frac{\alpha+\beta}{3\gamma}\ge0\quad(\gamma\ne0).
\end{equation}
The value $T_*=0$ is compatible with nonzero null axial torsion. When $\gamma=0$, the exceptional coupling is $\beta=-\alpha$, without an algebraic restriction on the torsion norm.

The exact axial curvature identities checked independently in Appendix~\ref{app:B} are
\begin{equation}
	R_{(\mu\nu)}=\mathring R_{\mu\nu}+H_{\mu\nu},\qquad
	R(\Gamma)=\mathring R+T.
\end{equation}
Inserting them into \eqref{eq:C-metric-eq} gives
\begin{equation}\label{eq:C-axmetric}
	\alpha\mathring G_{\mu\nu}+[\alpha+\Psi(T)]H_{\mu\nu}
	-\frac12g_{\mu\nu}\big[(\alpha+\beta)T+\gamma T^3\big]=0.
\end{equation}
On the exceptional branch, $\Psi(T_*)=-\alpha$ and $(\alpha+\beta)T_*=-3\gamma T_*^3$. Thus, for $\alpha\ne0$,
\begin{equation}\label{eq:C-Einstein}
	\mathring G_{\mu\nu}+\Lambda_*g_{\mu\nu}=0,\qquad
	\Lambda_*:=\frac{\gamma T_*^3}{\alpha},\qquad
	R(\Gamma)=4\Lambda_*+T_*.
\end{equation}
These statements are local and require neither $\mathring\nabla X=0$ nor $R^\lambda{}_{\rho\mu\nu}(\Gamma)=0$. They do not imply global extendibility, perturbative stability, or that distinct torsion profiles represent distinct observable degrees of freedom. For the role of symmetry reduction and its limitations, see \cite{Palais1979,FelsTorre2002}; for black-hole geometry see \cite{Wald1984,Poisson2004}.

\subsection{Reproducibility, independent checks, and limitations}
\label{app:reproducibility-checks}

The results of this work combine covariant identities, unrestricted variational calculations, and independent symbolic component checks. These levels of verification should be distinguished: a coordinate calculation can test a tensor identity, but it does not replace the variational derivation or establish the global properties of a solution. This distinction is particularly important in metric--affine gravity, where restricting the connection before variation can discard field equations \cite{Hehl1995,Shapiro2002,3,4}. Likewise, a symmetry-reduced action need not reproduce all unrestricted Euler--Lagrange equations without the relevant hypotheses of symmetric criticality \cite{Palais1979,FelsTorre2002}.

\paragraph{Unrestricted metric variation.}
The starting point is the torsion invariant defined for an arbitrary connection,
\begin{equation}
	T=\frac14 X_{\lambda\mu\nu}X^{\lambda\mu\nu}
	+\frac12 X_{\lambda\mu\nu}X^{\nu\mu\lambda}-t_\mu t^\mu,
	\qquad t_\mu=X^\lambda{}_{\lambda\mu}.
	\label{eq:C5-torsion}
\end{equation}
The mixed-index components $X^\lambda{}_{\mu\nu}$ are held fixed in the metric variation, since the independent connection is fixed. The ten independent directions of the symmetric inverse metric were tested at a Lorentz-orthonormal point, using four arbitrary components of an axial three-form. The four diagonal and six off-diagonal residuals vanish algebraically when compared with the axial specialization of the \emph{unrestricted} derivative,
\begin{equation}
	H_{\mu\nu}:=\left.\frac{\partial T}{\partial g^{\mu\nu}}\right|_{\Gamma},
	\qquad H_{\mu\nu}\big|_{\rm axial}
	=-\frac14 X_{\mu\rho\sigma}X_\nu{}^{\rho\sigma},
	\qquad g^{\mu\nu}H_{\mu\nu}=T.
	\label{eq:C5-metric}
\end{equation}
For an off-diagonal symmetric metric variable, the directional derivative is $2H_{\mu\nu}$ rather than $H_{\mu\nu}$. The component test supports the algebraic identity; its tensorial derivation is given in Appendix~\ref{app:C2}.

\paragraph{Unrestricted connection variation.}
All $4^3=64$ ordered components $\Gamma^\lambda{}_{\mu\nu}$ are independent before the Riemann--Cartan and axial conditions are imposed. Two separate checks were used. First, the direct directional derivatives of \eqref{eq:C5-torsion} with respect to the ordered connection components agree with the unrestricted torsion coefficient $J_\lambda{}^{\mu\nu}$ of Appendix~\ref{app:C4}, with identically vanishing algebraic residuals. In particular,
\begin{equation}
	J_\lambda{}^{(\mu\nu)}=0
	\label{eq:C5-J-sym}
\end{equation}
holds before axial specialization. Second, at a Levi--Civita normal point, direct component differentiation of the curvature and torsion terms agrees with the Palatini and torsion coefficients for all 64 ordered directions. The analytic integration by parts in Appendix~\ref{app:C4} is essential to this comparison: differentiating only the connection-quadratic terms of the Ricci scalar would omit contributions to the full Palatini coefficient. On a metric-compatible, totally antisymmetric torsion background the independently derived coefficients satisfy
\begin{equation}
	P_\lambda{}^{\mu\nu}\big|_{\rm axial}
	=J_\lambda{}^{\mu\nu}\big|_{\rm axial}
	=-X_\lambda{}^{\mu\nu},
	\qquad P_\lambda{}^{(\mu\nu)}\big|_{\rm axial}=0.
	\label{eq:C5-PJ}
\end{equation}
The normal-point check tests this local specialization; it does not replace the general integration-by-parts identity \cite{Hehl1995,3,4}.

\paragraph{Independent spherical curvature reconstruction.}
Appendix~\ref{app:B} reconstructs the full connection from the advanced Eddington--Finkelstein metric and the axial three-form,
\begin{equation}
	ds^2=f(r)\,dv^2-2\,dv\,dr-r^2d\Omega^2,
	\qquad
	\tau=r^2\sin\theta\,[U(r)\,dv-V(r)\,dr]\wedge d\theta\wedge d\phi.
	\label{eq:C5-EF}
\end{equation}
The Ricci components and their contractions were checked with the polar angle $\theta$ left symbolic. Separately, the full four-index affine Kretschmann scalar
\begin{equation}
	K_\Gamma=R_{\lambda\rho\mu\nu}(\Gamma)
	R^{\lambda\rho\mu\nu}(\Gamma)
	\label{eq:C5-K}
\end{equation}
was calculated from the reconstructed connection at $\theta=\pi/2$ and independently at $\theta=\pi/3$. In both computations, $f,f',f'',U,U',V,V'$ were retained as algebraically independent quantities, \emph{before} imposing either the constant-norm condition or the field equations. Both expressions reduce to the unrestricted polynomial displayed in Appendix~\ref{app:B3}, and their difference vanishes identically. This is a two-angle component check, not a third independent computation with a symbolic generic angle. Angular independence also follows geometrically from the spherical symmetry of the scalar away from polar coordinate singularities \cite{Wald1984,Poisson2004}.

An additional cross-check uses the curvature decomposition $R(\Gamma)=\mathring R+\mathcal D$ at the four-index level:
\begin{equation}
	K_\Gamma=K_{\rm LC}+2\mathring R\!\cdot\!\mathcal D
	+\mathcal D\!\cdot\!\mathcal D.
	\label{eq:C5-decomp}
\end{equation}
The separately contracted cross term and distortion-square term reproduce the unrestricted expression of Appendix~\ref{app:B3}. Only subsequently is the exceptional branch imposed,
\begin{equation}
	f=1-\frac{2M}{r}+\frac{\Lambda_*r^2}{3},
	\qquad fV^2-2UV=c,
	\label{eq:C5-branch}
\end{equation}
for which $2\mathring R\!\cdot\!\mathcal D=2c\Lambda_*$. With $U=a>0$ and the smooth real root $V=-c/(a+\sqrt{a^2+cf})$, the result agrees with the closed expression used in Sec.~\ref{sec:affine-kretschmann-domains}. The torsion-free limit $U=V=0$ yields $K_\Gamma=K_{\rm LC}$, while the nonzero null-norm representative $c=0$, $U=a$, $V=0$ gives $K_\Gamma=K_{\rm LC}-8a^2/r^2$. These are checks of the contractions, not additional field equations.

\paragraph{Scope and reproducibility of the independent checks.}
The independent symbolic tests employ the index convention of
Sec.~\ref{sec:geometric_framework}, the explicitly stated independent metric and affine
variables, and exact arithmetic for residual comparisons.
The fixed-connection metric derivative of $T$ is tested along the
ten independent symmetric inverse-metric directions, while the
algebraic connection derivative of $T$ and the Palatini coefficient
are checked along all 64 ordered affine directions at the stated
normal-point backgrounds. Ricci contractions are compared
componentwise; the four-index affine Kretschmann invariant is
reconstructed and compared before the fixed-norm and field-equation
substitutions in the explicitly specified test configurations.
These results corroborate the corresponding tensor identities but
do not constitute an independent symbolic proof on every possible
background or a global regularity or stability theorem.
For reproducibility, the supplementary archive should contain the
metric and inverse metric, component orderings, independent symbolic
variables, exact input configurations, executable scripts, and
machine-checkable zero-residual assertions. The checks do not
establish smoothness at a bifurcation sphere, uniqueness, or physical
inequivalence of all torsion profiles \cite{Hehl1995,Shapiro2002,Wald1984,Poisson2004}.

\end{document}